\documentclass[12pt]{article}
\usepackage{amsmath}
\usepackage{graphicx}
\usepackage{booktabs}
\usepackage{amsfonts}
\usepackage{hyperref}
\usepackage{lmodern}
\usepackage{amssymb}
\usepackage{tikz}
\usetikzlibrary{patterns}
\usetikzlibrary{snakes}

\newcommand{\be}{\begin{equation}}
\newcommand{\ee}{\end{equation}}

\newcommand{\bea}{\begin{eqnarray}}
\newcommand{\eea}{\end{eqnarray}}
\newcommand{\beqa}{\begin{eqnarray}}
\newcommand{\eeqa}{\end{eqnarray}}

\newcommand{\nn}{\nonumber}

\def\CC {{\cal C}}
\def\CD {{\cal D}}

\def\CF {{\cal F}}
\def\CG {{\cal G}}

\def\CM {{\cal M}}

\def\CR {{\cal R}}

\begin{document}

\setlength{\baselineskip}{7mm}
\begin{titlepage}
 \begin{flushright}
{\tt NRCPS-HE-01-2020}
\end{flushright}

\begin{center}
{\Large ~\\{\it   Maximally Chaotic Dynamical Systems \\ of Anosov-Kolmogorov \footnote{
Invited talk at the International Bogolyubov Conference "Problems of Theoretical and Mathematical Physics"  at the Steklov Mathematical Institute, Moscow-Dubna,  September 9-13, 2019.  }   
\vspace{1cm}

}

}

\vspace{1cm}

{\sl George Savvidy

\bigskip
\centerline{${}$ \sl Institute of Nuclear and Particle Physics}
\centerline{${}$ \sl NCSR Demokritos, Ag. Paraskevi,  Athens, Greece}
\bigskip

}
\end{center}
\vspace{20pt}

\centerline{{\bf Abstract}}
The maximally chaotic  K-systems are dynamical systems which have nonzero Kolmogorov entropy. On the other hand, the hyperbolic dynamical systems that fulfil the Anosov C-condition have exponential  instability of phase trajectories, mixing of all orders, countable Lebesgue  spectrum and  positive Kolmogorov entropy. The C-condition defines a rich class of maximally chaotic systems which span an open set in the space of all dynamical systems. The interest in Anosov-Kolmogorov C-K systems is associated with the attempts to understand the relaxation phenomena, the foundation of the statistical mechanics, the appearance of turbulence in fluid dynamics, the non-linear dynamics of the Yang-Mills field as well as the dynamical properties of gravitating N-body systems and the Black hole thermodynamics.  In this respect of special interest are C-K systems that are  defined on Reimannian manifolds of negative sectional curvature and on a high-dimensional  tori.  Here we shall review the classical- and quantum-mechanical properties of maximally chaotic dynamical systems,  the application of the C-K  theory to the investigation of the Yang-Mills dynamics and gravitational systems as well as their application in the Monte Carlo method.

\vspace{12pt}

\noindent

\end{titlepage}



\pagestyle{plain}
 
\section{\it Introduction}

It seems natural to define the maximally chaotic dynamical systems as systems that have nonzero Kolmogorov entropy \cite{kolmo,kolmo1}. A large class of maximally chaotic dynamical systems was constructed by Anosov \cite{anosov}.  These are the systems that fulfil the C-condition. The Anosov C-condition leads to the exponential instability of phase trajectories, to the mixing of all orders, countable Lebesgue  spectrum and  positive Kolmogorov entropy.  The uniqueness of the Anosov C-condition lies in the fact that it allows to define a rich class of maximally chaotic systems that span an open set in the space of all dynamical systems.  The examples of maximally chaotic systems were discovered and discussed in the earlier investigations by  Artin, Hadamard, Hedlund, Hopf, Birkhoff   and others \cite{Artin,Hadamard, hedlund, hopf,hopf1,Hopf,anosov1,Gibbs,Birkhoff,krilov,sinai3,turbul,kornfeld,arnoldavez} as well as in more recent  investigations \cite{rokhlin1,leonov,rokhlin,rokhlin2,smale,sinai2,sinai4,margulis,bowen0,bowen,bowen1,gines,Gutzwiller,Savvidy:2018ffh}. Here we shall introduce and discuss the classical- and quantum-mechanical properties of maximally chaotic dynamical systems,  the application of the C-K  theory to the investigation of the  gauge and gravitational systems as well as their application in the Monte Carlo method.  

In recent years the alternative concept of maximally chaotic systems was developed in series of publications \cite{Maldacena:2015waa,Gur-Ari:2015rcq,Cotler:2016fpe,Arefeva:1998,Arefeva:1998,Arefeva:1999,Arefeva:1999frh,Arefeva:2013uta} and references therein. It is based on the analysis of the quantum-mechanical properties of the black holes physics  and on the investigation of the so called out-of-time correlation functions. In general the two- and many-point thermodynamical correlation functions decay exponentially. It was observed that the thermodynamics of black holes exhibits extraordinary property of  fast relaxation and of the exponential growth of the out-of-time correlation functions.  Such chaotic behaviour has come to be referred to as "scrambling," and it has been conjectured that black holes are the fastest scramblers in nature.  The influence of chaos on the time dependent commutator of two observables  can develop no faster than exponentially with the exponent ${2\pi  \over \beta}t = 2\pi T t  $, which is growing  linearly in temperature $1/ \beta =T$ and time $t$.  This maximal linear growth is saturated in gravitational and dual to the gravity systems \cite{Maldacena:2015waa,Gur-Ari:2015rcq,Cotler:2016fpe}.  One of our aims is to calculate out-of-time correlation functions in the case of C-K systems  and to check if their quantum-mechanical correlation functions grow exponentially and if the exponent grows linearly with  temperature. That can help to understand better the concept of maximally  chaotic dynamical systems and their role in thermalisation phenomena.

This review is organised as follows. In the second section we shall discuss the  classification of the dynamical systems (DS)  by the increase of their statistical-chaotic properties \cite{kornfeld,arnoldavez}.   These are ergodic, mixing, n-fold mixing and finally the K-systems, which have mixing of all orders and nonzero Kolmogorov entropy.  This consideration defines the hierarchy of DS by their increasing chaotic/stochastic  properties, with maximally chaotic K-systems on the "top".  The question is: Do the maximally chaotic systems exist? The hyperbolic C-systems introduced by Anosov represent {\it a large class of  K-systems defined on the Riemannian manifolds of negative sectional curvature and on high-dimensional tori}. 

We shall consider the general properties of the C-systems in the third section. From the  C-condition  it follows that C-systems  have very strong instability of their trajectories and, in fact, the instability is as strong as it can be in principle \cite{anosov,anosov1}. The distance between infinitesimally close trajectories increases exponentially and on a closed phase space of the dynamical system 
this leads to the uniform distribution of almost all trajectories over the whole phase space.  The dynamical systems which fulfil the C-condition have very extended and rich  ergodic properties \cite{anosov}.   The C-condition, in most of the  cases, is a sufficient condition for the dynamical system to be a K-system as well. In this sense the C-systems provide extended and rich list of concrete examples of K-systems. The other important property of the C-systems is that in "between" the uniformly distributed trajectories there is  {a countable set of periodic trajectories.   The set of points on the periodic trajectories is everywhere dense in the phase space. The periodic trajectories and uniformly distributed trajectories are filling out the phase space of a C-system in a way very similar to the rational and irrational numbers on the real line. 

The hyperbolic geodesic flow on Riemannian manifolds of negative sectional curvature will be considered in fourth section \cite{anosov,hedlund, hopf,hopf1,Hopf}.  It was proven by Anosov that the geodesic flow on closed Riemannian manifold of negative sectional curvature fulfils the C-condition and  therefore defines a large class of maximally chaotic systems with nonzero Kolmogorov entropy.  This result provides a powerful tool for the investigation of the Hamiltonian systems.  If the time evolution of a classical physical system  under investigation can be reformulated as the geodesic flow on the Riemannian manifold of negative sectional curvature, then all ergodic/chaotic properties of the C-K systems can be ascribed to  that physical system.  The C-K systems have a tendency to approach the equilibrium state with exponential rate which is  proportional to the entropy.  The larger the entropy is, the faster a physical system tends to its equilibrium. 

In the fifth section we shall consider the classical and quantum dynamics of the Yang-Mills  fields \cite{Baseyan,Natalia,Asatrian:1982ga,SavvidyKsystem,Savvidy:1982jk,Savvidy:1984gi,Chirikov,Shchur,Nicolai,Maldacena:2015waa,Gur-Ari:2015rcq,Arefeva:1998,Arefeva:1999,Arefeva:1999frh,Arefeva:2013uta}. In the case of space homogeneous gauge fields the Yang-Mills equations become equivalent to the classical-mechanical system, the Yang-Mills classical mechanics (YMCM), which has  finite degrees of freedom. Using energy and momentum conservation integrals the system can be reduced to a system of lower dimension, and the fundamental question is if the residual system has additional hidden conserved integrals. The evolution of the YMCM can be formulated as the geodesic flow on a Riemannian manifold with the Maupertuis's metric. The investigation of the sectional curvature demonstrates that it is negative on the equipotential surface and generates exponential instability of the trajectories. The numerical integration also confirms this conclusion. The natural question which arrises here is to what extent the classical chaos influences the quantum-mechanical properties of the gauge fields. The corresponding quantum-mechanical system represents and defines  a  quantum-mechanical matrix system \cite{ Savvidy:1982jk,Savvidy:1984gi}.  We shall discuss its spectral properties and the traces of the classical chaos in its quantum-mechanical regime.   

The interesting application of the Anosov C-systems theory was found in the investigation of the relaxation phenomena in stellar systems  like globular clusters and galaxies \cite{body,garry}. Here again  one can use the  Maupertuis's metric in order to reformulate the evolution of N-body system in Newtonian  gravity as a geodesic flow on a Riemannian manifold. Investigation of the sectional curvature allows to estimate the average value of the exponential divergency of the phase trajectories and the relaxation time toward  the Maxwellian distribution of the stars velocities in elliptic galaxies  and  globular clusters \cite{body}. This time is by few orders of magnitude shorter than the Chandrasekchar binary relaxation time \cite{Chandrasekhar,Lang}. The  difference is rooted in the fact that in this approach one can take into account the long-range interaction of stars through their collective contribution into the sectional curvature which defines the relaxation time.    

Of special interest are continuous C-systems which are defined on the two-dimensional surfaces embedded into the hyperbolic Lobachevsky plane of constant negative curvature \cite{Poghosyan:2018efd}.  An example of such system has been defined in a brilliant article published in 1924 by the mathematician Emil Artin \cite{Artin}. The dynamical system is defined on the fundamental region of the Lobachevsky plane that is obtained by the identification of points congruent with respect to the modular group $SL(2,Z)$, a discrete subgroup of the Lobachevsky plane isometries $SL(2,R)$ \cite{Poincare,Poincare1,Fuchs, Ford}. The fundamental region in this case is a hyperbolic triangle, a non-compact region of a finite area. The geodesic trajectories are bounded to propagate on the fundamental hyperbolic triangle and are exponentially unstable.  In classical regime the exponential divergency of the geodesic trajectories resulted into the universal exponential decay of the classical correlation functions \cite{Poghosyan:2018efd,Collet,Pollicot,moore,dolgopyat,chernov}. The  Artin symbolic dynamics, the differential geometry and group-theoretical methods of Gelfand and Fomin \cite{Gelfand} are used to investigate the exponential decay rate of the classical correlation functions in the seventh section \cite{Poghosyan:2018efd}.

There is a great interest in considering quantisation of the hyperbolic dynamical systems and investigation of their quantum-mechanical properties. This subject is very closely related  to  the investigation of quantum mechanics of classically chaotic systems in gravity \cite{Maldacena:2015waa, Cotler:2016fpe,Gur-Ari:2015rcq}.  In the eighth section we shall study the behaviour of the correlation functions of the Artin hyperbolic dynamical system in the quantum-mechanical regime.  In order to investigate the behaviour of the correlation functions in the quantum-mechanical regime it is necessary  to know the spectrum of the system and the corresponding wave functions.  In the case of the modular group the energy spectrum has continuous part, which is originating from the asymptotically free motion inside an infinitely long channel extended in the vertical direction of the fundamental region, as well as infinitely many discrete energy states  corresponding to a bounded motion at the "bottom"  of the fundamental triangle \cite{maass,roeleke,selberg1,selberg2,bump, Faddeev,Faddeev1,hejhal2,winkler,hejhal,hejhal1}. The spectral problem has deep number-theoretical origin  and was partially solved in a series of pioneering articles \cite{maass,roeleke,selberg1,selberg2}.   It was solved partially because the discrete spectrum and the corresponding wave functions are not known analytically.   The general properties of the discrete spectrum have been derived by using Selberg trace formula \cite{selberg1,selberg2,bump, Faddeev,Faddeev1,hejhal2}.  Numerical calculations of the discrete energy levels were performed for many energy states   \cite{winkler,hejhal,hejhal1}.  In the eighth section we shall describe the quantisation of the Artin system and shall review the derivation of the  Maass wave functions describing  the continuous spectrum.

Having in hand the explicit expression of the wave functions one can analyse the quantum-mechanical  behaviour of the correlation functions in order to investigate the traces of the classical chaos in quantum-mechanical regime \cite{Babujian:2018xoy}.  In the ninth  
section we shall consider the correlation functions of the Louiville-like  operators and shall demonstrate  that all two- and four-point correlation functions  decay exponentially with time, with the exponents which depend on temperature.     
Alternatively to the exponential decay of  the correlation functions  the square of the commutator of the Louiville-like  operators  separated in time  grows exponentially \cite{Babujian:2018xoy}. This growth is reminiscent to the local exponential divergency of trajectories of the Artin system when it was considered in the classical regime. The exponential growth  of the commutator does not saturate the condition of maximal growth which was conjectured to be linear in temperature in the case of  the gravitational systems and BH thermodynamics \cite{Maldacena:2015waa,Gur-Ari:2015rcq,Cotler:2016fpe}.   In calculation of the quantum-mechanical correlation functions  a perturbative expansion is used in which the high-mode Bessel's functions are  considered as perturbations.  It has been found that calculations are stable with respect to these perturbations and do not influence the final results.  The reason is that in the integration region of the matrix elements  the high-mode Bessel's functions are exponentially small.  
 
In the tenth section we shall demonstrate that the Riemann zeta function zeros \cite{Riemann} define the position and the widths of the resonances of the quantised Artin hyperbolic system  
\cite{Savvidy:2018ffh}.  A possible relation of the zeta function zeros and quantum-mechanical spectrum was discussed in the passed,  David Hilbert seems to have proposed the idea of finding an eigenvalue   problem whose spectrum contains the zeros of $\zeta(s)$ \cite{Gutzwiller}.  The quantum-mechanical resonances have more complicated pole structure compared to that in the case of a pure discrete spectrum and can be adequately described in terms of the scattering S-matrix theory.  We shall use the S-matrix approach to analyse the scattering phenomenon in quantised Artin  system.  As it was discussed  above, the Artin dynamical system is defined on the fundamental region of the modular group on the Lobachevsky plane. It has a finite area and an infinite extension in the vertical direction that correspond to a cusp.   In quantum-mechanical regime the system can be associated with the narrow infinitely long waveguide stretched out to infinity along the vertical axis  and a cavity resonator attached to it at the bottom.  That suggests a physical interpretation of the Maass automorphic wave function in the form of an incoming plane wave of a given energy entering the resonator and  bouncing back to infinity.  As  the energy of the incoming wave comes close to the eigenmodes of the cavity a pronounced resonance behaviour shows up in the scattering amplitude.  The condition of the absence of incoming waves allows to find the position of the pole singularities \cite{Savvidy:2018ffh}. The  poles of the S-martrix are located in the energy complex plane and are expressed in terms of zeros $u_n$ of the Riemann zeta function $\zeta(\frac{1}{2} - i u_n) =0,~~~~  n=1,2,....$ as
\be
E = E_n - i {\Gamma_n \over 2}= ({u^2_n \over 4} + {3\over 16}) - i {u_n \over 2}, \nn
\ee
where $E_n$ is the energy and $\Gamma_n$ is the width of the n'th resonance. The conclusion is that the energy spectrum is quasi-discrete, consisting of smeared levels of width $\Gamma_n$ \cite{Savvidy:2018ffh}. 

In the last, eleventh, section we shall turn our attention to the investigation of the second class of the C-K systems that is defined on high-dimensional tori \cite{anosov}.  In order that the automorphisms of a torus fulfils  the C-condition it is necessary 
and sufficient that the evolution operator $T$ has no eigenvalues on a unit circle and the determinant  is equal to one. Therefore  $T$ is an automorphism of  the torus  onto itself. All trajectories with rational coordinates, and only they, are periodic trajectories of the automorphisms of a torus.  The entropy of the C-system on a torus is equal to the logarithmic sum of all eigenvalues  that lie outside of the unit circle \cite{anosov,smale,sinai2,margulis,bowen0,bowen,bowen1}:
$
h(T) = \sum_{\vert \lambda_{\beta} \vert > 1} \ln \vert \lambda_{\beta} \vert.
$
It was suggested in 1986 \cite{yer1986a} to use the C-K systems defined on a torus to generate high quality pseudorandom numbers for Monte-Carlo method   \cite{metropolis,neuman,neuman1,sobol,yer1986a,Demchik:2010fd,falcion}.   
Usually  pseudorandom numbers are generated by deterministic recursive rules
\cite{yer1986a,metropolis,neuman,neuman1,sobol}. 
Such rules produce pseudorandom numbers, and it is a great challenge to design 
pseudorandom number generators that produce high quality sequences. 
Although numerous RNGs introduced in the last  decades fulfil most of the 
requirements and are frequently used in simulations, each of them has some 
weak properties that influence the results \cite{pierr} and are less suitable for demanding 
MC simulations \cite{cern}.   The high entropy MIXMAX generator suggested in 
\cite{yer1986a,konstantin,Savvidy:2015jva,Savvidy:2015ida} was implemented into the Geant4/CLHEP and ROOT scientific toolkits at CERN \cite{hepforge,cern,root,geant}.

The Appendix $A$ contains the extended discussion of the C-condition, the definition of the exponentially expanding  and contracting foliations. 

In \cite{anosov} Anosov demonstrated how any C-cascade on a torus can be embedded into a certain  C-flow. The Appendix $B$ describes the Anosov construction that allows to embed a discrete time evolution on a torus  into an evolution that is continuous in time. The embedding was defined by a specific identification of the phase space coordinates  and  by construction of the corresponding  C-flow on a smooth Riemannian manifold of higher dimension. In Anosov construction the C-flow was not a geodesic flow.  Here we were interested in analysing the geodesic flow on the same Riemannian manifold. We present the calculation of the corresponding sectional curvatures  and demonstrate that  the geodesic flow has different dynamics  and  hyperbolic components.   

The Appendix $C$ describes the details of the computer implementation of the torus automorphisms, the computation of the periods of generated random sequences for Monte Carlo simulations. In a typical computer implementation of the torus automorphisms the initial vector will have rational 
components.  If the denominator is taken to be a prime number, then the recursion is realised on extended Galois field $GF[p^N]$   and  allows to find the period of the trajectories in terms of $p$ and the properties of the characteristic polynomial  of the evolution operator.  

The Appendix $D$ presents the derivation of the explicit formulas for the Kolmogorov entropy in case of torus automorphisms. The Appendix $E$ presents the  discussion of the properties of the periodic trajectories and their density distribution as a function of Kolmogorov entropy.

\section{\it  Hierarchy of Dynamical Systems and Kolmogorov Entropy}

In ergodic theory the dynamical systems (DS) are classified by the increase of their statistical-chaotic properties. {\it Ergodic systems} are defined as follows \cite{kornfeld,arnoldavez}.  Let $x=(q,p) $ be a point of the phase space $x \in M $ of the Hamiltonian systems.  The canonical coordinates are denoted as $q =(q_1,...,q_d)$ and $p=(p_1,...,p_d)$ are the conjugate momenta.  The phase space $M$ is equipped with a positive Liouville measure $d\mu(x)= \rho(q,p) dq_1...dq_d dp_1...dp_d$, which is invariant under the Hamiltonian flow.  The operator $T^t x = x_t$ defines the time evolution of the trajectory which was launched from the point  $x $ of the phase space. The ergodicity of the DS takes place if \cite{Birkhoff} \footnote{In what follows we shall be writing $dx$ instead of $d\mu(x)$ in order to compactify the expressions.}:   
\be
\lim_{t \rightarrow \infty} {1\over t} \int_{0}^{t} dt f(T^t x)  = \int_{M} f(x) dx ,
\ee
where $f(x)$ is a function/observable defined on the phase space $M $. So time averages equal to space averages in this case. It follows then that  
\be
\lim_{t \rightarrow \infty} {1\over t} \int_{0}^{t} dt   f(T^t x) g(x) dx  = \int_{M} f(x) dx \int_{M} g(x) dx .
\ee
Consider the function $f = \chi_A$ on the phase space which is equal to one on a set $A \subset M$
and to zero outside, similar function $g = \chi_B$ for a set $B \subset M$,   then
\be
\lim_{t \rightarrow \infty} {1\over t}~ \int_{0}^{t} dt~ \mu[ T^t A \cap  B ]  = \mu[A] \mu[B],
\ee
that is a part of the set $A$ which falls into the set $B$ {\it is in  average proportional} to their measures. The systems with stronger chaotic properties has been defined by Gibbs \cite{Gibbs,kornfeld}. The mixing takes place if for any two sets 
\be\label{mix1}
\lim_{t \rightarrow \infty} \mu[ T^t A \cap  B ]  = \mu[A] \mu[B],
\ee
that is a part of set $A$ which falls into the set $B$ {\it is  proportional} to their measures. Alternatively 
\be\label{mix2}
\lim_{t \rightarrow \infty} \int  f(T^t x) g(x) dx  = \int_{M} f(x) dx \int_{M} g(x) dx,
\ee
which  means  that the two-point correlation function tends to zero: 
\be\label{mix2}
\CD_{t}(f,g) =   
   \lim_{t \rightarrow \infty}  \langle f(T^t x) g(x) \rangle - \langle f(x)\rangle \langle g(x) \rangle =0,
\ee
and is known in physical language as the factorisation property of the two-point correlation functions. 
The n-fold mixing takes place if for any $n$  sets 
\be\label{mixn}
\lim_{t_n,...,t_1 \rightarrow \infty} \mu[ T^{t_n} A_n \cap ....T^{t_2} A_2 \cap T^{t_1} A_1 \cap  B ]  
= \mu[A_n]...   \mu[A_2] \mu[A_1] \mu[B]
\ee
or alternatively 
\be\label{mixnn}
\CD_{t}(f_n,...,f_1, g) =   
   \lim_{t_n,...,t_1 \rightarrow \infty}  \langle f_n(T^{t_n}x).....f_1(T^{t_1}x) g(x) \rangle - \langle f_n(x)\rangle .....\langle f_1(x)\rangle \langle g(x) \rangle =0.
\ee
A  class of dynamical systems which have even stronger chaotic properties was introduced by Kolmogorov in \cite{kolmo,kolmo1}.  These are the DS which have a non-zero  entropy, so called  quasi-regular DS, or simply  K-systems.  In order to define the Kolmogorov entropy let us consider a discrete time evolution operator $T^n x = x_n, ~n=0,1,2,..$.
Let $\alpha = \{A_i\}_{i \in I}$ ( $I$ is finite or countable)  be a  measurable partition of 
the phase space $M$ into the nonintersecting subsets $A_i$ which cover the whole phase space $M$, that is 
\be
\mu(M \setminus \bigcup_{i \in I} A_i)=0,~~~~\mu(  A_i \bigcap   A_j)=0, i \neq j~,
\ee
and define the entropy of the partition $\alpha$ as 
\be
h(\alpha) = - \sum_{i \in I} \mu(A_i) \ln \mu(A_i).
\ee
If two partitions $\alpha_1$ and $\alpha_2$ differ by a set of 
measure zero, then their entropies are equal.  The {\it refinement partition} $\alpha$ 
\be
\alpha = \alpha_1 \vee \alpha_2  \vee ... \vee \alpha_k
\ee
of the 
collection of partitions $\alpha_1,..., \alpha_k$ 
is defined as the intersection of all their composing sets  $A_i$:
\be
\alpha = \big\{ \bigcap_{i \in I} A_i~ \vert ~A_i \in \alpha_i~ for ~all~ i  \big\}.
\ee
The entropy of the partition $\alpha$ with respect to the automorphisms T 
is defined as a limit \cite{kolmo,kolmo1,sinai3,rokhlin1,rokhlin,rokhlin2}:
\be
h(\alpha, T)= \lim_{n \rightarrow \infty} {h(\alpha \vee T \alpha \vee ...\vee T^{n-1} \alpha) \over n},~~~~
n=1,2,...
\ee
This number is equal to the entropy of the refinement 
$
\beta = \alpha \vee T \alpha \vee ...\vee T^{n-1} \alpha 
$
which was generated  during the iteration of the partition $\alpha$ 
by the  automorphism $T$. Finally the entropy of the 
automorphism $T$ is defined as a supremum: 
\be\label{supremum}
h(T) = \sup_{\{ \alpha \}} h(\alpha,T),
\ee
where the supremum is taken over all partitions $\{ \alpha \}$ of  $M$. 
It was proven that the K-systems have mixing of all orders: K-mixing $\supset$ infinite mixing, $\supset$,..n-fold mixing,..$\supset$ mixing $\supset$ ergodicity \cite{kolmo,kolmo1,sinai3,rokhlin1,rokhlin,rokhlin2}.
The calculation of the entropy for a given dynamical system seems extremely 
difficult. The theorem proven by Kolmogorov \cite{kolmo,kolmo1} tells that 
if one finds the so called "generating  partition" $\beta$, then
\be
h(T) = h(\beta,T), 
\ee
meaning  that the supremum in  (\ref{supremum}) is  reached on a generating partition 
$\beta$. In some cases the construction of the generating partition $\beta$
allows an explicit  calculation of the entropy of a given dynamical system \cite{sinai4,gines}. 

In summary, the above consideration allows to define the hierarchy of DSs with their increasing chaotic properties and with the maximally chaotic K-systems on the "top".  The question is: Do maximally chaotic systems exist? The hyperbolic C-systems introduced by Anosov \cite{anosov} represent a large class of  K-systems and have additional ergodic properties.  
We shall consider  the C-systems in the next section. 

\section{\it  Hyperbolic Anosov C-systems} 

In the fundamental work on geodesic flows on closed
Riemannian manifolds $Q^n$ of negative curvature \cite{anosov} Dmitri Anosov
pointed out  that the basic property of the geodesic flow on such manifolds 
is the {\it uniform  instability of all trajectories}, 
which in physical terms means that {\it in  the 
neighbourhood  of every fixed trajectory the trajectories 
behave similarly to the trajectories in the neighbourhood of a saddle point} (see Fig. \ref{fig1}).
In other words, the hyperbolic instability of the 
dynamical system  $ T^t $ which is defined by the equations \footnote{It is 
understood that the phase space manifold   $M$ is equipped by the invariant   
Liouville  measure \cite{anosov}.} 
\be\label{hyperbolic}
\dot{x} = f(x)
\ee
takes place for all solutions $\delta x \equiv \omega$ of the deviation equation  
\be
 \dot{\omega} = {\partial f \over \partial x } \bigg\vert_{x(t)=T^t x} \omega
\ee
in the neighbourhood of each phase trajectory  $x(t)=T^t x$, where $x\in M$. 
The exponential instability of geodesics on Riemannian manifolds of 
constant negative curvature 
has been studied by many authors, beginning with Lobachevsky and Hadamard 
and especially  by  Artin \cite{Artin},  Hedlund  \cite{hedlund}, and Hopf \cite{hopf}.
The concept of exponential instability of a
dynamical system {\it trajectories}  appears to be extremely rich and Anosov suggested 
to elevate it into a fundamental property of a new class of dynamical systems
which he called C-systems\footnote{The letter C is used because these systems 
fulfil the "C condition" (\ref{ccondition})\cite{anosov}. }. 
The brilliant idea to consider dynamical systems 
which have  {\it local and homogeneous  hyperbolic instability of all trajectories }
is appealing to the intuition and  has  very deep  physical content.   
The richness of the concept 
is expressed by the fact  that  the C-systems 
occupy a nonzero volume  in the space of dynamical 
systems \cite{anosov}\footnote{This is in a contrast 
with the integrable systems, where under arbitrary small perturbation $\delta f(x)$ of (\ref{hyperbolic}) the integrability will be destroyed, as it follows  from the KAM theory.} and have a non-zero Kolmogorov entropy.  
\begin{figure}
\centering
\includegraphics[width=7cm]{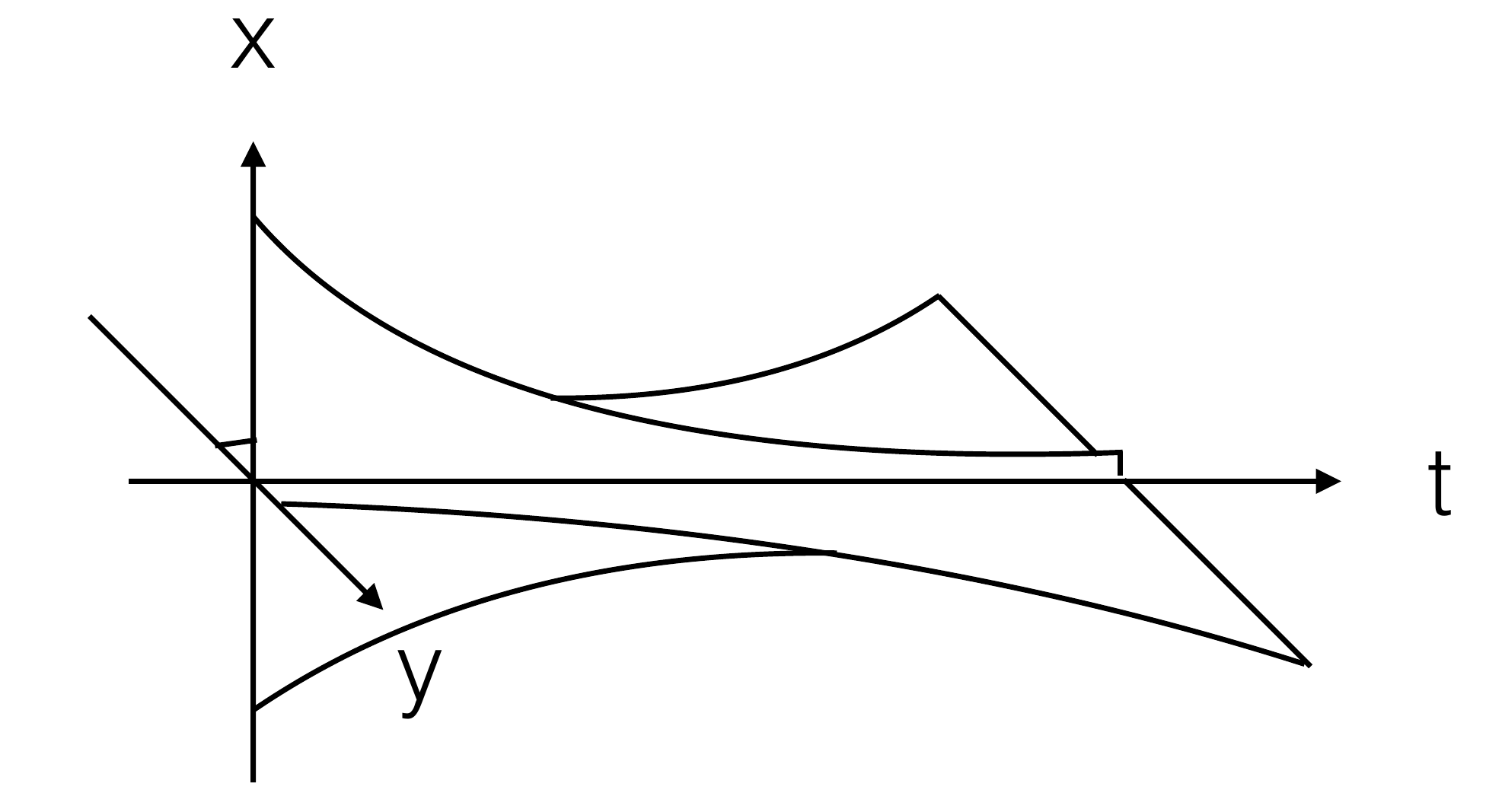}
\centering
\caption{The integral curves in the case of saddle point 
$\dot{x} =-x,~~~\dot{y}=y$ are exponentially contracting and expanding  near the solution $x=y=0$.  
A similar behaviour takes place in the neighbourhood of almost all trajectories 
of the C-system \cite{anosov} as one can get convinced inspecting the solutions (\ref{grows}) and (\ref{decay}) of the Jacobi variation equation (\ref{deviationequations1}) with negative sectional curvature  (\ref{anosovinequality1}). } 
\label{fig1}
\end{figure}
Anosov provided an extended list  of   C-K systems \cite{anosov}. 
The important examples of the C-K systems are:    {\it i) the geodesic flow on the Riemannian manifolds of  variable negative curvature and   ii)   C-cascades - the iterations of the hyperbolic automorphisms of tori}.   

In the forthcoming   sections we shall consider these maximally chaotic systems in details and the application of the C-K systems theory to the investigation of the Yang-Mills dynamics, the N-body problem in gravity and in the Monte Carlo method.  We shall consider the quantum-mechanical properties of the maximally chaotic dynamical systems as  well and in particular  the DSs which are defined on the closed surfaces of constant negative curvature imbedded into the Lobachevsky hyperbolic plane, the Artin DS \cite{Artin}.

\section{\it The Geodesic Flow on  Manifolds of Negative Curvature}

Let us consider the stability of the geodesic flow on a  Riemannian manifold $Q$ with local with coordinates $q^{\alpha} \in Q$ where $\alpha=1,2,....,3 N$.
The functions $q^{\alpha}(s) \in Q$ define a one-parameter  integral curve $\gamma(s)$  
on a Riemannian  manifold $Q$ and the corresponding  velocity vector
\be
u^{\alpha}= {d q^{\alpha} \over d s},~~~~~~\alpha=1,2,....,3 N~.
\ee
The proper time parameter $s$ along the $\gamma(s)$ is equal to  the length, while   
the Riemannian metric on $Q$ is defined as 
$$ds^2 = g_{\alpha\beta} d q^{\alpha} d q^{\beta},$$
and therefore 
\be\label{measure}
g_{\alpha\beta}u^{\alpha} u^{\beta}=1.
\ee
A one-parameter family of deformations assumed to 
form a congruence of world lines 
$
q^{\alpha}(s) \rightarrow q^{\alpha}(s,\upsilon).
$
In order to characterise the infinitesimal deformation of the curve $\gamma(s)$ it is
convenient to define a separation vector 
$
\delta q^{\alpha} = {\partial q^{\alpha} \over \partial \upsilon} d \upsilon,
$
where $\delta q^{\alpha}$ is a separation of points having equal distance from some arbitrary
initial points along two neighbouring curves. 

The  resulting phase space manifold $(q(s),u(s)) \in M$ has a bundle structure 
with the base $q \in Q $ and the  spheres 
$S^{3N-1}$ of unit tangent vectors $u^{\alpha}$ (\ref{measure}) as fibers.
The integral curve $\gamma(s)$ fulfils the geodesic equation 
\be\label{geodesicequation}
{d^2 q^{\alpha} \over d s^{2}}+ \Gamma^{\alpha}_{\beta\gamma} {d  q^{\beta} \over d s }
{d  q^{\gamma} \over d s }=0
\ee
and the {\it relative acceleration} depends only on the Riemann curvature:
\bea\label{deviationequations1}
{D^2 \delta q^{\alpha} \over ds^2}
 = - R^{\alpha}_{\beta\gamma\sigma} u^{\beta} \delta q^{\gamma} u^{\sigma} .
\eea
The above form of the Jacobi equation is difficult  to analyse, first 
of all because it
is written in terms of covariant derivatives $D   u^{\alpha}  = d u^{\alpha}  +
\Gamma^{\alpha}_{\beta\gamma} u^{\beta} d q^{\gamma} $.  And secondly because it is written in terms of separation of points on  trajectories  instead of distance between trajectories.  Following Anosov it is convenient to represent the  Jacobi equation in terms of simple derivatives. The norm of the deviation $\delta q$  has the form
$
 \vert \delta q  \vert^2   \equiv g_{\alpha\beta} \delta q^{\alpha} \delta q^{\beta}
$
and  its  second derivative is
\bea
{d^2\over ds^2} \vert \delta q  \vert^2  &=& 
2 g_{\alpha\beta}   \delta q^{\alpha} 
{D^2  \delta q^{\beta}\over ds^2} + 2 g_{\alpha\beta}  {D  \delta q^{\alpha}\over ds} 
{D  \delta q^{\beta}\over ds}  .
 \eea
Using (\ref{deviationequations1}) we shall get the Anosov form of the Jacobi equation 
\bea
 {d^2\over ds^2} \vert \delta q  \vert^2 &=& 
 -2   R_{\alpha \beta\gamma\lambda}  \delta q^{\alpha} u^{\beta} \delta q^{\gamma} u^{\lambda} +2 \vert  {D  \delta q \over ds} \vert^2=-2 K(q,u,\delta q) ~\vert u \wedge \delta q \vert^2 +2 \vert  \delta u  \vert^2 ~~~~~~~
\eea
where $K(q,u,\delta q)$ is the sectional curvature  in the two-dimensional directions 
defined by the velocity vector 
$u^{\alpha}$ and the deviation vector $\delta q^{\beta}:$
\be\label{sectional0}
 K(q,u, \delta q)  = { R_{\alpha\beta\gamma\sigma} \delta q^{\alpha} 
 u^{\beta} \delta q^{\gamma}  u^{\sigma} \over  \vert u \wedge \delta q  \vert^2 } . 
 \ee
One can decompose the deviation vector $\delta q$ into 
 longitudinal  and transversal components  
 $
 \delta q^{\alpha} = \delta q^{\alpha}_{\perp} + \delta q^{\alpha}_{\parallel},
 $ 
where $\delta q_{\parallel}$  describes a translation 
along the geodesic trajectories and has no physical interest, the transversal 
component $\delta q_{\perp}$ describes a physically relevant distance between original and infinitesimally close  trajectories $\vert u \delta q_{\perp} \vert =0$.  
Such decomposition allows to rederive the Jacobi equation 
only in terms of transversal deviation\footnote{The area spanned 
by the bivector is simplifies  $\vert u \wedge \delta q_{\perp} \vert^2= \vert u \vert^2  \vert   \delta q_{\perp}  \vert^2 - \vert u \delta q_{\perp} \vert^2 = \vert \delta q_{\perp} \vert^2$, because 
$ \vert u \vert^2  =1$ and $\vert u \delta q_{\perp} \vert =0$.}:
\bea\label{transversaldev}
{d^2\over ds^2} \vert \delta q_{\perp}  \vert^2 &=& 
-2 K(q,u,\delta q_{\perp}) ~\vert \delta q_{\perp} \vert^2 
+2 \vert   \delta u_{\perp}  \vert^2. 
\eea
Because the last term is positive definite the following inequality takes place for
{\it relative acceleration}: 
\bea\label{anosovinequality}
{d^2\over ds^2} \vert \delta q_{\perp}  \vert^2 \geq
-2 K(q,u,\delta q_{\perp}) ~\vert \delta q_{\perp} \vert^2. 
\eea
If the sectional curvature is negative and uniformly bound from above by a constant $\kappa$:
\be
K(q,u,\delta q_{\perp}) \leq - \kappa < 0, ~~~\text{where}~~~~~  \kappa = \min_{(q, u,\delta q_{\perp})} \vert K(q,u,\delta q_{\perp}) \vert 
\ee  
then 
\bea\label{anosovinequality1}
{d^2\over ds^2} \vert \delta q_{\perp}  \vert^2 \geq
2 \kappa ~\vert \delta q_{\perp} \vert^2.
\eea
The phase space of solutions of the second-order differential equation is divided into two separate sets  $X_q$ and $Y_q$\footnote{It follows from the variation equation (\ref{anosovinequality1})  and the boundary condition imposed on the deviation $\delta q_{\perp}$ and its first derivative ${d \over ds} \vert \delta q_{\perp}  \vert^2$  that for all $s$ the 
$
{d^2\over ds^2} \vert \delta q_{\perp}  \vert^2  >  0,
$
therefore  $\vert \delta q_{\perp}  \vert^2$ is a convex function and its graph is convex downward. Thus the variation 
equation   has no conjugate points because if $\delta q(s_1)=0$, $\delta q(s_2)=0$ and  $s_1 \neq s_2$
then  $\delta q(s) \equiv 0$ for $s_1 \leq s \leq s_2 $. The sets $X_q$ and $Y_q$ are defined as follows. The set $X_q$ consists of the vectors  $(\delta q(s), {d \delta q(s) \over d s}  )  \rightarrow 0$ when $s \rightarrow + \infty$ and the set $Y_q$ of the vectors $(\delta q(t), {d \delta q(s) \over d s}  )  \rightarrow 0$ when $s \rightarrow - \infty$. If $(\delta q(s), {d \delta q(s) \over d s}  )_{s=0} \in X_q$ then the first derivative is negative 
${d\over ds} \vert \delta q_{\perp}  \vert^2  < 0$  for all $s$.  As well if $(\delta q(s), {d \delta q(s) \over d s}  )_{s=0} \in Y_q$ then 
${d\over ds} \vert \delta q_{\perp}  \vert^2  > 0$  for all $s$ (see Appendix A).}
. The set $Y_q$ consists of the solutions with positive first derivative 
$$
{d \over ds} \vert \delta q_{\perp}(0)  \vert^2 > 0    
$$ 
and exponentially grows with $s \rightarrow + \infty$
\be\label{grows}
\vert \delta q_{\perp}(s) \vert \geq {1\over 2} \vert \delta q_{\perp}(0) \vert e^{\sqrt{2\kappa} s},
\ee
while the set $X_q$ consists of the solutions  with negative first derivative
$$
  {d \over ds} \vert \delta q_{\perp}(0)  \vert^2  < 0 
  $$ 
and decay exponentially with $s \rightarrow + \infty$
\be\label{decay}
\vert \delta q_{\perp}(s) \vert \leq {1\over 2} \vert \delta q_{\perp}(0) \vert e^{-\sqrt{2\kappa} s}.
\ee
This proves that the geodesic flow on closed Riemannian manifold of negative curvature fulfils the C-conditions, is therefore maximally chaotic and tends to equilibrium with exponential rate. We shall define a relaxation time as 
\be\label{relaxationtime}
\tau = 1/\sqrt{2\kappa}~,
\ee
 which is inversely proportional to the Kolmogorov entropy.

\section{\it The Yang-Mills Classical and Quantum Mechanics}

For space-homogeneous gauge fields $\partial_i A^a_k =0$, $i,k=1,2,3$ the Yang-Mills system reduces to a classical mechanical system  with the Hamiltonian of the form \cite{Baseyan,Natalia,SavvidyKsystem,Savvidy:1982jk,Savvidy:1984gi}
\be\label{YMclassical}
H= \sum_{i} {1\over 2} Tr \dot{A}_{i} \dot{A}_{i} + {g^2\over 4} \sum_{i,j}Tr [A_i,A_j]^2,
\ee
where the gauge field $A^a_{i}(t)$ depends only on time,  $i=1,2,3$, the index $a=1,...,N^2-1$ for $SU(N)$ group and in the Hamiltonian gauge $A_0=0$ the Gaussian constraint has the form
\be\label{hamiltonian} 
\CG=[\dot{A}_{i}, A_i]=0.
\ee
It is natural to call this system the Yang-Mills Classical Mechanics (YMCM). It is a mechanical system with $3 \cdot(N^2-1)$ degrees of freedom.  It is important to investigate classical equations of motion of this class of non-Abelian gauge fields, the properties of the separate solutions and of the system as a whole. In particular its integrability versus chaotic properties of the system. The YMCM has a number of conserved integrals: the space and isospin angular momenta  
\be\label{momenta}
m_i = \epsilon_{ijk} A^{a}_{j} \dot{A}^{a}_{k}, ~~~n^a = f^{abc} A^{a}_{i} \dot{A}^{a}_{i},~~~i=1,2,3~~~~a=1,...,N^2-1
\ee
in total $3 + (N^2-1)$ integrals, plus the energy integral (\ref{YMclassical}) (the  $n^a =0$ is the constraint (\ref{hamiltonian}) ). The question is if there exist additional conservation integrals. If the number of integrals coincides with the number of degrees of freedom, then the system is exactly integrable and its trajectories lie on high-dimensional  tori, if there are less integrals, then the trajectories lie on a manifold of a larger dimension, and if there is no conserved integrals at all, then the trajectories will cover the whole phase space. 
\begin{figure}
\centering
\includegraphics[width=5cm]{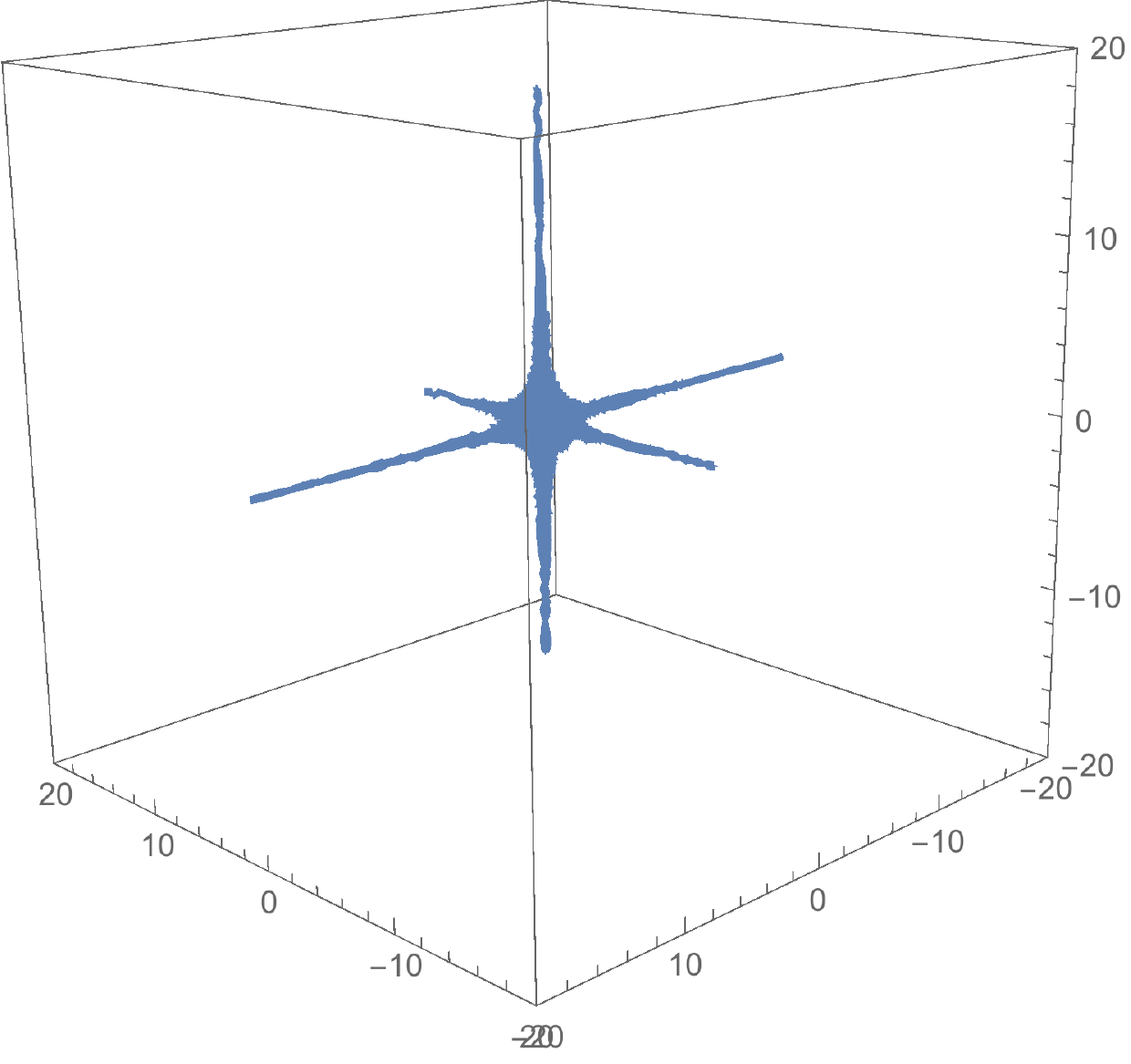}
\centering
\caption{A single trajectory of the YMCM  system integrated over a large time interval. The trajectory scatters on the equipotential surface  $ x^2_1 x^2_2 +x^2_2 x^2_3+x^2_3 x^2_1  = 1 $, densely filling the interior  region and making visible the six channels  of the equipotential surface on which a trajectory scatters.} 
\label{fig17}
\end{figure}

Let us consider in details the case of the $SU(2)$ gauge group by introducing the angular variables   which allow to separate   the angular  motion from the oscillations by using the substitution 
\be 
A = O_1 E O^T_2,
\ee
where $E=(x(t),y(t),z(t))$ is a diagonal matrix and $O_1,O_2$ are orthogonal matrices in Euler angular parametrisation. In this variables the Hamiltonian (\ref{hamiltonian}) will take the form 
\be\label{YMCM}
H_{FS}= {1\over 2}  (\dot{x}^2 + \dot{y}^2 + \dot{z}^2 )+  {g^2\over 2}( x^2y^2 +y^2z^2 +z^2x^2 ) + T_{YM},
\ee
where $T_{YM}$ is the rotational kinetic energy of the the  Yang-Mills "top" spinning in space and isospace. The question is whether the YMCM system (\ref{YMCM}) is an integrable system or not \cite{Natalia,SavvidyKsystem,Savvidy:1982jk}.  The general behaviour of the colour amplitudes  $(x(t),y(t),z(t))$  in time is characterised by rapid oscillations, decrease in some colour amplitudes and growth in others, colour "beats" \cite{Natalia} Fig.\ref{fig17}. The strong instability of the trajectories with respect to small variations of the initial conditions in the phase space led  to the conclusion that the system is stochastic and non-integrable. The search of the conserved integrals of the form $F(p_x,p_y,p_z,x,y,z)$ fulfilling the equation $\{ F, H_{FS}\} =0$ also confirms their absence, except the Hamiltonian itself. 
The evolution of the YMCM can be formulated as the geodesic flow on a Riemannian manifold with the Maupertuis's metric (see the details in the next section). The investigation of the sectional curvature demonstrates that it is negative on the equipotential surface and generates exponential instability of the trajectories Fig.\ref{fig17}.  The solutions of an YMCM system in an arbitrary coordinate system (after Lorentz boost) are nonlinear plane waves $A^a_{\mu}(k\cdot x)$ with a nonzero square of the wave vector $k^2=\mu^2$ \cite{Baseyan} chaotically oscillating in space-time.

The natural question which arrises here is to what extent the classical chaos influences 
the quantum-mechanical properties of the gauge fields. The significance of the answer to this question consists in the following. In field theory, e.g. in QED, the electromagnetic field is represented in the form of a set of harmonic oscillators whose quantum-mechanical properties (as of an integrable system) are well known, and the interaction between them is taken into account by perturbation theory. Such an approach excellently describes the experimental situation. In QCD the state of things is quite different. The properties of the YMCM as of a C-K-system, cannot be established to any finite order of the perturbation theory. Therefore to understand QCD it seems important to investigate the quantum-mechanical properties of the systems which in the classical limit are maximally chaotic.

The natural question arising now is what quantum-mechanical properties does the system with the Hamiltonian (\ref{YMclassical})  possess if in the classical limit $\hbar \rightarrow 0$ it is maximally chaotic. What is the structure of the energy spectrum and of the wave functions of quantised gauge system, if in the classical limit it is maximally chaotic. The Schr\"odinger equation for the gauge field theory in the $A_0 =0$ gauge has the following form; 
\be\label{fulleq}
{1\over 2} \int d^3 x [ - {\delta^2 \over \delta A^a_i \delta A^a_i } + H^a_i H^a_i ] \Psi[A]=  E \Psi[A],
\ee
where $H^a_i = {1\over 2} \epsilon_{ijk} G^a_{jk}(A)$ and the constraint equations are:
\be\label{fulleq1}
[\delta^{ab} \partial_i + g f^{acb} A^c_i] {\delta \over \delta A^b_i  } \Psi[A]=0.
\ee
In case of space-homogeneous fields the equations will reduce to the Yang-Mills quantum-mechanical system (YMQM) with finite degrees of freedom which defines a special class of quantum-mechanical matrix  models \cite{ Savvidy:1982jk,Savvidy:1984gi}.   At zero angular momentum  $\hat{m}_i = 0$ (\ref{momenta}) the YMQM Schr\"odinger equation takes the form   (equations (20) and (21) in  \cite{Savvidy:1984gi})
\be
\Big\{-{1\over 2} D^{-1} \partial_i D \partial_i + {g^2\over 2}(x^2_1 x^2_2 +x^2_2 x^2_3+x^2_3 x^2_1) \Big\}\Psi = E \Psi ,
\ee
where $D(x) = \vert (x^2_1 - x^2_2)(x^2_2 - x^2_3)(x^2_3 - x^2_1)\vert$. Using the substitution 
\be
\Psi(x) = {1\over \sqrt{D(x)}}~ \Phi(x)
\ee
and the fact that the $D(x)$ is a harmonic function $ \partial^2_i D(x)=0$ the equation can be reduced to the form 
\beqa
 -{1\over 2} \partial^2_i  \Phi +  {1\over 2} \sum_{i<j} \Big( {1\over (x_i-x_j)^2} +{1\over (x_i+x_j)^2}   + g^2  x^2_i x^2_j\Big) \Phi=   E \Phi .
\eeqa
The analytical investigation of this Schr\"odinger equation is a challenging problem because the equation cannot be solved by separation of variables as far as all canonical symmetries are already extracted and the residual system possess no continuous symmetries. Nevertheless some important properties of the energy spectrum can be  established by calculating the volume of the classical phase space defined by the condition $H(p,q) \leq E$. It follows that classical phase space volume is finite and the energy spectrum of YMQM system is discrete  \cite{ Savvidy:1982jk,Savvidy:1984gi}. Typically the classically chaotic systems have no degeneracy of the energy spectrum, the energy levels "repulse" from each other similar to the distribution of the eigenvalues of the matrices with randomly distributed elements \cite{Wigner,Mehta,Dyson}.

In the next section we shall consider the $N$-body problem in classical Newtonian gravity 
analysing the geodesic flow on a Riemannian manifold equipped  with the Maupertuis  metric. 

\section{\it Collective Relaxation of Stellar Systems}

The $N$-body problem in Newtonian gravity can be formulated as a geodesic flow on Riemannian manifold with the conformal Maupertuis  metric  \cite{body}
\be\label{metricM}
ds^2 = (E-U) d \rho^2 = W \sum^{3N}_{\alpha=1} (d q^{\alpha})^2,~~~~~
U=-G \sum_{a < b}{M_a M_b \over \vert \vec{r}_a - \vec{r}_a \vert },
\ee
where $W=E-U$ and $\{q^{\alpha}\}$ are the coordinates of the stars: 
\be
\{q^{\alpha}\} = \{  M^{1/2}_1 \vec{r}_1,.....,M^{1/2}_N \vec{r}_N      \},~~~~\alpha =1,...,3N.
\ee
The equation of the geodesics (\ref{geodesicequation}) on Riemannian manifold  with the metric 
$g_{\alpha\beta} = W \delta_{\alpha\beta}$ in (\ref{metricM}) has the form 
\be
{d^2 q^{\alpha} \over ds^2 } + {1\over 2W}\Big(   2 {\partial W \over \partial q^{\gamma}}     
{d q^{\gamma} \over ds} {d q^{\alpha} \over ds}- g^{\alpha\gamma}  {\partial W \over \partial q^{\gamma}}    g_{\beta\delta} {d q^{\beta} \over ds} {d q^{\delta} \over ds}     \Big) =0
\ee
and coincides  with the classical N-body equations  when the proper time interval $ds$ is replaced by the time interval $dt$  of the form $ds = \sqrt{2} W dt$. The Riemann curvature  in (\ref{deviationequations1})
for the Maupertuis  metric  has the form 
\bea
R_{\alpha\beta\gamma\delta} &=& {1\over 2W} [ W_{\beta\gamma} g_{\alpha\delta}- 
W_{\alpha\gamma} g_{\beta\delta} - W_{\beta\delta } g_{\alpha\gamma} +W_{\alpha\delta } g_{\beta\gamma}] -\nn\\
&-& {3\over 4 W^2} [ W_{\beta} W_{\gamma} g_{\alpha\delta} - 
W_{\alpha} W_{\gamma} g_{\beta\delta} - W_{\beta} W_{\delta } g_{\alpha\gamma} +W_{\alpha} W_{\delta } g_{\beta\gamma}] + \nn\\
&+&{1\over 4 W^2} [ g_{\beta \gamma} g_{\alpha\delta} - 
g_{\alpha \gamma} g_{\beta\delta} ] W_{\sigma}W^{\sigma},
\eea
where $W_{\alpha} = \partial W/ \partial q^{\alpha}$ , $W_{\alpha\beta} = \partial W/ \partial q^{\alpha}\partial q^{\beta}$ and the scalar curvature is 
\bea\label{scalarcurve}
R = 3N(3N -1) \Big[- {\triangle W \over 3N W^2} -({1\over 4} - {1\over 2N}){(\nabla W)^2 \over W^3} \Big]
\eea
and $\triangle W  = \partial^2 W/ \partial q^{\alpha} \partial q^{\alpha}$,~  $\nabla W  = (\partial W/ \partial q^{\alpha})( \partial W/ \partial q^{\alpha})$.
Let us now calculate the sectional curvature (\ref{sectional0}):
\bea
R_{\alpha\beta\gamma\sigma} \delta q^{\alpha} 
 u^{\beta} \delta q^{\gamma}  u^{\sigma}&=& {1\over 2W} [~ 2 \vert u W^{''} \delta q \vert \vert u \delta q \vert - 
\vert  \delta q  W^{''}  \delta q \vert  \vert u u \vert - \vert  u  W^{''} u \vert  \vert \delta q \delta q \vert~] -\nn\\
&-& {3\over 4 W^2} [~ 2 \vert u W^{'}\vert  \vert  W^{'} \delta q \vert \vert u \delta q \vert - 
\vert  \delta q  W^{'}  \vert   \vert   W^{'}  \delta q \vert   \vert u u \vert  -  \vert  u  W^{'}  \vert   \vert    W^{'}  u \vert   \vert \delta q \delta q \vert ~ ] + \nn\\
&+&{1\over 4 W^2} [ ~ \vert u \delta q \vert^2- 
\vert u u \vert  \vert \delta q \delta q \vert ~ ] ~\vert W^{'}W^{'} \vert. 
\eea
For the normal deviation $q_{\perp}$   we have $\vert u \delta q_{\perp} \vert =0 $ and taking into account that the velocity is normalised to unity $\vert u u \vert  =1$   we have 
\bea
R_{\alpha\beta\gamma\sigma} \delta q^{\alpha}_{\perp} 
 u^{\beta} \delta q^{\gamma}_{\perp}  u^{\sigma}&= &   \Big(~ {3\over 4 W^2} \vert  u  W^{'}  \vert^2     ~   
-{1\over 4 W^2} \vert W^{'}W^{'} \vert  -{1\over 2W}  \vert u W^{''} u \vert ~\Big)  \vert \delta q_{\perp}   \vert^2   \nn\\
&-& {1\over 2W}   ~    
\vert  \delta q_{\perp}  W^{''}  \delta q_{\perp} \vert   +{3\over 4 W^2} ~   
\vert  \delta q_{\perp}  W^{'}  \vert^2 .
\eea
Using the average value of the  velocity and deviation  taken in the form 
\be
\overline{u^{\alpha} u^{\beta}} = {1\over 3N} g^{\alpha\beta} \vert u u \vert,~~~~~\overline{\delta q_{\perp}^{\alpha} \delta q_{\perp}^{\beta}} = {1\over 3N} g^{\alpha\beta}  \vert \delta q_{\perp}   \vert^2
\ee
we shall get the following expression:
\bea\label{sectional2}
R_{\alpha\beta\gamma\sigma} \delta q^{\alpha}_{\perp} 
 u^{\beta} \delta q^{\gamma}_{\perp}  u^{\sigma}= \Big[-{1\over 3N} {\triangle W \over  W^2} -({1\over 4} - {1\over 2N}){(\nabla W)^2 \over W^3}\Big]  \vert \delta q_{\perp}   \vert^2 ,
\eea
which is proportional to the scalar curvature (\ref{scalarcurve}). 
As the number of stars in  galaxies is very large, $N \gg 1$,  we shall get that the dominant term in sectional curvature (\ref{sectional2}) is negative: 
\be
K(q,u, \delta q)  = { R_{\alpha\beta\gamma\sigma} \delta q^{\alpha} 
 u^{\beta} \delta q^{\gamma}  u^{\sigma} \over  \vert \delta q_{\perp}   \vert^2}=   - {1\over 4}  {(\nabla W)^2 \over W^3}  < 0.
\ee
Finally the deviation equation (\ref{anosovinequality}) will take the form 
\bea\label{anosovinequality3}
{d^2\over dt^2} \vert \delta q_{\perp}  \vert^2 \geq
  {(\nabla W)^2 \over W}  ~\vert \delta q_{\perp} \vert^2,
\eea
where we used the relation $ds =   \sqrt{2} W dt$. Thus the relaxation time can be defined as 
\be\label{relaxationtime1}
\tau = \sqrt{{W \over (\nabla W)^2 }}.
\ee
Now one can estimate the relaxation time of the elliptic galaxies and globular clusters
\be
\tau \approx 10^8 yr \Big({v \over 10 km/s}\Big) \Big( {n \over 1 pc^{-3}}\Big)^{-2/3} \Big( { M \over M_{\odot}  }\Big)^{-1}
\ee
by substituting the corresponding mean values for the velocities $v$, densities $n$ and masses of the stars $M$ \cite{body}. This time is by few orders of magnitude shorter than the Chandrasekchar binary relaxation time \cite{Chandrasekhar,garry,Lang}. 

\section{\it  Correlation Functions of Classical  Artin  System}

Of special interest are continuous C-systems which are defined on closed surfaces on the Lobachevsky plane of constant negative curvature. An example of such system has been defined in a brilliant article published in 1924 by the mathematician Emil Artin. The dynamical system is defined on the fundamental region of the Lobachevsky plane which is obtained by the identification of points congruent with respect to the modular group $SL(2,Z)$, a discrete subgroup of the Lobachevsky plane isometries $SL(2,R)$. The fundamental region in this case is a hyperbolic triangle. The geodesic trajectories  are bounded to propagate on the fundamental hyperbolic triangle.  The area  of the fundamental region is finite  and gets a topology of sphere by  "gluing" the opposite edges of the triangle as it is shown on Fig.\ref{fig5} and Fig.\ref{fig11}.   The  Artin symbolic dynamics, the differential geometry and group-theoretical methods of Gelfand and Fomin will be used to investigate the decay rate of the classical and quantum mechanical correlation functions.  The following three sections are devoted to the Artin system and are based on the results published in the articles \cite{Poghosyan:2018efd,Babujian:2018xoy,Savvidy:2018ffh}.

Let us start with the Poincare model of
the Lobachevsky plane, i.e. the upper half of the complex plane:
$H$=$\{z \in \mathbb{C}$, $\Im z >0\}$ supplied with the metric (we
set $z=x+i y$) 
\bea 
ds^2=\frac{dx^2+dy^2}{ y^2}\, 
\label{metric_hp}
\eea 
with the Ricci scalar $R=-2 $. Isometries 
of this space are given by $SL(2,\mathbb{R})$ transformations. 
The $SL(2,\mathbb{R})$ matrix ($a$,$b$,$c$,$d$ are real and $ad-bc=1$ )
\[
g=\left(
\begin{array}{cc}
a&b\\c&d
\end{array}
\right)
\] 
acts on a point $z$ by linear fractional substitutions
$
z\rightarrow \frac{az+b}{cz+d}~.
$
Note also that $g$ and $-g$ give the same transformation, hence 
the effective group is $SL(2,\mathbb{R})/\mathbb{Z}_2$.
We'll be interested in the space of orbits of a discrete subgroup
$\Gamma \subset SL(2,\mathbb{R})$ in $H$. Our main example will be 
the modular group $\Gamma=SL(2,\mathbb{Z})$. A nice choice of the 
fundamental region $\CF$ of $SL(2,\mathbb{Z})$ is displayed in Fig.\ref{fig1}.
The fundamental region $\CF$ of the  modular group consisting of those  points between the lines
$x=-\frac{1}{2}$ and $x=+\frac{1}{2}$ which lie outside the unit circle in  Fig.\ref{fig1}.
The modular triangle $\CF$ has two 
equal angles  $\alpha = \beta = \frac{\pi }{3}$ and the third one is equal to zero, $\gamma=0$, 
thus $\alpha + \beta + \gamma = 2 \pi /3 < \pi$.
The area  of the fundamental region is finite and equals to $\frac{\pi }{3}$ and gets a topology of sphere
by  "gluing" the opposite edges of the triangle. The invariant area element on the Lobachevsky plane is proportional to the square root of the determinant  of the metric (\ref{metric_hp}):
\be\label{me}
d \mu(z)= {dx dy \over y^2} \,.
\ee
Thus
$
\text{area}(\CF)=\int _{-\frac{1}{2}}^{\frac{1}{2}} dx
\int _{\sqrt{1-x^2}}^{\infty }\frac{dy}{y^2}  =\frac{\pi }{3}\,.
$
Following the Artin construction let us consider the model of the Lobachevsky plane realised 
in the upper half-plane $y>0$ of the complex plane  $z=x+iy \in \CC$
with  the Poincar\'e metric which is given by the line element (\ref{metric_hp}).

The Lobachevsky plane is a surface of a constant negative curvature, because its  curvature is equal to $R=g^{ik}R_{ik}= -2$  and it is twice the Gaussian curvature $K=-1$.
This  metric has two well known properties: 1) it is invariant with respect to all linear substitutions which form the group $g \in G$ of isometries of the Lobachevsky plane\footnote{$G$ is a subgroup of all M\"obius transformations. }:
 \beqa\label{real_frac_trans}
w= g \cdot z \equiv \left(
\begin{array}{cc}
a  & b   \\
c   & d  
\end{array} \right) \cdot z  \equiv \frac{a z +b}{c z +d},  
\eeqa
where $a, b,  c, d $ are {\it real coefficients of the matrix} $g $ and the determinant 
of $g$ is positive, $ a d -  b c  > 0 $.   The geodesic lines are either semi-circles orthogonal to the real axis  or rays perpendicular to the real axis. The equation for the geodesic lines on a curved surface has the form  (\ref{geodesicequation}),  where  the  Christoffer symbols  are evaluated for the metric (\ref{metric_hp}). The geodesic equations  take the form 
\beqa
&&\frac{d^2 x}{d t^2}-\frac{2}{ y} \frac{d x}{d t}\frac{d y}{d t}=0\,, ~~~~~
 \frac{d^2 y}{d t^2}+\frac{1}{ y}\left(\frac{d x}{d t}\right)^2-
\frac{1}{ y}\left(\frac{d y}{d t}\right)^2=0, \nn
\eeqa
and they have two solutions:
\beqa\label{traject}
& x(t)-x_0=r \tanh \left(t \right),~~~~  y(t)=\frac{r}{\cosh\left(t \right)}~~~&\leftarrow
\text{orthogonal semi-circles}~ \nn\\
& x(t)=x_0,~~~~~~~~~~~~~~~~~~~~~~  y(t)=e^t ~~~ &\leftarrow
\text{perpendicular rays }~.
\eeqa
Here $x_0 \in (-\infty, +\infty), t \in (-\infty, +\infty)$ and $r \in (0, \infty)$.
By substituting each of the above solutions into the metric (\ref{metric_hp}) one 
can get convinced that a point on the geodesics curve moves with a unit velocity (\ref{measure})
\be\label{unit}
{ds \over dt } =1.
\ee 
In order to construct a compact surface $\CF$ on the  Lobachevsky plane, one can identify all points in the upper half of the plane which are related to each other by the substitution (\ref{real_frac_trans})  with the integer coefficients and a unit determinant. These transformations form  a modular group $ SL(2,\mathbb{Z})$. Thus two points $z$ and $w$  are "identical"
if:
\begin{equation}\label{modular}
w =\frac{mz+n}{pz+q},~~~~
d=\left(
\begin{array}{cc}
m & n \\
p & q \\
\end{array} \right), ~~~~d \in SL(2,\mathbb{Z}) 
\end{equation}
with  integers $m$, $n$, $p$, $q$   constrained by the condition $mq-pn=1$.  The  $SL(2,\mathbb{Z})$  is 
the discrete subgroup of the isometry transformations $SL(2,\mathbb{R})$ of (\ref{real_frac_trans})\footnote{The modular group $SL(2,\mathbb{Z})$ serves as an example of the Fuchsian group \cite{Poincare,Fuchs}. Recall that Fuchsian groups are discrete subgroups of the group  of all isometry transformations $SL(2,\mathbb{R})$ of 
(\ref{real_frac_trans}). The Fuchsian group allows to tessellate the hyperbolic plane with regular polygons as faces, one of which can play the role of the fundamental region.}. 
The identification creates a regular tessellation of the  Lobachevsky plane by congruent hyperbolic triangles in Fig. \ref{fig5}.   The Lobachevsky plane is covered by the infinite-order triangular tiling. 
One of these triangles can be chosen as a fundamental region. That fundamental region $\CF$ of the above modular group (\ref{modular}) is the
well known "modular triangle", consisting of those  points between the lines
$x=-\frac{1}{2}$ and $x=+\frac{1}{2}$ which lie outside the unit circle in  Fig. \ref{fig5}.
The area  of the fundamental region is finite and equals to $\frac{\pi }{3}$. 
Inside the modular triangle $\CF$ there is exactly one representative
among all equivalent points of the Lobachevsky plane with the exception 
of the points on the triangle edges which are opposite to each other. 
These points can be identified in order to form a {\it closed surface} $\bar{\CF}$
by  "gluing" the opposite edges of the modular triangle together.
On  Fig. \ref{fig5} one can see the pairs of points on the edges of the triangle  
which are identified. Now one can consider the behaviour of the geodesic trajectories defined on the  surface $\bar{\CF}$ of constant negative curvature. 
\begin{figure}
\centering
\begin{tikzpicture}[scale=2]
\clip (-4.3,-0.4) rectangle (4.5,2.8);
\draw[step=.5cm,style=help lines,dotted] (-3.4,-1.4) grid (3.4,2.6);
\draw[,->] (-2.3,0) -- (2.4,0); \draw[->] (0,0) -- (0,2.3);
\foreach \x in {-2,-1.5,-1,-0.5,0,0.5,1,1.5,2}
\draw (\x cm,1pt) -- (\x cm,-1pt) node[anchor=north] {$\x$};
\foreach \y in {1}
\draw (1pt,\y cm) -- (-1pt,\y cm) node[anchor=east] {$\y$};
\draw (0,0) arc (0:180:1cm);
\draw (1,0) arc (0:180:1cm);
\draw (2,0) arc (0:180:1cm);
\draw (-1,0) arc (0:95:1cm);
\draw (1,0) arc (0:-95:-1cm);
\draw (1.5,0)--(1.5,2.1);
\draw (0.5,0)--(0.5,0.86602540378);
\draw (-0.5,0)--(-0.5,0.86602540378);
\draw (-1.5,0)--(-1.5,2.1);
\draw[ ultra thick, blue] (0.5,0.86602540378)--(0.5,2.1);
\draw[ ultra thick, blue] (-0.5,0.86602540378)--(-0.5,2.1);
\draw[ultra thick, blue] (0.5,0.86602540378) arc (60:120:1cm);
\draw (-0.25,1.8) node{ $\CF$};
\draw (0,2.4) node{ $\CD$};
\draw [thin,gray](2.35,0) arc (0:180:1.5cm);
\draw[fill] (0.1,1.29903810568) circle [radius=0.02];
\draw [->,] ((0.1,1.29903810568) to ((0.2,1.36);
\draw (0.21,1.15) node{ $(x,y)$};
\draw (0.15,1.45) node{ $\vec{v}$};
\draw (0,2.4) node{ $\CD$};
\draw (-0.6,0.7) node{ $A$};
\draw (0.4,0.7) node{ $B$};
\draw (-0.1,0.85) node{ $C$};
\draw[fill] (-0.5,0.86602540378) circle [radius=0.02];
\draw[fill] (0,1) circle [radius=0.02];
\draw[fill] (0.5,0.86602540378) circle [radius=0.02];
\draw[fill] (-0.5,1.7) circle [radius=0.025];
\draw[fill] (0.5,1.7) circle [radius=0.025];
\draw (-0.6,1.7) node{ $a$};
\draw (0.6,1.7) node{ $a$};
\draw[fill] ( -0.4,0.915) circle [radius=0.025];
\draw[fill] ( 0.4,0.915) circle [radius=0.025];
\draw (-0.35,0.85) node{ $b$};
\draw (0.3,0.85) node{ $b$};
\draw [<-,thin,  teal,dashed] (1.3,1.43) to [out=90,in=120] (2,1.6) ;
\draw  [thin,  teal] (2.1,1.6) node{ $K$};
\end{tikzpicture}
\caption{The  fundamental region $\CF$ is the hyperbolic triangle $ABD$,  the vertex D is at infinity of the $y$ axis. The edges of the triangle are the arc $AB$,  the rays $AD$   and $BD$. The points on the edges $AD$ and $BD$ and the points of the arks $AC$ with $CB$ 
should be identified by the transformations $w=z+1$ and $w = -1/z$ in order to form the Artin {\it  surface} $\bar{\CF}$
by  "gluing" the opposite edges of the modular triangle together Fig.\ref{scattering}.  The modular transformations (\ref{modular}) of the fundamental  region $\CF$ create a regular tessellation of the  whole Lobachevsky plane by congruent hyperbolic triangles. 
 K is the geodesic trajectory passing through the point ($x,y$) of $\CF$ in the $\vec{v}$ direction.}
\label{fig5} 
\end{figure}
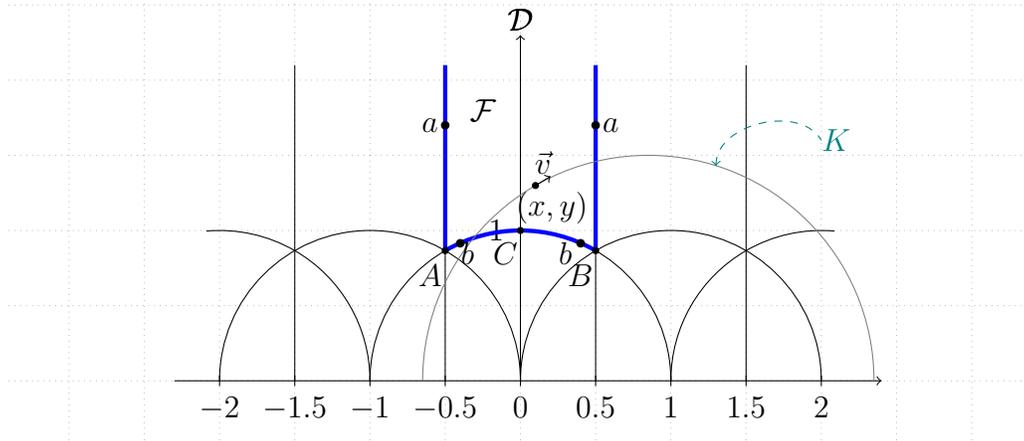
Let us consider an arbitrary point $(x,y) \in \CF$ and the 
velocity vector $\vec{v} = (\cos \theta, \sin \theta)$. These are the coordinates of the 
phase space $(x,y,\theta) \in \CM$, and they uniquely determine  the geodesic trajectory  as 
the orthogonal circle $K$ in the whole Lobachevsky plane.  As this trajectory "hits" the edges 
 of the fundamental region $\CF$ 
and goes outside of it,  one should apply the modular transformation (\ref{modular}) to that parts of the circle $K$ which lie  outside of $\CF$ in order to return them back to the $\CF$. 
That algorithm will define the whole trajectory on $\bar{\CF}$ for $t \in (-\infty, +\infty)$.

Let us describe the time evolution of the physical observables $\{f(x,y, \theta)\}$ which are defined on the phase space  $(x,y,\theta) \in \CM$, where $z =x+iy \in \bar{\CF}$ and $\theta \in S^1$ is a
direction of a unit velocity vector.   For that one should know a  time evolution of geodesics on the phase space $\CM$. The simplest motion on  the ray  $CD$ in Fig.\ref{fig5},  
  is given by  the solution (\ref{traject})
$x(t)=0,~~  y(t)=e^{t} $ and can be represented as a group transformation (\ref{real_frac_trans}):
\beqa\label{transformation1}
  z_1(t) =g_{1}(t) \cdot i  = \left(
\begin{array}{cc}
e^{t/2} & 0 \\
0 & e^{-t/2} \\
\end{array} \right) \cdot i  = i e^{t}~,~~~  t \in (-\infty, +\infty).
\eeqa
The other motion on the circle of a unit radius, the arc $ACB$ on  Fig.\ref{fig5}, is given by the transformation 
\beqa\label{transformation2}
  z_2(t) =g_{2}(t) \cdot i  = \left(
\begin{array}{cc}
\cosh(t/2) & \sinh(t/2) \\
\sinh(t/2) & \cosh(t/2) \\
\end{array} \right) \cdot i  = {i \cosh(t/2) +  \sinh(t/2) \over i \sinh(t/2) + \cosh(t/2)} .
\eeqa
Because the isometry group $SL(2,\mathbb{R})$ acts transitively on the Lobachevsky plane, any  geodesic 
can be  mapped into any other  geodesic through the action of the  group element $g \in SL(2,\mathbb{R}) $ (\ref{real_frac_trans}), thus 
the  generic trajectory can be represented  in the following form: 
\be\label{transformation3}
z(t)= g g_{1}(t) \cdot i = \begin{pmatrix}
 a e^{t/2}  & b e^{-t/2} \\
  c e^{t/2} & d e^{-t/2}
 \end{pmatrix} \cdot  i , ~~~z(t)= { i a e^{ t} + b  \over i c e^{ t} + d }~.
\ee
This provides a convenient description of the time evolution of the geodesic flow on the whole Lobachevsky plane with a unit velocity vector (\ref{unit}). In order to project this motion into  the
closed surface  $\bar{\CF}$ one should identify the group elements $g \in SL(2,\mathbb{R})$ which are 
 connected by the modular transformations $SL(2,\mathbb{Z})$. For that one can use  
a parametrisation of the group elements  $g \in SL(2,\mathbb{R})$ defined in \cite{Gelfand}. Any element $g$ can be defined by the parameters   $(\tau, \omega_2)$
\be\label{phasespace}
(\tau, \omega_2), ~~~\tau \in \CF ,~~~~\omega_2 = {e^{i \theta} \over  \sqrt{ y}} ,~~~
0 \leq \theta \leq 2 \pi,
\ee
where $\tau=x+iy$ are the coordinates in the fundamental region $\CF$ and the 
angle $\theta $ defines the direction of the unit velocity vector $\vec{v} = (\cos \theta, \sin \theta)$  at the point $(x,y)$ (see Fig.\ref{fig5}).   The functions $\{f(x,y, \theta)\}$ on the phase space $  \CM$ can be written 
as depending on  ($ \tau, \omega_2$)  and the invariance of  the functions  
with respect to the modular transformations  (\ref{modular}) takes the form\footnote{This defines the automorphic functions, the generalisation of the trigonometric, hyperbolic, elliptic  and other periodic functions \cite{Poincare,Ford}.}
\be\label{periodic}
f( \tau',\omega'_2) = f\Big( { m \tau+ n \over p\tau +q } ,(p \tau +q) \omega_2 \Big) = f (\tau,\omega_2).
\ee
The evolution of the  function $\{f(\tau, u)\}$, where  $ u= e^{-2i \theta }$,  under the geodesic flow $g_1(t)$  (\ref{transformation1}) is defined   by the mapping 
 \be\label{evolu1}
 \tau' = 
 {\tau \cosh( t/2 )  + u ~\overline{\tau} ~\sinh( t/2 )  \over \cosh( t/2 )   + u ~\sinh( t/2 ) },~~~~~
u' = 
{u \cosh( t/2 )  + \sinh( t/2 )   \over  \cosh( t/2 )    + u \sinh( t/2 ) },
\ee 
The evolution of the  observables under the geodesic flow $g_2(t)$ (\ref{transformation2}) has a  
similar form, except of an additional factor $i$ in front of the variable  $u$.
These expressions allow to define the transformation of the functions  $\{ f(x,y,u)\}$
under the time evolution as $f(x,y,u) \rightarrow  f(x',y',u')$ where
\be\label{coor1}
x'=x'(x,y,u,t),~~~y'=y'(x,y,u,t),~~~u' = u'(u, t). 
\ee
By using the Stone's theorem this transformation of functions can be 
expressed as an action of a one-parameter 
group of  unitary operators $U_{t}$:
\be\label{geoflow}
U_{t} f(g ) = f(g g_t ).
\ee
Let us calculate  transformations which are induced by $g_1(t)$  and $g_2(t)$  
 in (\ref{transformation1})-(\ref{transformation2}).  The time evolution   is given by the equations (\ref{evolu1}) and (\ref{coor1}):
\beqa\label{evolu}
&U_1(t) f(\tau,u) = f\Big({\tau \cosh( t/2 )  + u ~\overline{\tau} ~\sinh( t/2 )  \over \cosh( t/2 )   + u ~\sinh( t/2 ) } ,~ { u \cosh( t/2 )  + \sinh( t/2 )   \over  \cosh( t/2 )    + u \sinh( t/2 ) }\Big),\nn\\
&U_2(t) f(\tau,u) = f\Big(
{\tau \cosh( t/2 )  + i u ~\overline{\tau} ~\sinh( t/2 )  \over \cosh( t/2 )   + i u ~\sinh( t/2 ) },~{ u \cosh( t/2 )   -i \sinh( t/2 )   \over  \cosh( t/2 )    + i  u \sinh( t/2 ) } \Big).
\eeqa
A one-parameter family of unitary operators $U_t$ can be represented as an exponent of the 
self-adjoint  operator $U_t=\exp(i  H  t)$, thus we have 
\begin{equation}
U_{t} f(g )= e^{i  H  t} f(g ) = f(g g_t )
\end{equation}
and by differentiating it over the time $t$ at $t=0$ we shall get 
$
   Hf =-i \frac{ \mathrm d}{ \mathrm dt}U_t f \vert_{t=0}, 
$
that allows to calculate the operators $H$ corresponding to the  $U_1(t)$ and $U_2(t)$. 
Differentiating over time in (\ref{evolu})  we shall get for $H_1$ and $H_2 $:
    \begin{eqnarray}
   2  H_1  =\frac{y }{u}\left(\frac{\partial }{\partial x}+i \frac{\partial }{\partial y}\right)-i \frac{\partial }{\partial u} - u y \left( \frac{\partial }{\partial x}-i\frac{\partial }{\partial y}\right) 
+i u^2 \frac{\partial }{\partial u}  \nn\\
 2 i H_2  =\frac{y }{u}\left(\frac{\partial }{\partial x}+i \frac{\partial }{\partial y}\right)-i \frac{\partial }{\partial u} + u y \left( \frac{\partial }{\partial x}-i\frac{\partial }{\partial y}\right)
-i u^2 \frac{\partial }{\partial u} . 
    \end{eqnarray}
Introducing annihilation and creation operators  $ H_{-}=H_1 -iH_2 $ and  $ H_{+}=H_1 +iH_2 $
 yields 
\begin{eqnarray}
&H_{+}=
\frac{y }{u}\left(\frac{\partial }{\partial x}+i \frac{\partial }{\partial y}\right)-i \frac{\partial }{\partial u},~~~ 
&H_{-}=- u y \left( \frac{\partial }{\partial x}-i\frac{\partial }{\partial y}\right)
+i u^2 \frac{\partial }{\partial u} 
\end{eqnarray}
and by calculating the commutator $[H_+, H_-]$ we shall get 
$
H_{0}=u \frac {\partial }{\partial u}$
and  their  $sl(2,R)$ algebra is:
\begin{eqnarray}
\left[H_+,H_-\right]=2 H_0, ~~
\left[H_0,H_+\right]=-H_+,~~
\left[H_0,H_-\right]=H_-.
\end{eqnarray}
We can also calculate the expression for the invariant Casimir operator: 
\be
H= {1\over 2} (H_+H_- + H_-H_+) - H^2_0 = -y^2(\partial^2_x + \partial^2_y) + 2 i y\partial_x  u\partial_u   = 
-y^2(\partial^2_x + \partial^2_y) -  y\partial_x  \partial_{\theta}.  
\ee
Consider a class of functions which fulfil the following two equations: 
\begin{eqnarray}
H_0 f(x,y,u) = -{N \over 2}  f(x,y,u) ,  ~~~~~
    H_{-} f(x,y,u)=0 ,  
    \end{eqnarray}
where $N$ is an integer number. The first equation has  the solution $f_N(x,y,u)= ({ 1  \over  u y})^{N/2} \psi(x,y)=\omega^N_2 \psi(x,y)$ and by substituting it into the second one we shall get 
\be
N \psi(\tau,\overline{ \tau }) +(\overline{ \tau }-\tau)\frac{\partial \psi(\tau,\overline{ \tau })}{\partial \tau } =0.
\ee
By taking  $\psi(\tau,\overline{ \tau }) = (\overline{ \tau }-\tau)^N \Phi(\overline{ \tau },\tau) $ we shall get 
the equation $\frac{\partial \Phi}{\partial \tau } =0 $, that is, $\Phi$
is a anti-holomorphic function and  $f(\omega_2,\tau,\overline{ \tau })$ takes the 
form\footnote{The factors  $(2 i)^N$ have been  absorbed  by the redefinition of $\Phi$.}
 \begin{eqnarray}\label{classfunc}
     f(\omega_2,\tau,\overline{ \tau })=\omega^N_2 (\overline{\tau}-\tau)^N \Phi
     (\overline{\tau}) = {1 \over \overline{\omega_2}^N }\Phi
     (\overline{\tau}).
     \end{eqnarray}
The invariance under the action of the modular transformation (\ref{periodic}) will take 
the form
\be
\Phi ({\frac{m \overline{\tau} + n  }{p \overline{\tau} + q  }}) =  \Phi(\overline{\tau}) 
(p \overline{\tau} + q )^N 
\ee
and $\Phi(\overline{\tau})$  is a theta  function of weight $N$ \cite{Poincare,Ford}. 
The invariant integration measure on the group $G$ is given by the formula 
 \cite{Hopf,Gelfand}
\begin{equation}
d \mu = \frac{dx dy}{y^2} d\theta
\end{equation}
and the invariant product of functions on the  phase space $(x,y,\theta) \in \CM$  will be given by the integral 
\beqa
(f_1 ,f_2)
&=& \int_{0}^{2\pi} d \theta \int_{\CF}f_1(\theta,x,y) \overline{f_2(\theta,x,y) }
{d x d y \over y^2}.
\eeqa
It was demonstrated that the functions on the phase space are of the form (\ref{classfunc}),
where $\tau = x + i y$ and ${(\tau-\overline{\tau})\over 2 i }= y$, thus  the expression for 
the scalar product  will takes the following form: 
\beqa
(f_1 ,f_2)
&=& \int_{0}^{2\pi} d \theta \int_{\CF} \Phi_1 (\overline{\tau})  ~  \overline{\Phi_2 (\overline{\tau})}  {1 \over   \vert \omega_2 \vert^{2N} }
  {d x d y \over y^2}=  2\pi  \int_{\CF} \Phi_1 (\overline{\tau})  ~  \overline{\Phi_2 (\overline{\tau})}  y^{N-2}
  d x d y ,
\eeqa
where $N \geq 2$. This expression for the scalar product allows to calculate the two-point correlation functions under the geodesic flow  (\ref{transformation1})-(\ref{transformation3}).

A  correlation function can be defined as an integral over a pair of  functions 
in which the first one is stationary and the second one evolves with the 
geodesic flow:
\beqa
\CD_{t}(f_1,f_2) &=&   \int_{\CM}f_1(g) \overline{f_2(g g_t) } d \mu .
\eeqa
By using (\ref{evolu1}) and (\ref{coor1})  one can represent the integral in the following form    \cite{Savvidy:1982jk}:
\beqa
\CD_{t}(f_1,f_2) &=&    
= \int_{0}^{2\pi}\int_{\CF}f_1[x,y,\theta] ~\overline{f_2[x'(x,y,\theta,t), y'(x,y,\theta,t), \theta'(\theta,t)] }
{d x d y \over y^2}d \theta.
\eeqa
From   (\ref{geoflow}), (\ref{evolu})  and (\ref{evolu1}), (\ref{evolu}) it follows that
\beqa
&f_1(\omega_2,\tau,\overline{ \tau }) = {1 \over  \overline{\omega_2}^N  } \Phi_1 (\overline{\tau}),~~~\\
&\overline{f_2(\omega'_2,\tau',\overline{ \tau' })} ={1 \over   \omega_2^N  (\cosh( t /2)  +  e^{-2i\theta} \sinh( t /2) )^N }  
\overline{ \Phi_2 \Big( {\overline{\tau} \cosh( t/2 )  + e^{-2i\theta} ~\tau ~\sinh( t/2 ) 
 \over \cosh( t/2 )   + e^{-2i\theta} ~\sinh( t/2 )  }\Big)}.\nn
\eeqa
Therefore the correlation function takes the following form:
\beqa
\CD_t(f_1,f_2) 
&=&   \int_{0}^{2\pi} d \theta \int_{\CF}    \Phi_1 (\overline{\tau})  \overline{\Phi_2 (\overline{\tau'})}  ~ {y^{N-2 } d x d y \over    (\cosh( t/2 )   +  e^{-2i\theta} ~\sinh( t/2 )  )^N }  .  \nn
\eeqa
The upper bound on the correlations functions is 
\beqa
 \vert \CD_t(f_1,f_2) \vert   \leq   
    \int_{0}^{2\pi} d \theta \int_{\CF}  \vert    \Phi_1 (\overline{\tau})  \overline{\Phi_2 (\overline{\tau'})}  \vert ~\vert    {y^{N-2 } d x d y \over    (\cosh( t/2 )   +  e^{-2i\theta} ~\sinh( t/2 )  )^N }  \vert ~ \nn
\eeqa
and  in the limit $t \rightarrow + \infty$ the correlation function exponentially decays:
\beqa
\vert \CD_t(f_1,f_2) \vert  & \leq &     ~ M_{\Phi_1 \Phi_2}(\epsilon)~  e^{-{N\over 2}\vert  t \vert }~~.
\eeqa
If the surface metric is of a general form, $ds^2 = {dx^2 +dy^2 \over K  y^2},
$  with curvature $K < 0 $, then in the last formula the exponential factor will be 
\beqa
\vert \CD_t(f_1,f_2) \vert  & \leq &     ~ M_{\Phi_1 \Phi_2}(\epsilon)~  e^{-{N\over 2} K \vert t \vert}~
\eeqa
and the characteristic time decay   (\ref{relaxationtime}), (\ref{relaxationtime1})  \cite{yer1986a,Savvidy:2018ygo}  will take the form
\be
\tau_0 = {2\over N K }.
\ee
The decay time  of the correlation functions 
is shorter when the surface has a larger negative curvature or, in other words, when the divergency  
of the trajectories is stronger. 
 
The earlier investigation of the correlation functions of  Anosov geodesic flows
was performed in 
\cite{Collet,Pollicot,moore,dolgopyat,chernov} by using different approaches including 
Fourier series for the $SL(2, R)$ group, zeta function for the geodesic flows, 
relating the poles of the Fourier transform of the correlation functions to the spectrum of an associated Ruelle operator,  the methods of unitary representation theory,  spectral properties of the 
corresponding Laplacian and others.  In our analyses we have used the time evolution equations,
the properties of automorphic functions on $\CF$ and estimated a decay exponent in terms of 
the space curvature and the transformation properties of the functions \cite{Poghosyan:2018efd}.

\section{\it Quantum-mechanical Artin System} 
 
In the previous section we described the behaviour of the correlation functions/observables which are defined on the phase space of the Artin system and demonstrated the exponential decay of the classical correlation functions with time. In this section we shall describe the quantum-mechanical properties of classically Artin system which is maximally chaotic  in its classical regime.  
There is a great interest in considering quantisation of the hyperbolic dynamical systems and investigating their quantum-mechanical properties \cite{Savvidy:1982jk,Savvidy:1984gi}. Here we shall study the behaviour of the correlation functions of the Artin hyperbolic dynamical system in quantum-mechanical regime \cite{Babujian:2018xoy,Savvidy:2018ffh}. This subject is very closely related  with  the investigation of quantum mechanics of classically non-integrable systems.  

{\it In classical regime the exponential divergency of the geodesic trajectories resulted into the universal exponential decay of its classical correlation functions} \cite{Savvidy:2018ygo,Poghosyan:2018efd}. 
In order to investigate the behaviour of the correlation functions in quantum-mechanical regime it is necessary  to know the spectrum of the system and the corresponding wave functions.  In the case of the modular group the energy spectrum has continuous part, which is originating from the asymptotically free motion inside an infinitely long "y -channel"  extended in the vertical direction of the fundamental region as well as infinitely many discrete energy states  corresponding to a bounded motion at the "bottom"  of the fundamental triangle. The spectral problem has deep number-theoretical origin  and was partially solved in a series of pioneering articles \cite{maass,roeleke,selberg1,selberg2}.   It was solved partially because the discrete spectrum and the corresponding wave functions are not known analytically.   The general properties of the discrete spectrum have been derived by using Selberg trace formula \cite{selberg1,selberg2,bump, Faddeev,Faddeev1,hejhal2}.  Numerical calculation of the discrete energy levels were performed for many energy states   \cite{winkler,hejhal,hejhal1}.  

Here we shall describe the quantisation of the Artin system \cite{Babujian:2018xoy,Savvidy:2018ffh}.  The derivation of the Maass wave functions \cite{maass} for the continuous spectrum will be reviewed in details. We shall use  the Poincar\'e representation for Maass non-holomorphic automorphic wave functions. By introducing a natural physical variable $\tilde{y}$ for the distance in the vertical direction on the fundamental triangle  $\int dy/y =  \ln y = \tilde{y} $   and the corresponding momentum $p_y$ we shall 
 represent the Maass wave functions (\ref{alterwave}) in the form which is appealing to the physical intuition: 
\bea\label{alterwave0}
&\psi_{p} (x,\tilde{y})  
= e^{-i p \tilde{y}  }+{\theta(\frac{1}{2} +i p) \over \theta(\frac{1}{2} -i p)}   \, e^{  i p \tilde{y}}  +{ 4  \over  \theta(\frac{1}{2} -i p)}   \sum_{l=1}^{\infty}\tau_{i p}(l)
K_{i p }(2 \pi  l e^{\tilde{y}} )\cos(2\pi l x).~~~~
\eea
The first two terms describe the incoming and outgoing plane waves. The plane wave  $e^{-i p \tilde{y}  }$ incoming from infinity of the $y$ axis on Fig. \ref{scattering}  ( the vertex $\CD$)  elastically scatters on the boundary $ACB$ of the fundamental triangle  $\CF$ and reflected backwards $e^{  i p \tilde{y}}$.  The reflection amplitude is a pure phase and is given by the expression in front of the outgoing plane wave $e^{  i p \tilde{y}}$: 
\be\label{phase}
{\theta(\frac{1}{2} +i p) \over \theta(\frac{1}{2} -i p)} = \exp{[i\, \varphi(p)]}.
\ee
The rest of the wave function describes the standing waves $\cos(2\pi l x)$  in the $x$ direction between boundaries $x=\pm 1/2$
with the amplitudes $K_{i p }(2 \pi  l y )$, which are exponentially decreasing with index $l$. The continuous energy spectrum is given by the formula  
\be
E=   p^2  + \frac{1}{4} .
\ee
The wave functions of the discrete spectrum have the following form \cite{maass,roeleke,selberg1,selberg2,hejhal,hejhal1, winkler}:
\bea\label{wavedisc0}
\psi_n(z) &=&   \sum_{l=1}^{\infty} c_l(n) \,
 K_{i u_n }(2 \pi  l e^{\tilde{y}} ) 
\left\{  \begin{array}{ll} 
\cos(2\pi l x) \\
\sin(2\pi l x)   \\
\end{array} \right. , 
\eea
where the spectrum $E_n = {1\over 4} + u^2_n$ and the coefficients $c_l(n)$  are not known analytically but were computed numerically for many 
values of $n$ \cite{hejhal,hejhal1,winkler}.  

Having in hand the explicit expression of the wave function one can analyse a quantum-mechanical  behaviour of the correlation functions defined in \cite{Maldacena:2015waa}: 
\bea\label{basicopera1}
&\CD_2(\beta,t)=  \langle   A(t)   B(0) e^{-\beta H}   \rangle ,~~~~ 
 \CD_4(\beta,t)=  \langle  A(t)   B(0) A(t)   B(0)e^{-\beta H}   \rangle \\
\label{basicopera2}
&C(\beta,t) =   -\langle [A(t),B(0)]^2 e^{-\beta H} \rangle~,
\eea
where the operators $A$ and $B$ are chosen to be of  the Louiville type \cite{Babujian:2018xoy}:
\be\label{basicopera}
A(N)=  e^{-2 N \tilde{y}},~~~N=1,2,.....
\ee
Analysing  the basic matrix elements of the Louiville-like  operators (\ref{basicopera}) we shall demonstrate   that all two- and four-point correlation functions (\ref{basicopera1}) decay exponentially with time, with the exponents which depend on temperature Fig.\ref{twopointfunc} and Fig.\ref{fourpointfunc}.  These exponents define the decorrelation time $t_d(\beta)$. 

Alternatively to the exponential decay of  the correlation functions (\ref{basicopera1}) the square of the commutator of the Louiville-like  operators  separated in time (\ref{basicopera2}) grows exponentially Fig.\ref{timeevolutionofcommutator} \cite{Babujian:2018xoy}. This growth is reminiscent of the local exponential divergency of trajectories in the Artin system when it was considered in the classical regime \cite{Poghosyan:2018efd}. The exponential growth Fig.\ref{timeevolutionofcommutator} of the commutator (\ref{basicopera2}) does not saturate the condition of maximal growth (\ref{exponentc1t})  
\be\label{exponentc12}
C(\beta,t) \sim K(\beta) \,e^{ {2\pi  \over  \chi(\beta)} t }
\ee
of the correlation functions which is conjectured to be linear in temperature   $T=1/\beta$:
\be\label{exponentc1t}
C(\beta,t) \sim K(\beta) \,e^{ {2\pi  \over  \beta} t }, ~~~~\chi(\beta) = \beta.
\ee
In calculation of the quantum-mechanical correlation functions  a perturbative expansion was used in which the high-mode Bessel's functions in (\ref{alterwave0}) and (\ref{wavedisc0}) are  considered as perturbations.  It has been found that calculations are stable with respect to these perturbations and do not influence the final results.  The reason is that in the integration region of the matrix elements (\ref{basicmatrix}) the high-mode Bessel's functions are exponentially small.

\begin{figure}[htbp]
	 \hspace*{-4cm} 
	\begin{tikzpicture}[scale=1.9]
	\clip (-4.3,-0.4) rectangle (4.5,2.8);
	\draw[,->] (-2.1,0) -- (2.1,0)  node[anchor=north west] {$x$ };
	\draw[dashed] (0,0) -- (0,1.2);
	\draw[dashed,->]  (0,1.8)--(0,2.2)  node[anchor=south east] {$y$};
	\foreach \x in {-1,-0.5 ,0,0.5,1}
	\draw (\x cm,1pt) -- (\x cm,-1pt) node[anchor=north] {$\x$};
	\foreach \y in {1}
	\draw (1pt,\y cm) -- (-1pt,\y cm) node[anchor=east] {$\y$};
	\draw (1,0) arc (0:180:1cm);
	\draw (0.5,0)--(0.5,0.86602540378);
	\draw (-0.5,0)--(-0.5,0.86602540378);
	\draw[ ultra thick, black] (0.5,0.86602540378)--(0.5,2.1);
	\draw[ ultra thick, black] (-0.5,0.86602540378)--(-0.5,2.1);
	\draw[ultra thick, black] (0.5,0.86602540378) arc (60:120:1cm);
	\draw (-0.6,0.7) node{ $A$};
	\draw (0.4,0.7) node{ $B$};
	\draw (-0.1,0.85) node{ $C$};
	\draw [->,snake=snake,
	segment amplitude=.6mm,
	segment length=1.5mm,
	line after snake=1mm] (-0.3,2) -- (-0.3,1.3);
	\draw [->,snake=snake,
	segment amplitude=.6mm,
	segment length=1.5mm,
	line after snake=1mm] (0.3,1.3)-- (0.3,2);
	\draw [->,snake=snake,
	segment amplitude=.6mm,
	segment length=1.5mm,
	line after snake=1mm] (0.1,1.5) -- (0.2,1.7);
	\draw [->,snake=snake,
	segment amplitude=.6mm,
	segment length=1.5mm,
	line after snake=1mm] (-0.1,1.5) -- (-0.2,1.7);
	\draw [->,snake=snake,
	segment amplitude=.6mm,
	segment length=1.5mm,
	line after snake=1mm] (-0.1,1.4) -- (-0.2,1.2);
	\draw [->,snake=snake,
	segment amplitude=.6mm,
	segment length=1.5mm,
	line after snake=1mm] (0.1,1.4) -- (0.2,1.2);
	\draw (1,1.8) node{ $\frac{\theta({1\over 2}+i p)}{\theta({1\over 2}-i p)}e^{i p \tilde{y}}$};
	\draw (-1,1.8) node{ $e^{-i p \tilde{y}}$};
	\draw (0.08,2.3) node{ $\CD$};
	\end{tikzpicture}
	 \includegraphics[angle=0,width=3.5cm]{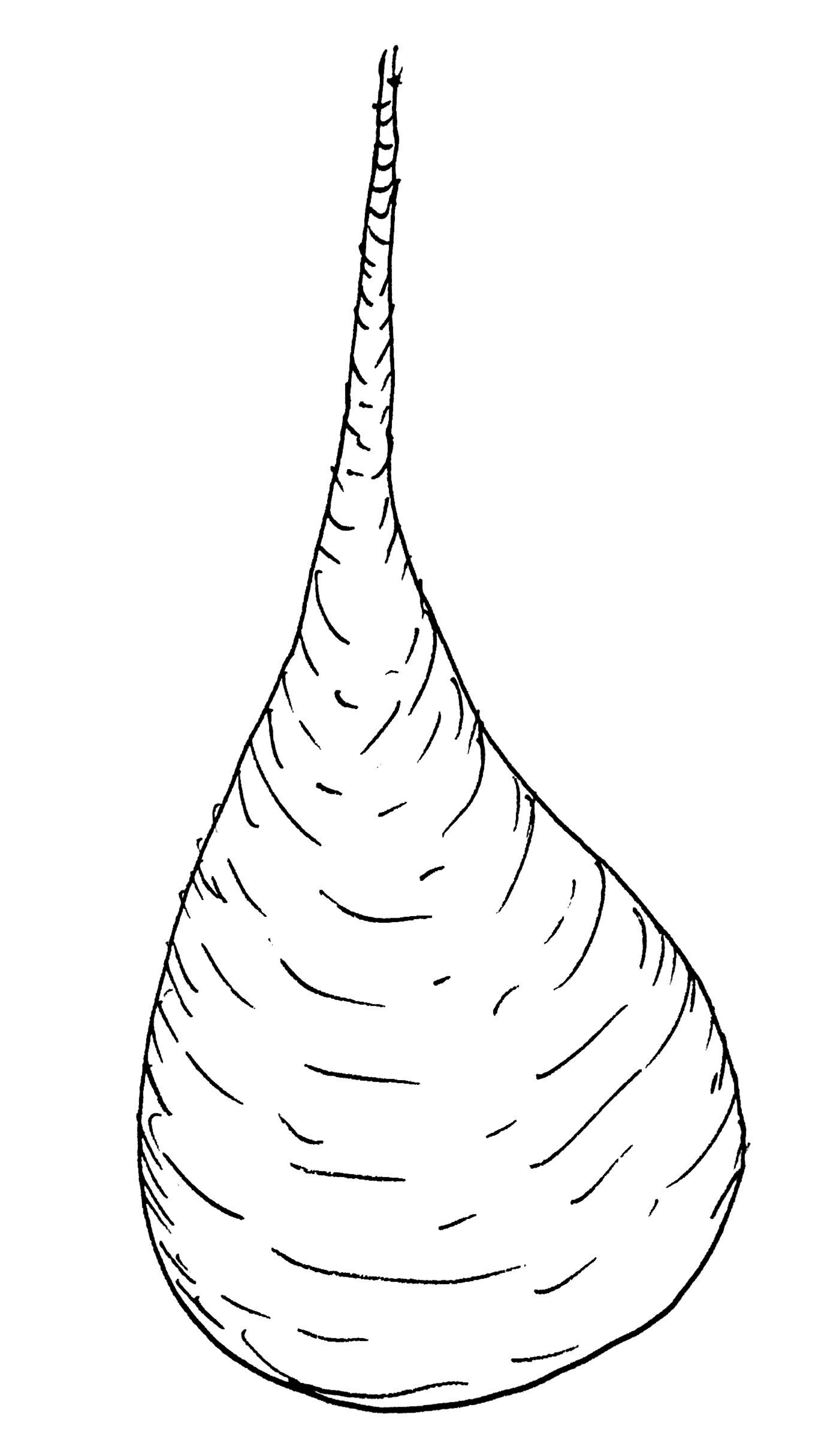}

	\caption{ The incoming and outgoing plane waves. The plane wave  $e^{-i p \tilde{y}  }$ incoming from infinity of the $y$ axis,    the vertex $\CD$,  elastically scatters on the boundary $ACB$ of the fundamental triangle  $\CF$ on Fig. \ref{fig5}, the bottom of the Artin surface $\bar{\CF}$.  The reflection amplitude $\theta(\frac{1}{2} +i p) / \theta(\frac{1}{2} -i p)$ is a pure phase and is given by the expression in front of the outgoing plane wave $e^{  i p \tilde{y}}$.
The rest of the wave function describes the standing waves in the $x$ direction between boundaries $x=\pm 1/2$ with the amplitudes, which are exponentially decreasing. On the right figure is an artistic image of the Artin surface, as far as it cannot be smoothly embedded  into $R^3$.}
	
	\label{scattering}
\end{figure}

Let us consider the geodesic flow on $\CF$  described by the  action (\ref{metric_hp})
\bea
S=\int ds=\int \frac{\sqrt{\dot{x}^2+\dot{y}^2}}{ y}\,\,dt~
\eea 
and the equations of motion
\bea
 \frac{d}{dt}\,\, \frac{\dot{x}}{\,y\sqrt{\dot{x}^2+\dot{y}^2}}=0, ~~~~~~~~
 \frac{d}{dt}\,\, \frac{\dot{y}}{\,y\sqrt{\dot{x}^2+\dot{y}^2}}
+\frac{\sqrt{\dot{x}^2+\dot{y}^2}}{\,y^2}=0.
\label{EOM_generic}
\eea 
 Notice the invariance of the action and of the equations under time reparametrisations  
$t\rightarrow t(\tau)$. The presence of this local gauge symmetry 
indicates that we have a constrained dynamical system \cite{Faddeev:1969su}. 
A convenient  gauge fixing which
specifies the time parameter $t$ to be proportional  to the proper time, is archived by imposing 
the condition  
\bea
{ \dot{x}^2+\dot{y}^2 \over y^2}=2 H \,,
\label{gauge}
\eea
where $H$ is a constant. 
In this gauge the equations (\ref{EOM_generic}) will take the form \cite{Faddeev:1969su}
\bea
\frac{d}{dt}\,\,( \frac{\dot{x}}{\,y^2 })=0,~~~~~ \frac{d}{dt}\,\, (\frac{\dot{y}}{\,y^2})
+\frac{2H }{\,y}=0.
\label{EOM_generic1}
\eea 
Defining the canonical momenta as $
p_x = \frac{\dot{x}}{\, y^2},  ~ 
  p_y= 
\frac{\dot{y}}{\,y^2},
$ conjugate to the coordinates 
$(x,y)$,    
one can get the geodesic equations   (\ref{EOM_generic1})  in the Hamiltonian form: 
 \bea
 \dot{ p_x}=0,~~~~
\dot{p_y} =-
\frac{2H }{\,y} .
\label{EOM_generic2}
\eea 
The Hamiltonian will take the form
 \bea
 H = {1\over 2}y^2 (p^{2}_{x} +p^{2}_{y})
\label{hamiltonian1}
\eea 
and the corresponding equations will take the following form: 
\bea
&& \dot{x} = \frac{\partial H }{\, \partial p_x} = y^2 p_x,~\dot{y} = \frac{\partial H }{\, \partial p_y} = y^2 p_y
\\
&&\dot{p_x} = - \frac{\partial H }{\, \partial x} = 0,~\dot{p_y} = -\frac{\partial H }{\, \partial y} = - y(p^2_x +p^2_y)= -\frac{2H }{\,y},\nn
\eea
and they coincide with  (\ref{EOM_generic2}).
The advantage of the gauge (\ref{gauge}) is that the Hamiltonian (\ref{hamiltonian1}) coincides with the constraint.

Now it is fairly standard to quantize this Hamiltonian system by replacing in (\ref{hamiltonian1})   
$
p_x=-i\frac{\partial}{\partial x},  
p_y=-i\frac{\partial}{\partial y}
$
and considering time independent  Schr\"odinger equation
$
H\psi = E \psi.
$
The resulting equation explicitly reads:
\bea
-y^2(\partial_x^2+\partial_y^2)\psi= E \psi.
\eea 
On the lhs one easily recognises  the  Laplace operator \cite{maass,roeleke,selberg1,selberg2,bump,Faddeev,Faddeev1,hejhal2}(with an 
extra minus sign) in Poincare metric (\ref{metric_hp}). It is easy to see that the Hamiltonian is positive semi-definite Hermitian operator: 
 \bea
-\int \psi^*(x,y )\, y^2(\partial_x^2+\partial_y^2) \, \psi(x,y ) {d x dy \over y^2}  
=  \int (\vert \partial_x \psi(x,y)\vert^2+ \vert \partial_y \psi(x,y)\vert^2) d x dy  \geq 0.~~~~
\label{ semiposti}
\eea 
It is convenient to introduce parametrization of the energy $E=s(1-s)$ and 
to rewrite the Schr\"odinger equation as
\bea
-y^2(\partial_x^2+\partial_y^2)~ \psi(x,y)  = s(1-s) ~\psi(x,y).
\label{Laplace_eq}
\eea 
As far as $E$ is real and semi-positive and parametrisation is symmetric with respect to $s \leftrightarrow 1-s $ it follows that the parameter $s$ should be chosen within the range
\be\label{srange}
s \in  [1/2 , 1  ]~~ \text{or } ~~ s=1/2 +i u ,~~~u ~\in~ [0,\infty].
\ee
One should impose the "periodic" boundary condition on the wave function  with respect to the modular group 
\bea\label{invariance}
\psi(\frac{a z+b}{c z+d})=\psi(z),~~~\left(
\begin{array}{cc}
a&b\\c&d
\end{array}
\right) \in SL(2,Z)
\eea
in order to have  the wave function which is properly defined on the fundamental 
region $\bar{\CF}$ shown in Fig. \ref{fig5} .
Taking into account that the transformation $T:z\rightarrow z+1$ 
belongs to $SL(2,Z)$, one has to impose the periodicity condition 
$\psi(z)=\psi(z+1)$ and get  the Fourier expansion
$
\psi(x,y)=\sum_{n=-\infty}^\infty f_n(y)\exp(2\pi i n x).
$  
Inserting this into Eq. (\ref{Laplace_eq}), for the Fourier 
component $f_n(y)$ one can get
$
\frac{d^2f_n(y)}{dy^2}+(s(1-s)-4\pi^2n^2)f_n(y)=0~.
$ 
For the case $n\neq 0$ the solution which exponentially decays at 
large $y$ reads  
$
f_n(y)=\sqrt{y} K_{s-\frac{1}{2}}(2\pi n|y|)
$
and for $n=0$ one simply gets
$
f_0(y)=c_0 y^s+c^{'}_0 y^{1-s}.
$
Thus the solution can be represented in the form \cite{maass,roeleke,selberg1,selberg2,bump,Faddeev,Faddeev1,hejhal2}
\bea\label{solution3}
\psi(x,y) =  c_0 y^s+c^{'}_0 y^{1-s}  
+ \sqrt{y}\sum_{n=-\infty \atop n \neq 0}^\infty c_n  K_{s-\frac{1}{2}}(2\pi n|y|) \exp(2\pi i n x),
\eea
where the coefficients $c_0, c^{'}_0, c_n$ should be defined in such a way that the wave function will fulfil  the boundary conditions (\ref{invariance}). Thus one should  impose also the invariance with respect to the second generator of the modular group $SL(2,Z)$, that is,  with respect to the  transformation $S:z\rightarrow -1/z$ ~:
$
\psi(z)=\psi(-1/z).
$
This functional equation defines the coefficients $c_0, c^{'}_0, c_n$.   Another option is to start from a particular solution and perform summation over all nonequivalent shifts of the argument
by the elements of $SL(2,Z)$, that is using the Poincar\'e  series representation \cite{Poincare,Poincare1,maass,roeleke,selberg1,selberg2,bump,Faddeev,Faddeev1,hejhal2}. Let us demonstrate this strategy by using  the simplest solution (\ref{solution3}) with $c_0 =1, c^{'}_0=0$:
\[
\psi(z)= y^s = (\Im z)^s\, .
\] 
The $\Gamma_{\infty}$ is the
subgroup of $\Gamma = SL(2,Z)$ generating shifts $z\rightarrow z+n$, 
$n\in Z$.  Since $y^s$ is already invariant with respect to $\Gamma_{\infty}$, 
one should perform summation over the conjugacy classes 
$\Gamma_{\infty} \backslash \Gamma $.  There is  a bijection between the set of mutually 
prime pairs  $(c,d)$ with $(c,d)\neq (0,0)$ and the set of conjugacy 
classes $\Gamma_\infty\backslash \Gamma $. The fact that the integers
$(c,d)$ are mutually prime integers means that their greatest common divisor 
(gcd) is equal to one: $gcd(c,d)=1$.  As a result, it is defined by the classical Poincar\'e  series representation \cite{Poincare,Poincare1} and  
for the sum of our interest we get 
\bea
\psi_{s}(z) \equiv \frac{1}{2}\sum_{\gamma \in \Gamma_\infty\backslash \Gamma}
(\Im (\gamma z))^s   
=\frac{1}{2}\sum_{(c,d) \in \mathbb{Z}^2 \atop gcd(c,d)=1} \frac{y^s}{((c x+d)^2+c^2y^2)^s}~,
\label{sum1}
\eea   
where, as explained above, the sum on r.h.s. is taken over all 
mutually prime pairs $(c,d)$.  To evaluate the sum one should multiply  both sides of the
eq. (\ref{sum1}) by 
$
\sum_{n=1}^{\infty}\frac{1}{n^{2s}}
\equiv \zeta(2s)
$ \cite{maass}
so that the wave function will be expressed in terms of the Eisenstein series:  
\bea\label{solution4}
\zeta(2s) \, \psi_{s}(z) =\frac{1}{2} 
\sum_{(m,k) \in \mathbb{Z}^2 \atop (m,k)\neq(0,0)}
\frac{y^s}{((m x+k)^2+m^2y^2)^s} .
\eea
The evaluation of the sum can be now performed explicitly, and it allows to represent the (\ref{solution4}) in the following form:
\bea
\zeta(2s) \, \psi_{s}(x,y) = \zeta(2s) y^s + \frac{\sqrt{\pi}\Gamma(s-\frac{1}{2})\zeta(2s-1)}
{\Gamma(s)}\,y^{1-s} +\nn\\
+\sqrt{y} \frac{4 \pi^s}{\Gamma(s)} \sum_{l=1}^{\infty}\tau_{s-\frac{1}{2}}(l)
K_{s-\frac{1}{2}}(2 \pi  ly )\cos(2\pi l x) ,
\eea 
where the modified Bessel's $K$ function is given by the expression
$
K_{i u}(y) = {1\over 2} \int^{\infty}_{-\infty} e^{- y \cosh t} e^{i u t}   dt
$
and 
$
\tau_{i p}(n )=\sum_{a \cdot b=n}\left(\frac{a}{b}\right)^{ip} .
$
By using Riemann's reflection relation
\bea 
\zeta (s)=\frac{\pi ^{s-\frac{1}{2}} \Gamma \left(\frac{1-s}{2}\right)}
{ \Gamma \left(\frac{s}{2}\right)}\,\zeta (1-s)
\eea
and introducing the function
\bea\label{thata1}
\theta(s)=\pi ^{-s} \zeta (2 s) \Gamma (s)
\eea  
we get an elegant expression of the eigenfunctions obtained by Maass \cite{maass}:
\bea\label{elegant}
\theta(s) \psi_{s}(z) =\theta (s)y^s+\theta(1-s)\,y^{1-s}  
+4\sqrt{y} \sum_{l=1}^{\infty}\tau_{s-\frac{1}{2}}(l)
K_{s-\frac{1}{2}}(2 \pi  ly )\cos(2\pi l x) .\qquad 
\eea 
This wave function is well defined in the complex $s$ plane and has a simple pole at $s=1$.
The physical continuous spectrum was defined in (\ref{srange}), where  $s=\frac{1}{2} +iu $, $u \in [0,\infty]$, therefore  
\be\label{eigenvalues}
E= s(1-s)= \frac{1}{4} +u^2 .
\ee
The continuous spectrum wave functions $\psi_s(x,y)$ are delta function normalisable \cite{maass,roeleke,selberg1,selberg2,Faddeev,bump}.
The wave function (\ref{elegant}) can be  conveniently represented also in the form 
\bea
\psi_{ \frac{1}{2} +i u}(z) = y^{ \frac{1}{2} +i u}+{\theta(\frac{1}{2} -i u) \over \theta(\frac{1}{2} +i u)}   \,y^{\frac{1}{2} -i u}  
+{4\sqrt{y} \over \theta(\frac{1}{2} +i u)}   \sum_{l=1}^{\infty}\tau_{i u}(l)
K_{i u }(2 \pi  ly )\cos(2\pi l x) , 
\eea
where 
$
K_{-i u}(y ) =K_{i u }( y),~~~~~\tau_{-i u}(l) =\tau_{i u}(l)~. 
$
The physical interpretation of the wave function becomes  more transparent if one introduce the new variables 
\be\label{newvariab}
\tilde{y} = \ln y,~~~~ p= -u,~~~~E=   p^2  + \frac{1}{4}, 
\ee
as well as  the alternative normalisation of the wave function $\psi_{p} (x,\tilde{y})  \equiv
y^{- \frac{1}{2}} \psi_{ \frac{1}{2} +i u}(z)  $
\bea\label{alterwave}
\psi_{p} (x,\tilde{y})    
 = e^{-i p \tilde{y}  }+{\theta(\frac{1}{2} +i p) \over \theta(\frac{1}{2} -i p)}   \, e^{  +i p \tilde{y}}  + { 4  \over  \theta(\frac{1}{2} -i p)}  \sum_{l=1}^{\infty}\tau_{i p}(l)
K_{i p }(2 \pi  l e^{\tilde{y}} )\cos(2\pi l x) .
\eea
The first two terms describe the incoming and outgoing plane waves. The plane wave  $e^{-i p \tilde{y}  }$ incoming from infinity of the $y$ axis on Fig.\ref{scattering}  ( the vertex $\CD$)  elastically scatters on the boundary $ACB$ of the fundamental region $\CF$.  The reflection amplitude is a pure phase and is given by the expression in front of the outgoing plane wave $e^{  i p \tilde{y}}$ 
\be\label{phase}
{\theta(\frac{1}{2} +i p) \over \theta(\frac{1}{2} -i p)} = \exp{[i\, \varphi(p)]}.
\ee
The rest of the wave function describes the standing waves $\cos(2\pi l x)$  in the $x$ direction between boundaries $x=\pm 1/2$
with the amplitudes $K_{i p }(2 \pi  l y )$, which are exponentially decreasing with index $l$.

In addition to the continuous spectrum the system (\ref{Laplace_eq})  has a discrete spectrum  
\cite{maass,roeleke,selberg1,selberg2,Faddeev,bump}.
The number of discrete states is infinite: $E_0=0 < E_1 < E_2 < ....\rightarrow \infty$, the spectrum is extended to infinity -  unbounded from above -  and lacks any accumulation  point  except infinity. 
 The wave functions of the discrete spectrum have the form \cite{maass,roeleke,selberg1,selberg2, winkler,hejhal,hejhal1}
\bea\label{wavedisc}
\psi_n(z) &=&   \sum_{l=1}^{\infty} c_l(n) \,
\sqrt{y}\, K_{i u_n }(2 \pi  l y ) 
\left\{   \begin{array}{ll} 
\cos(2\pi l x) \\
\sin(2\pi l x)   \\
\end{array} \right. , 
\eea
where the spectrum $E_n = {1\over 4} + u^2_n$ and the coefficients $c_l(n)$  are not known analytically, but were computed numerically for many 
values of $n$ \cite{winkler,hejhal,hejhal1}.  Having explicit expressions of the wave functions one can analyse the quantum-mechanical  behaviour of the correlation functions, which we shall investigate in the next sections. 

\section{ \it Quantum Mechanical Correlation Functions }

The two-point correlation function is defined as: 
\bea
&\CD_2(\beta,t)=  \langle     A(t)   B(0) e^{-\beta H}   \rangle = 
  \sum_{n}  \langle  n \vert e^{i H t } A(0) e^{-i H t } B(0) e^{-\beta H} \vert n  \rangle=\nn\\
 &=\sum_{n,m} e^{i (E_n -E_m)t - \beta E_n}   \langle  n \vert   A(0)\vert m  \rangle  \langle m\vert  B(0)  \vert n  \rangle.
\eea
The energy eigenvalues (\ref{eigenvalues}) are parametrised by $n = {1\over 2} +i u$, $E_n =\frac{1}{4} + u^2 $ and $m = {1\over 2} +i v$, $E_m =\frac{1}{4} + v^2 $, thus \cite{Babujian:2018xoy}
\bea
&\CD_2(\beta,t)=\int^{+\infty}_{0} \int^{+\infty}_{0} du \,  dv  ~ e^{i (u^2 -v^2)t - \beta( \frac{1}{4} + u^2)}  
  \\
&\int_{\CF}\psi_{ \frac{1}{2} -i u}(z)  \, A \, \psi_{ \frac{1}{2} +i v }(z) \,  d\mu(z) 
\int_{\CF}\psi_{ \frac{1}{2} -i v }(w)  \, B \, \psi_{ \frac{1}{2} +i u}(w) \,  d\mu(w)~,\nn
\eea
where the complex conjugate  function is $\psi^*_{ \frac{1}{2} +i u}(z) =\psi_{ \frac{1}{2} -i u}(z) $.
Defining the basic matrix element as
\bea\label{basicmatrix}
A_{uv} = 
\int_{\CF}  \psi_{ \frac{1}{2} -i u}(z)  \, A \, \psi_{ \frac{1}{2} +i v }(z) \,  d\mu(z)  = 
\int^{1/2}_{-1/2} dx  \int^{\infty}_{\sqrt{1-x^2}} {dy \over y^2}  \psi_{ \frac{1}{2} -i u}(z)  \, A \, \psi_{ \frac{1}{2} +i v }(z)   
\eea
for the two-point correlation function one can get 
\bea
\CD_2(\beta,t)=  \int^{+\infty}_{-\infty} e^{i (u^2 -v^2)t - \beta( \frac{1}{4} + u^2)}   
A_{uv}\,  B_{vu} \, du dv.~~~
\eea
In terms of the new variables (\ref{newvariab}) the  basic matrix element (\ref{basicmatrix}) will take the form 
\bea\label{basicmatrixelement}
 A_{p q}  
= \int^{1/2}_{-1/2} dx  \int^{\infty}_{{1\over 2}\log(1-x^2)}  dy  
  \psi^*_{p} (x,y)  \,  ( e^{- \frac{1}{2} y}   A \,   e^{  \frac{1}{2} y})  \, \psi_{ q} (x,y).~~~~~ 
 \eea
The matrix element (\ref{basicmatrix}), (\ref{basicmatrixelement}) plays a fundamental role in the investigation of the correlation functions because all correlations can be expressed through it.  One should choose also appropriate observables 
$A$ and  $B$.  The operator $y^{-2} $  seems very appropriate for two reasons. Firstly, the convergence of the integrals over the fundamental region $\CF$ will be well defined. Secondly, this operator is reminiscent of 
the exponentiated Louiville  field  since $y^{-2} = e^{-2 \tilde{y}}$ . Thus the interest is in calculating  the matrix element  (\ref{basicmatrixelement})
for the observables in the form of the  Louiville-like operators \cite{Babujian:2018xoy}:
\be\label{Louiville-like}
A(N)=  e^{-2 N y} 
\ee  
with matrix element  
\bea
&A_{p q}(N) =\int^{1/2}_{-1/2} dx  \int^{\infty}_{{1\over 2}\log(1-x^2)}  dy  ~ \psi^*_{p} (x,y)  \,   e^{- 2 N y}  \, \psi_{ q} (x,y) ,\nn\\&  ~~ N=1,2,...
 \eea
 The other interesting observable is  $A = \cos(2\pi N x), N=1,2,... $.
The evaluation of the above matrix elements is convenient to perform using a perturbative expansion in which the part of the wave function (\ref{alterwave}) containing the Bessel's functions and the contribution of the discrete spectrum (\ref{wavedisc}) is considered as a perturbation. These terms of the perturbative expansion are small and don't change the physical behaviour of the correlation functions.  The reason behind this fact is that  in the integration region $\Im z \gg 1, \Im w \gg 1$ of the matrix element (\ref{basicmatrix}) the Bessel's functions  decay exponentially. Therefore the contribution of these high modes is small (analogues to the so called mini-superspace approximation in the Liouville theory).   In the first approximation of the wave function (\ref{alterwave}) for the matrix element one can get \cite{Babujian:2018xoy}
 \bea
& A_{pq}(N)= {\, _2F_1 \left(  \frac{1}{2}, N+  i {p-q \over 2 }  ;    \frac{3}{2};   \frac{1}{4}          \right)  \over   2 N +i (p-q)}  
+{\, _2F_1 \left(  \frac{1}{2}, N+i {p+q \over 2 }  ;    \frac{3}{2};   \frac{1}{4}          \right)  \over   2 N+ i (p+q)} e^{-i \varphi(q)}  \nn\\
&+{\, _2F_1 \left(  \frac{1}{2}, N- i {p+q \over 2 }  ;    \frac{3}{2};   \frac{1}{4}          \right)  \over   2 N -i (p+q) } e^{i \varphi(p)} + 
{\, _2F_1 \left(  \frac{1}{2}, N-i {p-q \over 2 }  ;    \frac{3}{2};   \frac{1}{4}          \right)  \over   2 N+ i (p-q )} e^{i (\varphi(p)-\varphi(q))}, 
\eea
where the reflation phase $\varphi(p)$ was defined in (\ref{phase}). Thus  
\bea\label{twopointcorrel}
\CD_2(\beta,t)=  \int^{+\infty}_{-\infty} e^{i (p^2 -q^2)t - \beta( \frac{1}{4} + p^2)}   
A_{pq}(N)\,  A_{qp}(M) \, dp dq~. 
\eea
 \begin{figure}[h]
 \centering
        \includegraphics[width=0.3\textwidth]{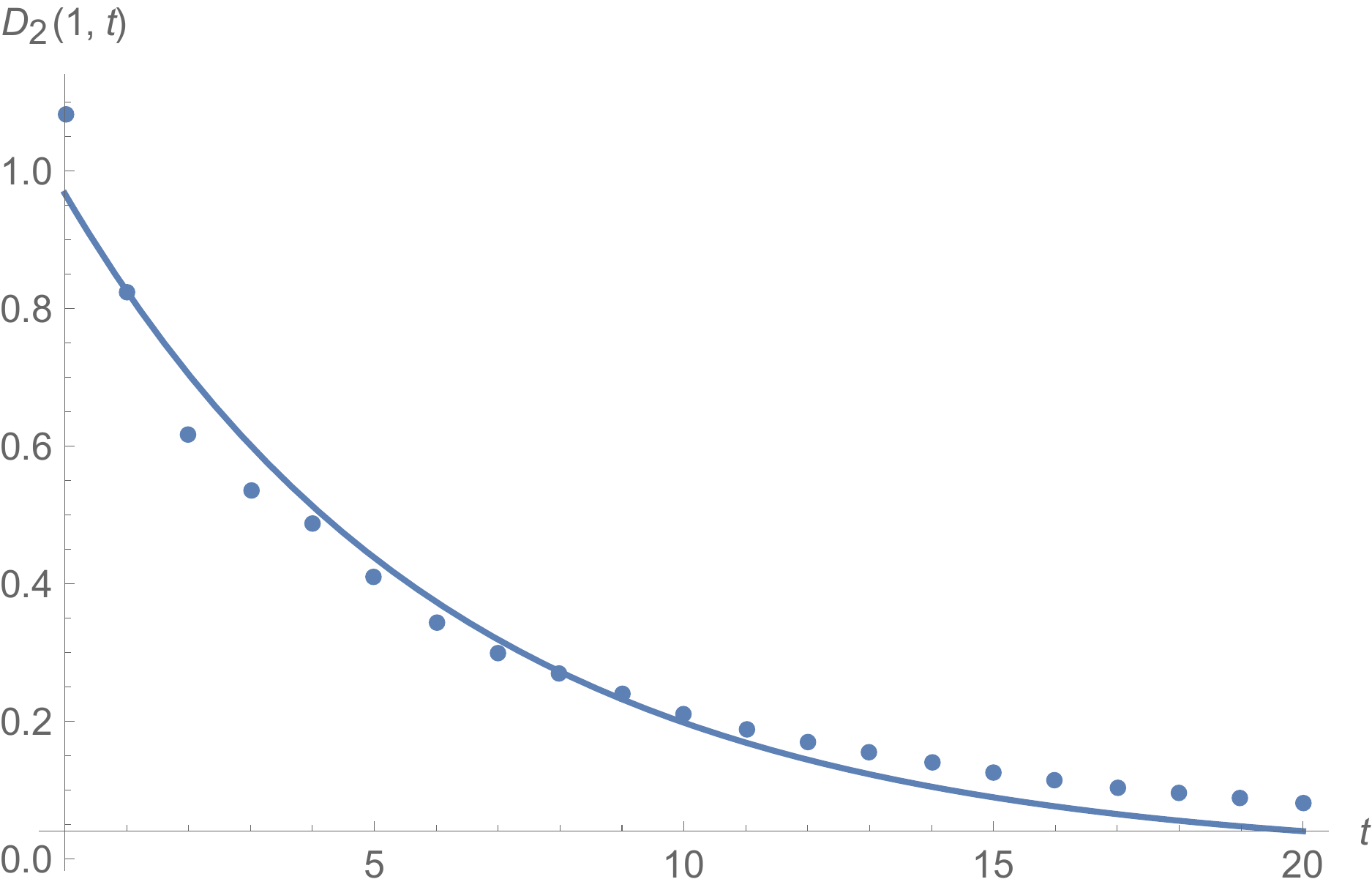}~~~~
         \includegraphics[width=0.3\textwidth]{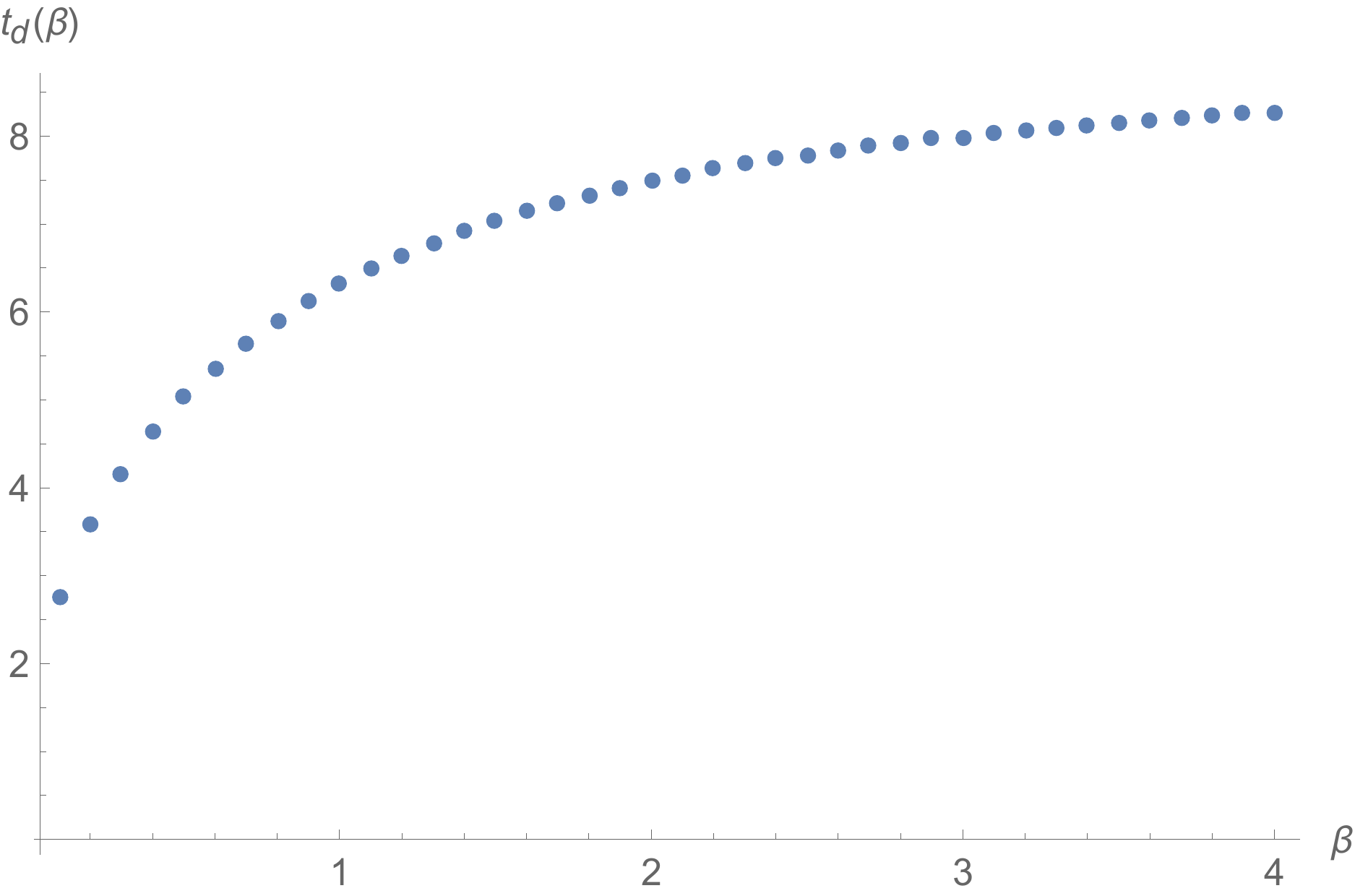}~~~~
         \includegraphics[width=0.3\textwidth]{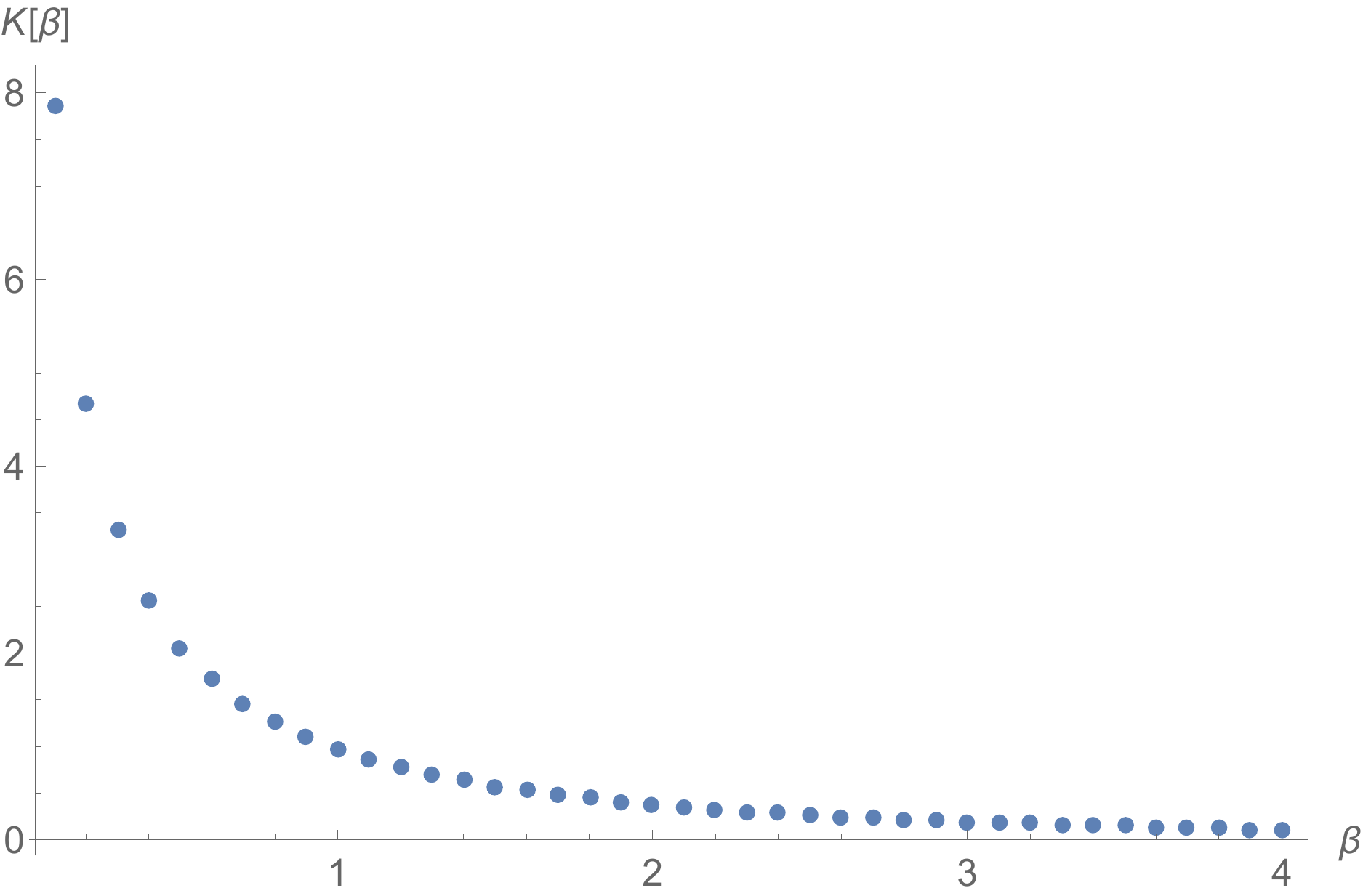}
        \caption{The exponential decay of the two-point correlation function $\CD_2(\beta,t)$ as a function of time  at   temperature $\beta =1$.  The points are fitted by the curve $K(\beta) \exp{(- t / t_d(\beta))}$. The exponent $t_d(\beta)$ has a well defined high and low temperature limits. The limiting values in dimensionless units are $t_d(0) \approx 0.276$ and $t_d(\infty) \approx 0.749$. The temperature dependence of $K(\beta)$ is shown on the l.h.s. graph.
}
\label{twopointfunc}
\end{figure}
The correlation function is for two Louiville-like  fields in the power $N$ and $M$ respectively.
This expression is very convenient for the analytical and numerical analyses.   
It is expected that the two-point correlation function decay exponentially  \cite{Maldacena:2015waa}
\be\label{twopoint}
\CD_2(\beta,t)  \sim K(\beta)~e^{-{t \over t_d(\beta)}}, 
\ee
where $ t_d(\beta)$ is the decorrelation time and defines one of the characteristic time scales in the quantum-mechanical system. The exponential decay of the two-point correlation function with time at different  temperatures is shown on Fig.\ref{twopointfunc}. The dependence of the exponent $ t_d(\beta)$ and of the prefactor $K(\beta)$ as a function of temperature are presented in Fig.\ref{twopointfunc}.   
 As one can see, at high and low temperatures the decorrelation time tends to the fixed values.   The corresponding limiting values in dimensionless units are  shown on the Fig.\ref{twopointfunc}.

It was conjectured in the literature \cite{Maldacena:2015waa} that the classical chaos can be diagnosed in thermal quantum systems by using an out-of-time-order correlation functions as well as by the square of the commutator of the operators which are separated in time. The out-of-time four-point correlation function of interest was defined in \cite{Maldacena:2015waa} as follows:
\bea\label{outoftime}
&\CD_4(\beta,t)=  \langle   A(t)   B(0) A(t)   B(0)e^{-\beta H}   \rangle= 
 \sum_{n,m,l,r} e^{i (E_n -E_m+E_l - E_r)t - \beta E_n} \nn\\
 & \langle  n \vert   A(0)\vert m \rangle \langle m\vert  B(0)   \vert l \rangle
\langle l \vert   A(0)\vert r \rangle  \langle r\vert  B(0)   \vert n \rangle.   \nn
\eea
The other important observable  is the square of the commutator of the Louiville-like  operators  separated in time \cite{Maldacena:2015waa} 
\be\label{commutatorL}
C(\beta,t) = \langle [A(t),B(0)]^2 e^{-\beta H} \rangle~.
\ee 
The energy eigenvalues we shall parametrise as $n = {1\over 2} +i u$, $m = {1\over 2} +i v$,$l = {1\over 2} +i l$ and $r = {1\over 2} +i r$, thus  from (\ref{outoftime}) we shall get \cite{Babujian:2018xoy}
\bea\label{fourmatrix}
&\CD_4(\beta,t)= \int^{+\infty}_{-\infty} e^{i (u^2 -v^2 + l^2  - r^2)t - \beta( \frac{1}{4} + u^2)}   
  A_{uv}\,  B_{vl} \, A_{lr}\,  B_{ru} \,  du dv dl dr ~. 
 \eea
 \begin{figure}
 \centering
 \includegraphics[angle=0,width=5cm]{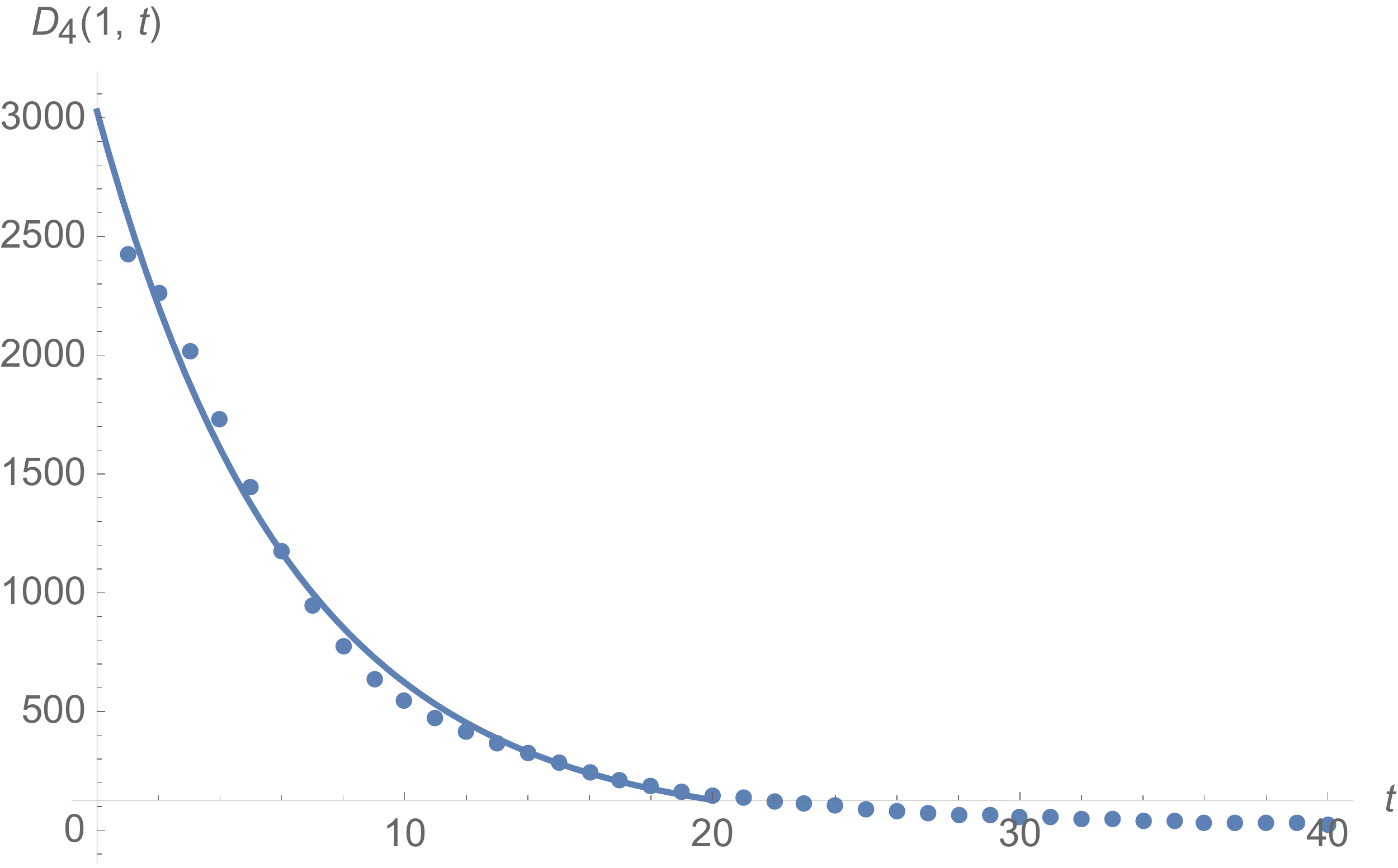}~~~~~~~~~
  \includegraphics[angle=0,width=5cm]{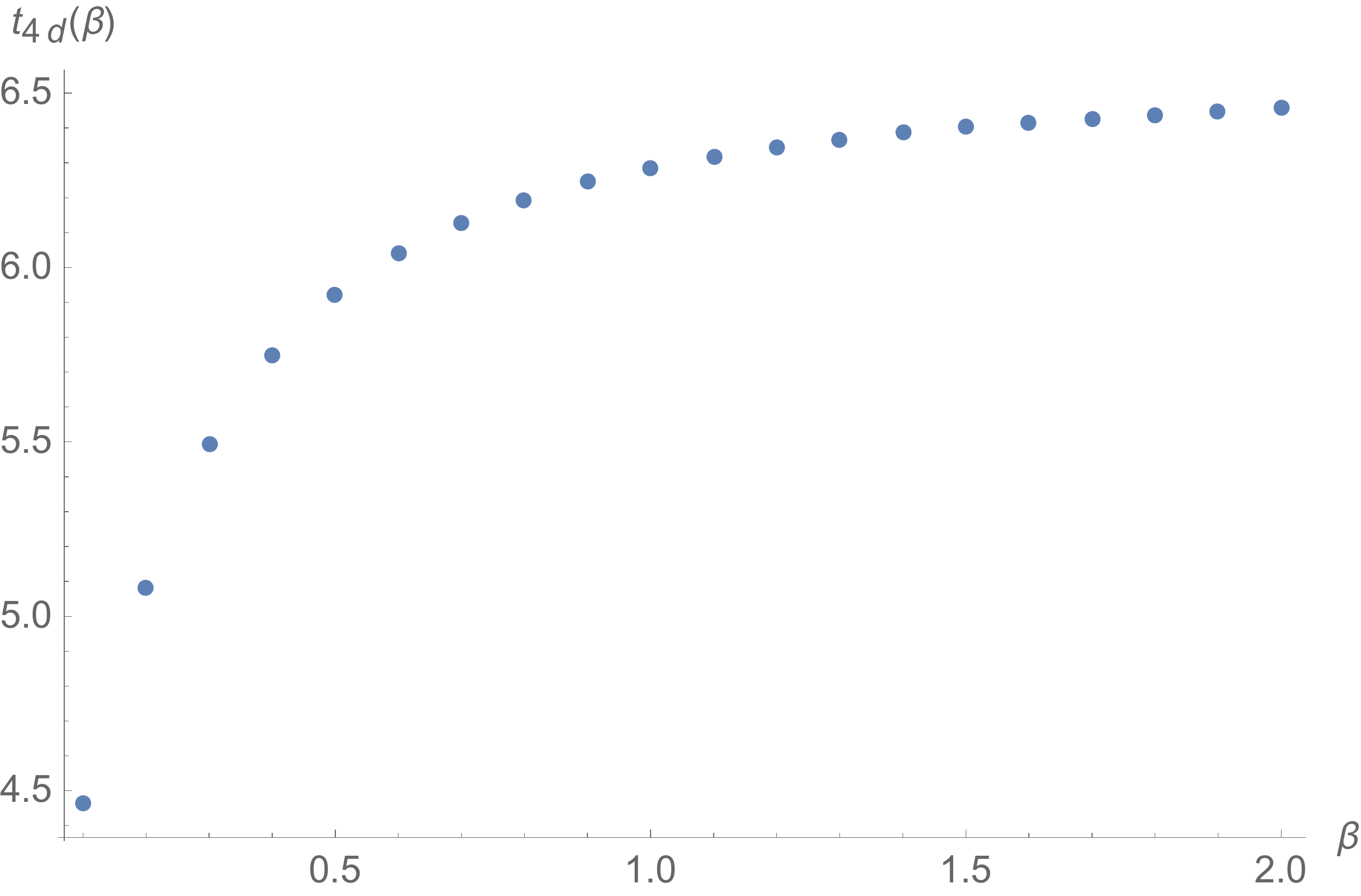}
\caption{  The exponential decay of the correlation function $\CD_4(\beta,t)$ as a function of time at $\beta =1$. The rest of the functions $\CD'_4(\beta,t),\CD''_4(\beta,t),\CD'''_4(\beta,t)$ demonstrate a similar exponential  decay $  \sim ~\exp{(-{t \over t_{4d}(\beta)} )}$. The temperature dependence of the exponent $t_{4d}(\beta)$ has a well defined high and low temperature limits and is shown on the r.h.s. graph. 
The corresponding limiting values of the function $t_{4d}(\beta)$ in dimensionless units are $t_{4d}(0)=0,112$ and $t_{4d}(\infty)=0,163$. The behaviour of the exponent $t_d(\beta)$ of the two-point correlation function is shown on the Fig.\ref{twopointfunc}.
}
\label{fourpointfunc}
\end{figure}
In terms of the variables (\ref{newvariab})   the four-point correlation function (\ref{fourmatrix})  will take the following form: 
\bea
&\CD_4(\beta,t)=  \int^{+\infty}_{-\infty} e^{i (p^2 -q^2 + l^2  - r^2)t - \beta( \frac{1}{4} + p^2)}   
 A_{pq}(N)\,  A_{ql}(M)\,  A_{lr}(N)\,  A_{rp}(M) \,   dp dq  dl dr .  ~~~~~~~~
\eea
As it was suggested in \cite{Maldacena:2015waa}, the most important correlation function indicating the traces of the classical chaotic dynamics in quantum regime is (\ref{commutatorL}) 
\bea\label{commutatorL1}
C(\beta,t) = - \CD_4(\beta,t) + \CD'_4(\beta,t) +\CD''_4(\beta,t)-\CD'''_4(\beta,t). 
\eea
In the case of the Artin system one can get \cite{Babujian:2018xoy}
\bea\label{fourpointfunct1}
 \CD'_4(\beta,t) +  \CD''_4(\beta,t) =   
  2 \int^{+\infty}_{-\infty} e^{- \beta( \frac{1}{4} + p^2)}  ~\cos{ ( q^2 - r^2)t }  \nn\\
 ~~~~~~~~~~~~~~~~~A_{pq}(N)\,  A_{ql}(M)\,  A_{lr}(N)\,  A_{rp}(M) \,   dp dq  dl dr   
\eea
and 
\bea\label{fourpointfunct2}
\CD_4(\beta,t) +  \CD'''_4(\beta,t)  
= 2 \int^{+\infty}_{-\infty} e^{- \beta( \frac{1}{4} + p^2)}  ~\cos{(p^2 -q^2 + l^2  - r^2)t }   \nn\\
 ~~~~~~~~~~~~~~~~~A_{pq}(N)\,  A_{ql}(M)\,  A_{lr}(N)\,  A_{rp}(M) \,   dp dq  dl dr. 
\eea
The Fig.\ref{fourpointfunc} shows the behaviour of the four-point correlation $\CD_4(\beta,t)$ as the function of the temperature and time.  All four correlation functions decay exponentially: 
\be\label{fourpoint}
\CD_4(\beta,t)  \sim K(\beta)~e^{-{t \over t_{4d}(\beta)}}.
\ee
The  four-point correlation functions $\CD_{4}(\beta,t)$ do not have a simple exprssion in terms of the two-point  correlation functions $\CD_2(\beta,t)$, as one can see from the data presented on Fig.\ref{twopointfunc} and Fig.\ref{fourpointfunc}. 
\begin{figure}
 \centering
 \includegraphics[angle=0,width=4cm]{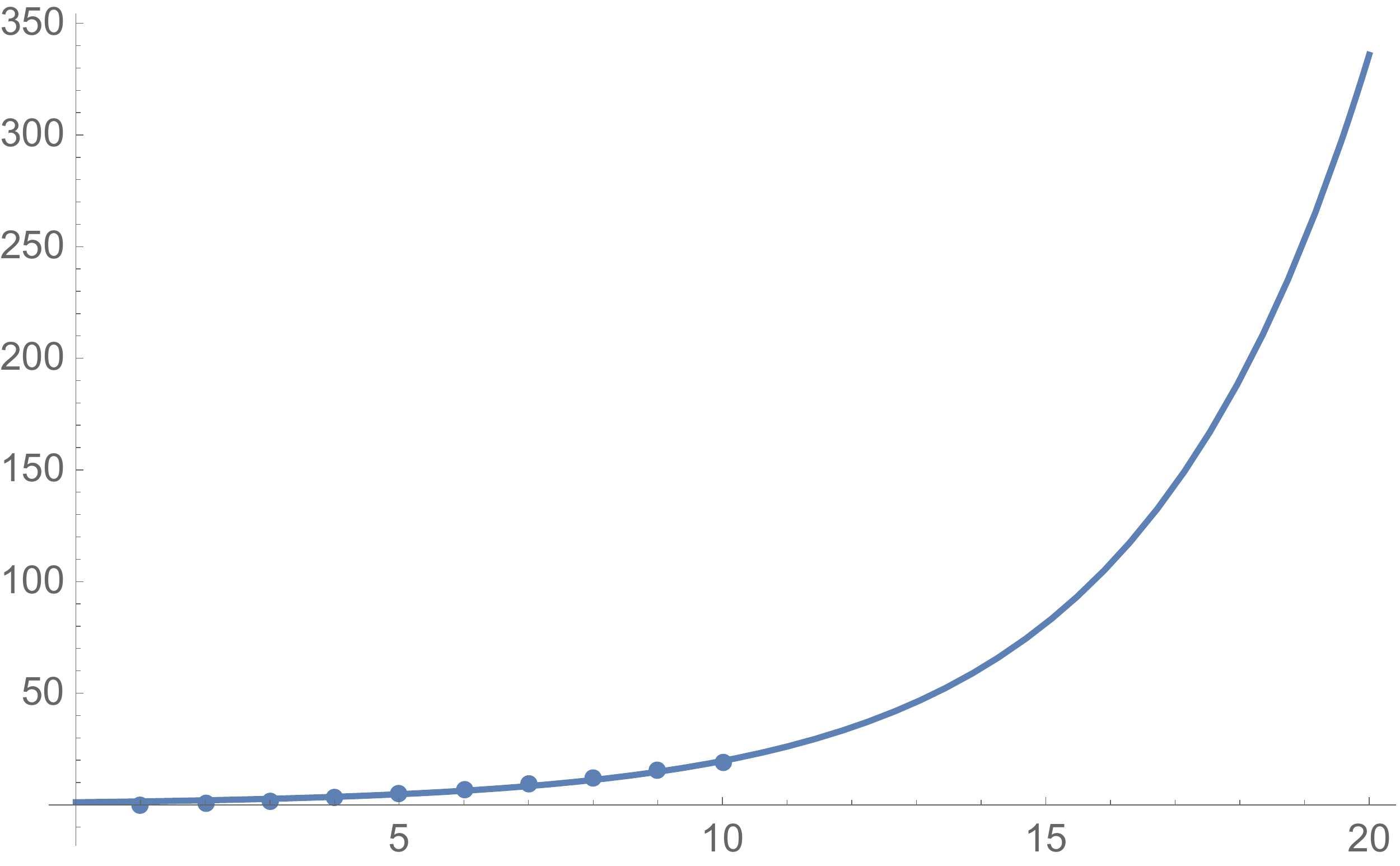}~~~~~~
  \includegraphics[angle=0,width=4cm]{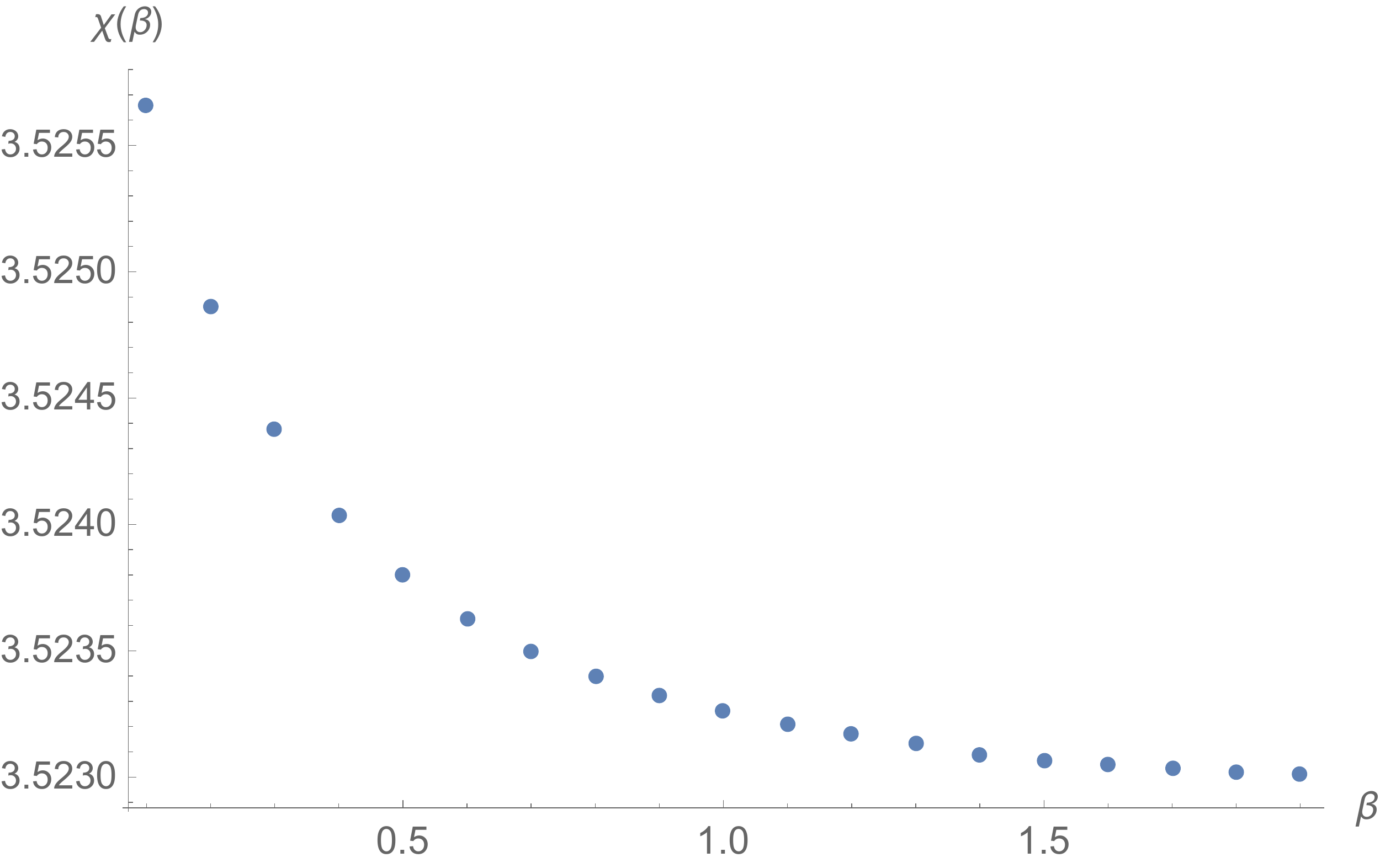}
\caption{  Time evolution of the correlation function $C(\beta,t)$ (\ref{commuta}) at temperature $\beta =0.1$.  For the short  time intervals the function $C(\beta,t)$ exponentially increases with time. This behaviour is reminiscent to the exponential divergency of the classical  trajectories in hyperbolic dynamical systems.  As one can see, the exponent  $\chi(\beta)$ which defines the behaviour of the correlation function of the operators separated in time in  the commutator (\ref{commuta}), (\ref{exponentc1}) slowly decreases with $\beta$. Such behaviour of the correlation function $C(\beta,t)$ does not saturate the maximal growth future (\ref{lineargrowth}) which is linear in $\beta$. 
}
\label{timeevolutionofcommutator}
\end{figure}
Turning to the investigation of the commutator (\ref{commutatorL}), (\ref{commutatorL1}) it is convenient to represent it in the following form \cite{Babujian:2018xoy}:  
\bea\label{commuta}
&C(\beta,t) =2 \int^{+\infty}_{-\infty} e^{- \beta( \frac{1}{4} + p^2)} 
~\{ \cos{ ( q^2 - r^2)t }  - \cos{ (p^2 -q^2 + l^2  - r^2)t }  \}\nn\\
&A_{pq}(N)\,  A_{ql}(M)\,  A_{lr}(N)\,  A_{rp}(M) \,   dp dq  dl dr  ,
\eea
where  (\ref{fourpointfunct1}) and (\ref{fourpointfunct2}) have been  used.  It was conjectured in \cite{Maldacena:2015waa}  that the influence of chaos on the commutator $C(\beta,t)$  can develop no faster than exponentially: 
\be \label{lineargrowth}
C(\beta,t) \approx f(\beta)\, e^{{2\pi  \over \beta} t} ,~~ 
\ee
with the exponent ${2\pi  \over \beta}t = 2\pi T t  $, which is growing linear in temperature $1/ \beta =T$ and time $t$. 
Calculating the function $C(\beta,t)$ one can check if in case of classically chaotic Artin  system the  grows is exponential: 
\be\label{exponentc1}
C(\beta,t) \sim f(\beta) \,e^{ {2\pi  \over  \chi(\beta)} t }, 
\ee
and if the exponent $\chi(\beta)$ grows linearly with  $\beta$. 

The results of the integration are presented on the Fig.\ref{timeevolutionofcommutator}. This beautifully confirms the fact that the correlation function $C(\beta,t)$ indeed grows exponentially with time  as it takes place in its classical counterpart. As one can see, the exponent  $\chi(\beta)$ defining the behaviour of the commutator $C(\beta,t)$   in (\ref{commuta}) and (\ref{exponentc1}) slowly decreases with $\beta$. Such behaviour of the commutator $C(\beta,t)$ does not saturate the maximal growth of the correlation function which is linear in $\beta$.

In order to check if the results are sensitive to the truncation of the high modes of the Maass wave function (\ref{elegant}) one can  include the high modes into the integration of the  basic matrix element $A_{uv}$  in (\ref{basicmatrix}).  It has been found that  their influence on the behaviour of the correlation functions is negligible. The numerical values of the exponents $t(\beta)$ and $\chi(\beta)$ are changing in the range of few percentage and do not influence the results.  In summary,  all two and four-point correlation functions decay exponentially.   The commutator $C(\beta,t)$ in (\ref{commuta}) and (\ref{exponentc1}) grows exponentially with exponent which is almost constant Fig.\ref{timeevolutionofcommutator}. This behaviour does not saturate the condition of the maximal growth (\ref{lineargrowth}).

\section{\it  Artin Resonances and Riemann Zeta Function Zeros  }

Here we shall demonstrate that the Riemann zeta function zeros define the position and the widths of the resonances of the quantised Artin dynamical system \cite{Savvidy:2018ffh}.  As it was discussed in previous sections 
the Artin dynamical system is defined on the fundamental region of the modular group on the Lobachevsky plane. It has a finite area and an infinite extension in the vertical direction that correspond to a cusp Fig.\ref{fig11}.  In classical regime the geodesic flow on this non-compact surface of constant negative curvature represents one of the most chaotic dynamical systems, has mixing of all orders, Lebesgue spectrum and non-zero Kolmogorov entropy.  In quantum-mechanical regime the system can be associated with the narrow infinitely long waveguide stretched out to infinity along the vertical axis  and a cavity resonator attached to it at the bottom.  That suggests a physical interpretation of the Maass automorphic wave function in the form of an incoming plane wave of a given energy entering the resonator,  bouncing and scattering to infinity.  As  the energy of the incoming wave comes close to the eigenmodes of the cavity a pronounced resonance behaviour shows up in the scattering amplitude \cite{Savvidy:2018ffh}.

\begin{figure}
 \centering
 \includegraphics[angle=0,width=4cm]{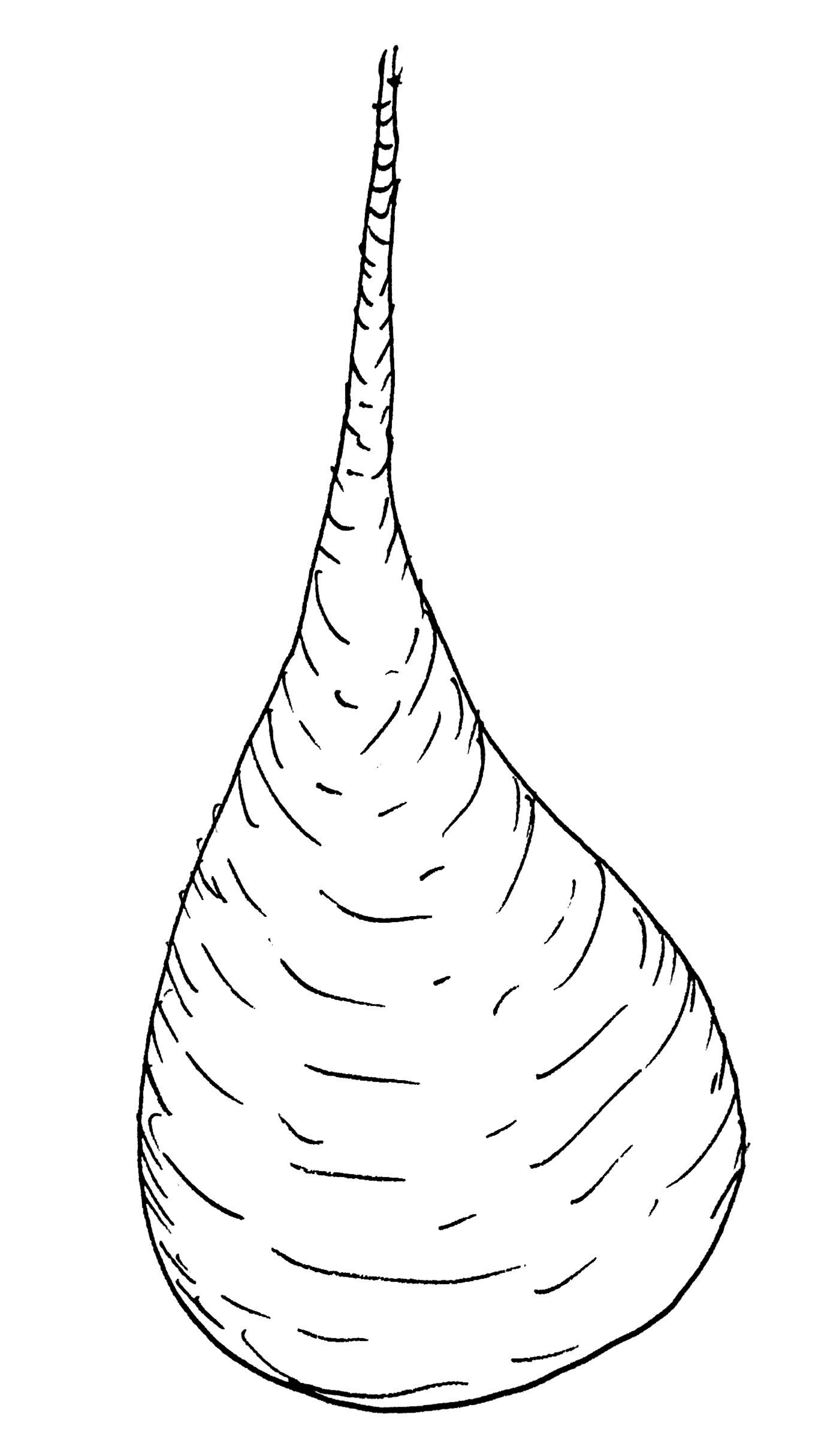}
\caption{ The Arin  system is defined on a non-compact surface $\bar{\CF}$ of constant negative curvature which has a topology of sphere with a cusp on the north pole which is stretched to infinity. The deficit angles on the vertices of the Artin surface can be computed using the formula $2\pi - \alpha$, thus $\int K \sqrt{g} d^2 \xi = (-1) {\pi \over 3} + (2\pi - 2{ \pi \over 3})+ (2\pi - 2 { \pi \over 2}) + (2\pi - 0 ) =4\pi$.}
\label{fig11}
\end{figure} 

We already presented above   (\ref{alterwave})  the Maass wave function \cite{maass} in terms of the  natural physical variable $\tilde{y}$, which is the distance in the vertical direction on the Lobachevsky plane  $  \ln y = \tilde{y} $,   and of the corresponding momentum $p$ \cite{Babujian:2018xoy}.  The plane wave  $e^{-i p \tilde{y}  }$ incoming from infinity $\CD$ of the $y$ axis on Fig.\ref{fig5}, Fig.\ref{scattering} and Fig.\ref{fig11}  elastically scatters on the boundary $ACB$ of the fundamental triangle  $\CF$.  The reflection amplitude is a pure phase and is given by the expression in front of the outgoing plane wave $e^{  i p \tilde{y}}$ :
\be\label{Smatrix}
S={\theta(\frac{1}{2} +i p) \over \theta(\frac{1}{2} -i p)}  =\exp{[ 2\, i\, \delta(p)]}.
\ee
The other terms  of the wave function describes the standing waves $\cos(2\pi l x)$  in the $x$ direction between the boundaries $x=\pm 1/2$
with the amplitudes $K_{i p }(2 \pi  l e^{\tilde{y} })$, which are exponentially decreasing with index $l$. The continuous energy spectrum is given by the formula   \cite{Babujian:2018xoy}
\be\label{energy}
E=   p^2  + \frac{1}{4} .
\ee
{\it In physical terms the system can be described as a narrow infinitely long waveguide stretched out to infinity along the vertical dierection  and a cavity resonator attached to it at the bottom  $ACB$ } 
(see Fig.\ref{scattering} and Fig.\ref{fig11}).  In order to support this interpretation we can calculate the area of the Artin surface which is below the fixed coordinate $y_0= e^{\tilde{y}_0}$:
\be
\text{Area}(\CF_0)= \int _{-\frac{1}{2}}^{\frac{1}{2}} dx
\int _{\sqrt{1-x^2}}^{y_0 }\frac{dy}{y^2}  =\frac{ \pi }{3}- 2 e^{-\tilde{y}_0}\, = \text{Area}(\CF)-  e^{-\tilde{y}_0}\, ,
\ee 
and confirm that the area above  the ordinate $\tilde{y}_0$ is exponentially small: $ e^{-\tilde{y}_0}$. The horizontal ( $dy =0$) size of the Artin surface also decreases exponentially in the vertical direction:
\be\label{111}
L_0=2 \int  ds =2 \int \frac{\sqrt{dx^2+dy^2}}{ y}=  \int _{-\frac{1}{2}}^{\frac{1}{2}} {dx\over y_0 } =  e^{-\tilde{y}_0}.
\ee 
One can suggest therefore the following physical interpretation of the Maass wave function (\ref{alterwave}): The incoming plane wave $e^{- i p \tilde{y}}$ of energy $E=p^2  + \frac{1}{4} $ enters the "cavity resonator", 
bouncing  back into the outgoing plane wave at infinity $e^{i p \tilde{y}}$.  As the energy of the incoming wave $E =p^2  + \frac{1}{4}$ close to the eigenmodes of the cavity  resonator one should expect a pronounced resonance behaviour of the scattering amplitude \cite{Savvidy:2018ffh}. 
\begin{figure}
 \centering
 \includegraphics[angle=0,width=6cm]{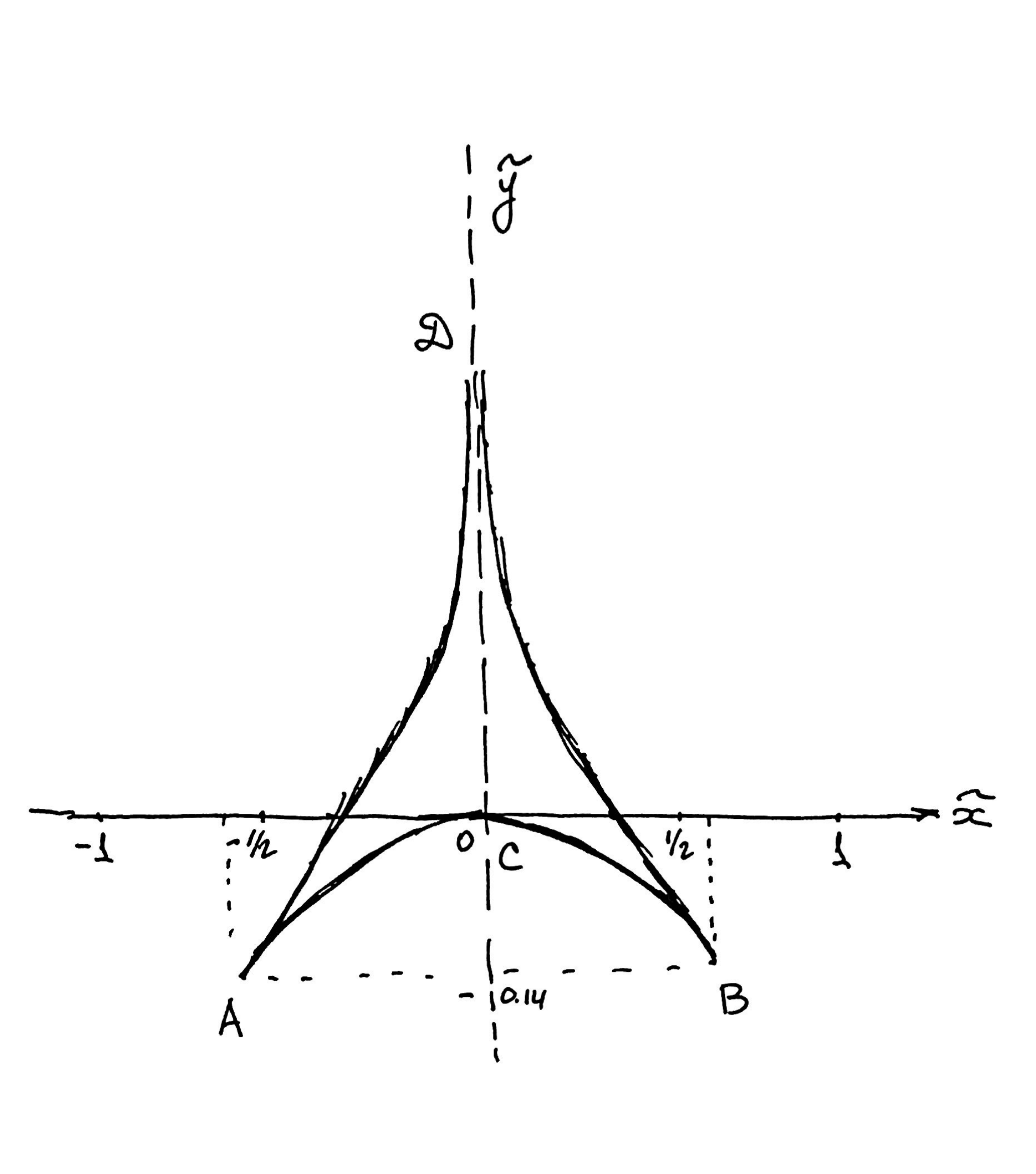}
\caption{ The system can be described as a narrow (\ref{111}) infinitely long waveguide stretched to infinity along the vertical dierection  and a cavity resonator attached to it at the bottom $ACB$.}
\label{fig10}
\end{figure}

To trace such behaviour let us consider the analytical continuation of the Maass wave function (\ref{alterwave}) to the complex  energies $E$. The analytical continuation of the scattering 
amplitudes as a function of the energy $E$ considered as a complex variable allows to establish important spectral properties of the quantum-mechanical system. In particular,  the method of analytic continuation allows to determine the real and complex S-matrix poles. The real  poles on the physical sheet correspond  to the  discrete energy levels and the complex poles on the second sheet below the cut correspond  to the resonances in the quantum-mechanical system  \cite{landauqmech} .  The asymptotic form of the wave function  can be represented in the following form: 
\be\label{reswave4}
\psi = A(E)\, e^{i p \tilde{y}} + B(E)\, e^{-i p \tilde{y}} , ~~~p = \sqrt{E-1/4}.
\ee
In order to make the functions $A(E) $ and $B(E) $ single-valued one should cut the complex plane 
along the real axis \cite{landauqmech} starting from $E=1/4$. The complex plane with a cut so defined a physical sheet.  To the left from the cut,  at  energies  $E_0 < 1/4$,  the wave function takes the following form: 
\be\label{reswave1}
\psi = A(E)\, e^{- \sqrt{\vert E-1/4 \vert } \tilde{y}} + B(E)\, e^{ \sqrt{\vert E-1/4 \vert} \tilde{y}}, 
\ee
where the exponential factors are real and one of them decreases and the other one increases at $\tilde{y} \rightarrow \infty$.  The  bound states are characterised  by the fact that the corresponding wave function tends to zero at infinity $\tilde{y} \rightarrow \infty$.  
This means that the second term in (\ref{reswave1})   should be absent, and a discrete energy level $E_0 < 1/4$ corresponds to a zero of the $B(E)$ function \cite{landauqmech}: 
\be\label{zeros2}
B(E_0) =0.~~~
\ee
Because the energy eigenvalues are real, all zeros of $B(E)$ on the physical sheet are real. 
Now consider a system which is unbounded and its energy spectrum has a continuous part \cite{landauqmech}. The energy spectrum can be quasi-discrete, consisting of smeared levels of  a width $\Gamma$.  In describing  such states one should describe the wave packet moving to infinity, thus only outgoing waves should be presence at infinity. This boundary condition involves complex quantities and the energy eigenvalues in general are also complex \cite{landauqmech}. With such boundary conditions the Hermitian energy operators can have complex eigenvalues of the form \cite{landauqmech}
\be\label{compeigen1}
E = E_0 - i {\Gamma \over 2},
\ee
where $E_0$ and $\Gamma$ are both real and positive.  
\begin{figure}
 \centering
 \includegraphics[angle=0,width=8cm]{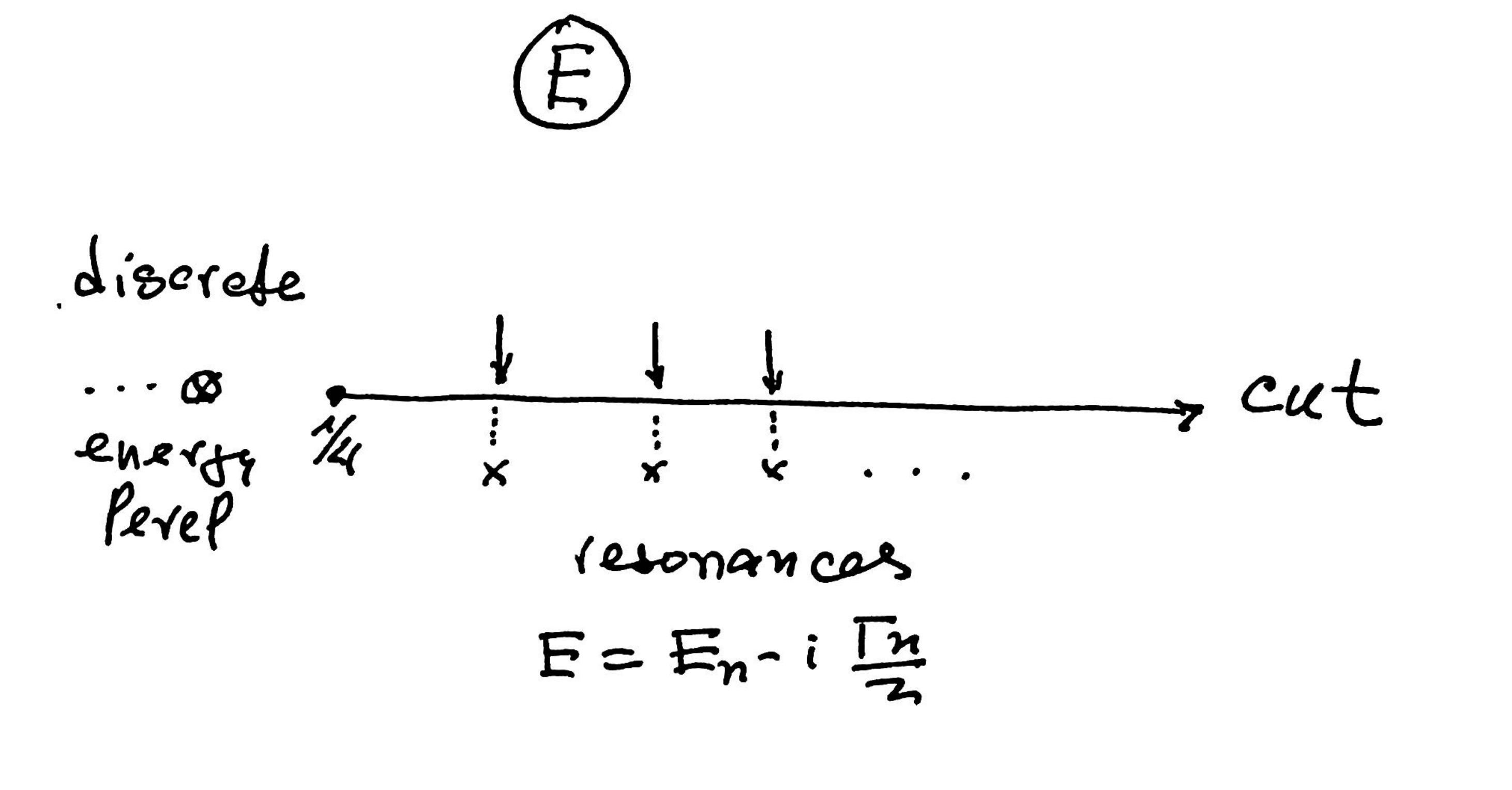}
\caption{ The resonances  $E_n - i {\Gamma_n \over 2}$ are located under the right hand side of the real axis.}
\label{fig4}
\end{figure}
The condition which defines the complex energy eigenvalues (\ref{compeigen1}) reduces to the requirement  that at $E = E_0 - i {\Gamma \over 2}$ the incoming wave $ e^{-i p \tilde{y}} $ in (\ref{reswave4}) should be absent \cite{landauqmech}:     
\be\label{zerosquasi1}
B(E_0 - i {\Gamma \over 2}) =0.
\ee
The point $E_0 - i {\Gamma \over 2}$ is located under the right hand side of the real axis, see 
Fig.\ref{fig4}. In order to reach that point without leaving the physical sheet one should move from the upper side of the cut  anticlockwise.  However in that case, the phase of the wave function changes its  sign and the outgoing wave transforms into the incoming wave. In order to keep the outgoing character of the wave function one should cross  the cut strait into the  second sheet Fig.\ref{fig4}.  Expanding the function 
$B(E)$ near the quasi-discrete energy level (\ref{compeigen1}) as $B(E) =(E - E_0 + {i \Gamma \over 2}) b +...$ one can get 
\be
\psi~~ \approx~~ b^* (E - E_0 - {i \Gamma \over 2})  e^{i p \tilde{y}} +b  (E - E_0 + {i  \Gamma \over 2})  e^{-i p \tilde{y}}
\ee
and the S-matrix  will take the following form \cite{landauqmech}
\be\label{resonphase}
S=e^{2 i \delta} = {E - E_0 - i \Gamma / 2 \over E - E_0 +i \Gamma / 2} e^{2 i \delta_0},
\ee
where $e^{2 i \delta_0} =  b^* / b $. One can observe that moving throughout the resonance region the phase is changing by $\pi$.

Let us now consider the asymptotic  behaviour of the wave function (\ref{alterwave}) at large $\tilde{y}$.  The conditions (\ref{zeros2}) and  (\ref{zerosquasi1}) of the absence of incoming wave takes the form \cite{Savvidy:2018ffh}:
\be
 \theta(\frac{1}{2} -i p) =0
\ee
and due to  (\ref{thata1}):
\bea 
\theta(\frac{1}{2} -i p)=  {\zeta (1- 2 i p) \Gamma (\frac{1}{2} -i p) \over  \pi ^{\frac{1}{2} -i p} }=0.
\eea  
The solution of this equation can be expressed in terms of zeros  of the Riemann zeta function
\cite{Riemann}:
\be
\zeta(\frac{1}{2} - i u_n) =0,~~~~  n=1,2,....~~~~u_n > 0.
\ee
Thus one should solve the equation 
\be
1- 2 i p_n = \frac{1}{2} - i u_n~.
\ee 
The location of poles is therefore at the following values of the complex momenta 
\be\label{poles}
p_n = {u_n \over 2} - i \,{1 \over 4} \, ,  ~~~~~  n=1,2,..... 
\ee
and at the corresponding complex energies (\ref{energy}) :
\be\label{resonances34} 
E = p^2_n +{1 \over 4}~ =~ ({u_n \over 2} -   \,{1 \over 4}\,i)^2  +{1 \over 4} ~= ~{u^2_n \over 4} + {3\over 16}
- i \, {u_n \over 4}.
\ee
Thus one can observe that there are resonances (\ref{compeigen1})
\be
E = E_n - i {\Gamma_n \over 2}
\ee
at the following energies and of the corresponding widths (\ref{resonances34}) \cite{Savvidy:2018ffh}:
\be\label{exactresonance}
E_n = {u^2_n \over 4} + {3\over 16},~~~~~~\Gamma_n = {u_n \over 2} . 
\ee
The ratio of the width to the energy tends to zero  \cite{Riemann}:
\be
{ \Gamma_n  \over E_n} = {u_n \over 2}/ ({u^2_n \over 4} + {3\over 16}) \approx {2 \over u_n} 
\rightarrow ~~0
\ee
and the resonances become infinitely narrow. The ratio of the width to the energy spacing between nearest levels is  
\be
{\Gamma_n \over  E_{n+1} - E_n } = {2 u_n \over (u_{n+1} + u_n)(u_{n+1} - u_n)} \approx 
{1 \over u_{n+1} - u_n}  .
\ee
As far as the zeros of the zeta function have the property to  "repel", the difference $u_{n+1} - u_n$ can vanish with small probability \cite{Turing,Gourdon}. One can conjecture the following representation of the S-matrix (\ref{Smatrix}):
\be\label{Smatirxphase}
S=e^{2 i\, \delta} =  {\theta(\frac{1}{2} +i p) \over \theta(\frac{1}{2} -i p)} = \sum^{\infty}_{n=1}{E - E_n - i \Gamma_n / 2 \over E - E_n +i \Gamma_n / 2}~ e^{2 i \delta_n}
\ee
with yet unknown phases $\delta_n$. In order to justify the above representation of the S-matrix one can find the location of the poles on the second Riemann sheet by using expansion of the S-matrix
(\ref{Smatrix}) at the "bumps" which occur  along the real axis at energies  
\be \label{approxresonance}
E_n  = {u^2_n \over 4}  +{3\over 16}~.
\ee
The expantion will take the following form:
\bea\label{Smatirxphase1}
S\vert_{E \approx E_n} &  
= &  {\theta(\frac{1}{2} +i \sqrt{E - {1\over 4}} ) \over 
\theta( \frac{1}{2} -i  \sqrt{E - {1\over 4} }) } \vert_{E \approx E_n}~=~{ \theta(\frac{1}{2} +i \sqrt{E_n - {1\over 4}} )  +   \theta^{'}(\frac{1}{2} 
+i \sqrt{E_n - {1\over 4}} ) ~ (E - E_n)
 \over  \theta(\frac{1}{2} -i \sqrt{E_n - {1\over 4}} )  +   \theta^{'}(\frac{1}{2} -i \sqrt{E_n - {1\over 4}} ) ~  (E - E_n)} \nn\\
&=& 
 {    E - E_n + \theta(\frac{1}{2} +i \sqrt{E_n - {1\over 4}} ) /\theta^{'}(\frac{1}{2} 
 +i \sqrt{E_n - {1\over 4}} ) 
 \over    E - E_n + \theta(\frac{1}{2} -i \sqrt{E_n - {1\over 4}} )/\theta^{'}(\frac{1}{2} 
 -i \sqrt{E_n - {1\over 4}} ) } ~~ {\theta^{'}(\frac{1}{2} +i \sqrt{E_n - {1\over 4}} )   \over \theta^{'}(\frac{1}{2} -i \sqrt{E_n - {1\over 4}} )   }  \nn\\ 
 &\equiv &  ~~~
 {    E - E^{'}_n - i \Gamma^{'}_n/2   
 \over    E - E^{'}_n + i \Gamma^{'}_n/2  }~~  e^{2 i \delta^{'}_n}~,\nn\\
 \eea
where
\bea
E^{'}_n - i \Gamma^{'}_n/2 = E_n - {\theta(\frac{1}{2} -i \sqrt{E_n - {1\over 4}} ) 
\over \theta^{'}(\frac{1}{2} -i \sqrt{E_n - {1\over 4}} ) } ,~~~~ e^{2 i \delta^{'}_n}=
 {\theta^{'}(\frac{1}{2} +i \sqrt{E_n - {1\over 4}} )   \over \theta^{'}(\frac{1}{2} 
 -i \sqrt{E_n - {1\over 4}} )   } , 
\eea 
 thus
\bea\label{approxresonance1}
E^{'}_n - E_n = - \Re {\theta(\frac{1}{2} -i \sqrt{E_n - {1\over 4}} ) 
\over \theta^{'}(\frac{1}{2} -i \sqrt{E_n - {1\over 4}} ) } ,~~~~~~~~- i  \Gamma^{'}_n/2 = - \Im  {\theta(\frac{1}{2} -i \sqrt{E_n - {1\over 4}} ) 
\over \theta^{'}(\frac{1}{2} -i \sqrt{E_n - {1\over 4}} ) } 
\eea
and all quantities $E^{'}_n$ , $ \Gamma^{'}_n/2$ and $\delta^{'}_n$ are real.  Considering the first ten zeros of the zeta function which are known numerically \cite{Turing,Gourdon} one can calculate the position of the resonances and their widths using the approximation formulas (\ref{approxresonance1}) and get convinced that the energies and the widths of the resonances given by the exact formula (\ref{exactresonance}) and  the one given by the approximation formulas   (\ref{approxresonance1}) are consistent  within the two precent deviation.  Alternative attempts of the physical interpretation of the zeros of the Riemann zeta function, as well as the P\'olya-Hilbert conjecture and further references can be found in \cite{Schumayer:2011yp}.

\section{\it  C-cascades and MIXMAX Random Number Generator}

In this section we shall turn our attention to the investigation of the second class of the C-K systems defined on high dimensional tori \cite{anosov}. The automorphisms of a torus are generated by the linear transformation  
\bea\label{cmap}
x_i \rightarrow \sum^{n}_{j=1} T_{ij} x_j,~~~~(mod ~1),
\eea
where the integer matrix $T$ has a determinant equal to one $Det~ T =1$. 
In order for the automorphisms of the torus (\ref{cmap})  {\it to fulfil  the C-condition it is necessary 
and sufficient that the matrix $T$ has no eigenvalues on the unit circle.}  
Thus the  spectrum $\{ \Lambda = {\lambda_1},...,
\lambda_n \}$ of the matrix $T$ should fulfil the following 
two conditions \cite{anosov}: 
\bea\label{mmatrix}
1)&~Det~ T=  {\lambda_1}{\lambda_2}....{\lambda_n}=1\nn\\
2)&\vert {\lambda_i} \vert \neq 1, ~~~~~~~~~~~\forall i.
\eea
Because the determinant of the matrix $T$ is equal to one,
the Liouville's measure $d\mu = dx_1...dx_m$ is invariant under the action of $T$.
The inverse matrix $T^{-1}$ is also an integer matrix because $Det~ T=1$.
Therefore $T$ is an automorphism of  the torus   onto itself. All trajectories with rational coordinates $(x_1,...,x_n)$, and only they, 
are periodic trajectories of the automorphisms of the  torus (\ref{cmap}).
The above conditions (\ref{mmatrix}) on the eigenvalues of the matrix $T$ are  sufficient 
to prove that the system belongs to the class of  Anosov C-systems and therefore has {\it mixing} properties defined 
above  (\ref{mix1})- (\ref{mixnn}). Because the C-systems have {\it mixing of 
all orders} \cite{anosov}  it follows that the C-systems are exhibiting  the decay of the 
 correlation functions of any order. The entropy of the Anosov automorphisms on a torus (\ref{cmap}), (\ref{mmatrix}) can be calculate and is equal to the sum \cite{anosov,smale,sinai2,margulis,bowen0,bowen,bowen1}:
\be\label{entropyofT}
h(T) = \sum_{\vert \lambda_{\beta} \vert > 1} \ln \vert \lambda_{\beta} \vert.
\ee
\begin{figure}
 \centering
\includegraphics[width=7cm]{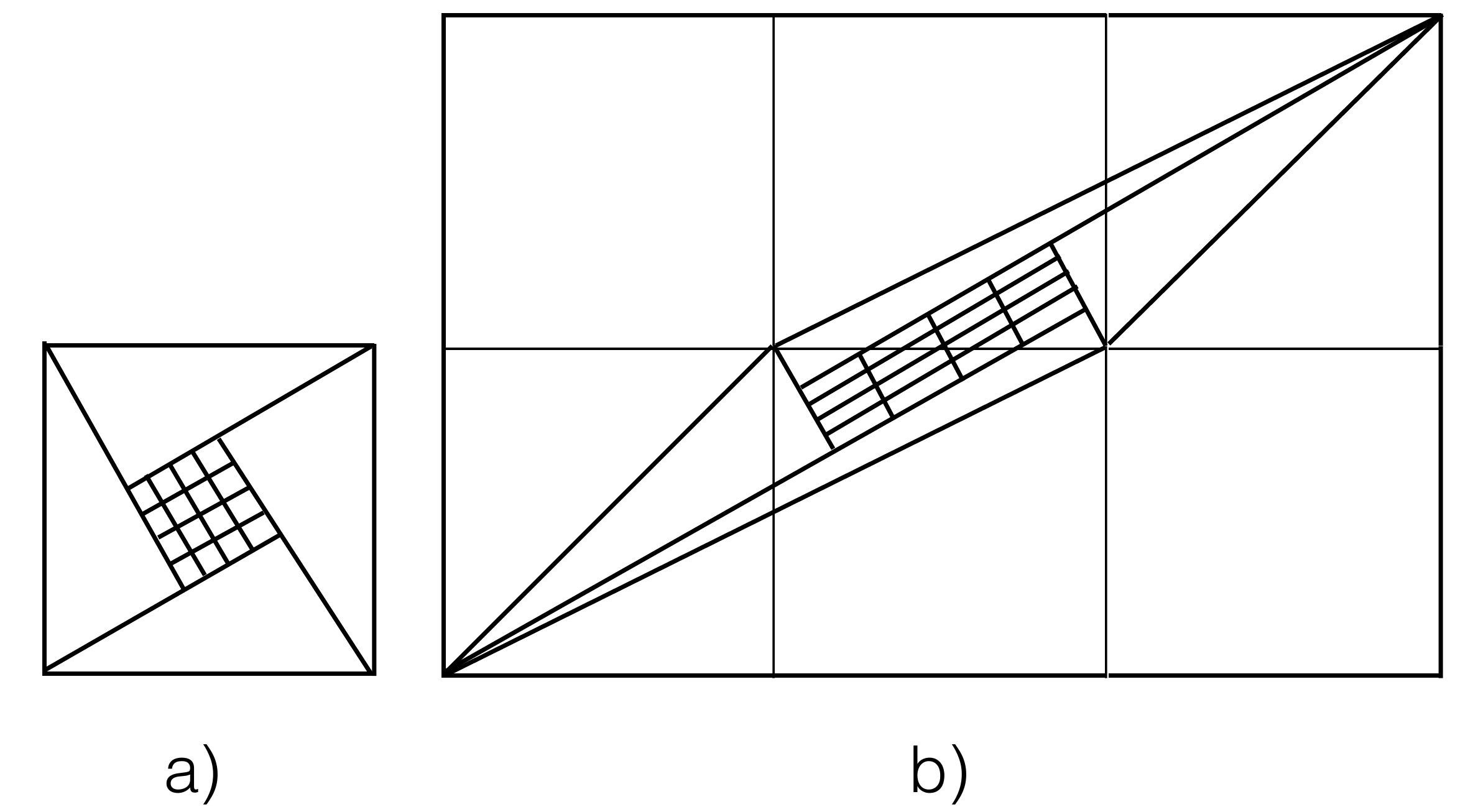}
\centering
\caption{The automorphisms on a two-torus.   
The a) depicts the parallel lines along the eigenvectors  
and b) depicts their positions after the action of the automorphism.} 
\label{fig2}
\end{figure}
{\it Thus the entropy $h(T)$ directly depends on the spectrum  of the operator $T$.  
This fact allows to characterise and compare the chaotic properties of  
dynamical C-systems quantitatively computing and comparing their entropies. }

A strong instability of trajectories of a dynamical C-system leads to the appearance   
of statistical properties in its behaviour \cite{leonov}.  As a result the time average  $
 \bar{f_N}(x) ={1 \over N} \sum^{N-1}_{n=0} f(T^n x)
$
of the function $f(x)$ on phase space $M$  behaves  as a superposition of quantities which are statistically weakly dependent. 
Therefore for the C-systems on a torus  it was  demonstrated that
the fluctuations of the time averages from the phase space integral 
$
\langle f \rangle= \int_{M} f(x) d x 
$
multiplied by $\sqrt{N}$ 
have at large $N \rightarrow \infty$ the Gaussian distribution \cite{leonov}:
\be\label{gauss}
\lim_{N\rightarrow \infty}\mu \bigg\{ x	:\sqrt{N}   \left(  \bar{f_N}(x)  - \langle f \rangle \right) < z    \bigg\}
= {1 \over \sqrt{2 \pi} \sigma_f}\int^{z}_{-\infty} e^{-{y^2 \over 2 \sigma^2_f}} dy.
\ee
The quantity  
$
\sqrt{N}   \Bigg(  \bar{f_N}(x)  -  \langle f \rangle  \Bigg)
$
converges in distribution to the normal random variable with standard deviation  $\sigma_f$  
\be
\sigma^2_f = \sum^{+ \infty}_{n=-\infty} [\langle f(x) f(T^n x) \rangle-
\langle f(x)  \rangle^2 ].
\ee
We were able to express it in terms of entropy 
\be
\sigma^2_f =\sum^{+ \infty}_{n=-\infty} { M^2  \over   128 \pi^4  } ~  e^{-4 n h(T) }=
{M^2 \over 128 \pi^4} {e^{4 h(T)} +1\over e^{4 h(T)} -1 }.
\ee
During the Meeting Igor ask me if the C-K systems have temperature? It seems to me that in accordance with the above result one can associate the $\sigma_f$ with the temperature if one compare the above Gaussian distribution with Gibbs distribution $kT = \sigma_f$.

It follows from the Anosov results that these hyperbolic C-systems are K-systems as well and are therefore maximally chaotic.  It was suggested in 1986 in \cite{yer1986a} to use the C-K systems defined on a torus to generate high quality pseudorandom numbers for Monte-Carlo method. 
The modern powerful computers open a new era for the application of the Monte-Carlo Method  \cite{metropolis,neuman,neuman1,sobol,yer1986a,Demchik:2010fd,falcion} 
 for the simulation of physical systems with many degrees of freedom and of 
higher complexity. The Monte-Carlo simulation is an important computational 
technique in many areas of natural sciences, and it has significant application 
in particle and nuclear physics, quantum physics, statistical physics, 
quantum chemistry, material science, among many other multidisciplinary applications. 
At the heart of the Monte-Carlo (MC) simulations are pseudo Random Number Generators (RNG).

Usually  pseudo random numbers are generated by deterministic recursive rules
\cite{yer1986a,metropolis,neuman,neuman1,sobol}. 
Such rules produce pseudorandom numbers, and it is a great challenge to design 
pseudo random number generators that produce high quality sequences. 
Although numerous RNGs introduced in the last  decades fulfil most of the 
requirements and are frequently used in simulations, each of them has some 
weak properties which influence the results \cite{pierr} and are less suitable for demanding 
MC simulations which are performed for the high energy experiments at CERN
and other research centres. 
The RNGs are essentially used in high energy experiments at CERN  for the design of the efficient particle detectors and for the statistical analysis of the experimental data \cite{cern}. 
\begin{figure}
  \centering
\includegraphics[width=8cm]{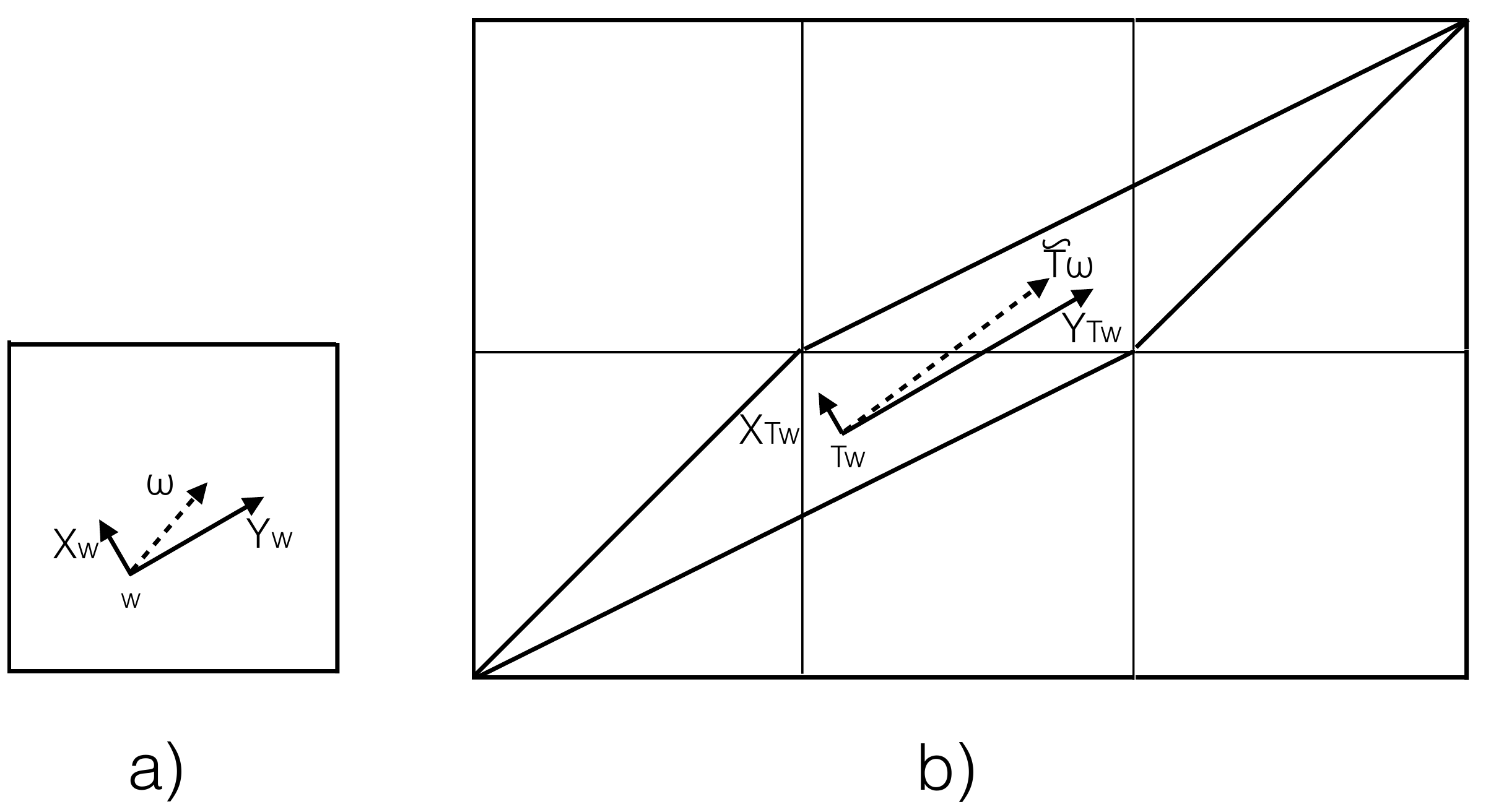}
\centering
\caption{The tangent vector $\omega \in R_{x}$ at  $x \in M^2$  is decomposable  
into the sum $R_{x}= X_{x} \bigoplus Y_{x}$
where the spaces $X_x$ and $Y_x$ are defined by the  
eigenvectors of the $2 \times 2 $ matrix $T(2,0)$ (\ref{eq:matrix}).  
 It is exponentially contracting  the distances on  $ X_{x}  $ and expanding the 
distances on  $ Y_{x} $ (details are given in Appendix A).} 
\label{fig3}
\end{figure}

In order to fulfil these demanding requirements it is necessary to have a solid theoretical 
and mathematical background on which the RNG's are based. RNG  should have a long period, 
be statistically robust, efficient, portable and have a possibility to change and 
adjust the internal characteristics in order to make RNG suitable for concrete 
problems of high complexity. In \cite{yer1986a} it was suggested that  
Anosov C-systems \cite{anosov},  defined on a high dimensional 
torus, are excellent candidates for the pseudo-random number generators.
The C-system chosen in \cite{yer1986a} was the one which realises 
linear automorphism  $T$  defined  in (\ref{cmap}).   For convenience 
in this section the dimension $n$ of the phase space $M$ is denoted by  $N$.
A particular matrix chosen in \cite{yer1986b} was defined for all $N \geq 2$. The operators $T(N,s)$  
are parametrised by the integers  $N$ and $s$ 
\be
\label{eq:matrix}
T(N,s) = 
   \begin{pmatrix} 
      1 & 1 & 1 & 1 & ... &1& 1 \\
      1 & 2 & 1 & 1 & ... &1& 1 \\
      1 & 3+s & 2 & 1 & ... &1& 1 \\
      1 & 4 & 3 & 2 &   ... &1& 1 \\
      &&&...&&&\\
      1 & N & N-1 &  N-2 & ... & 3 & 2
   \end{pmatrix}
\ee
Its  entries are all integers  $T_{ij} \in \mathbb{Z}$ and 
 $Det~ T =1$. The spectrum and the value of the Kolmogorov entropy can be calculated. 
 \begin{figure}
 \centering
\includegraphics[width=4cm]{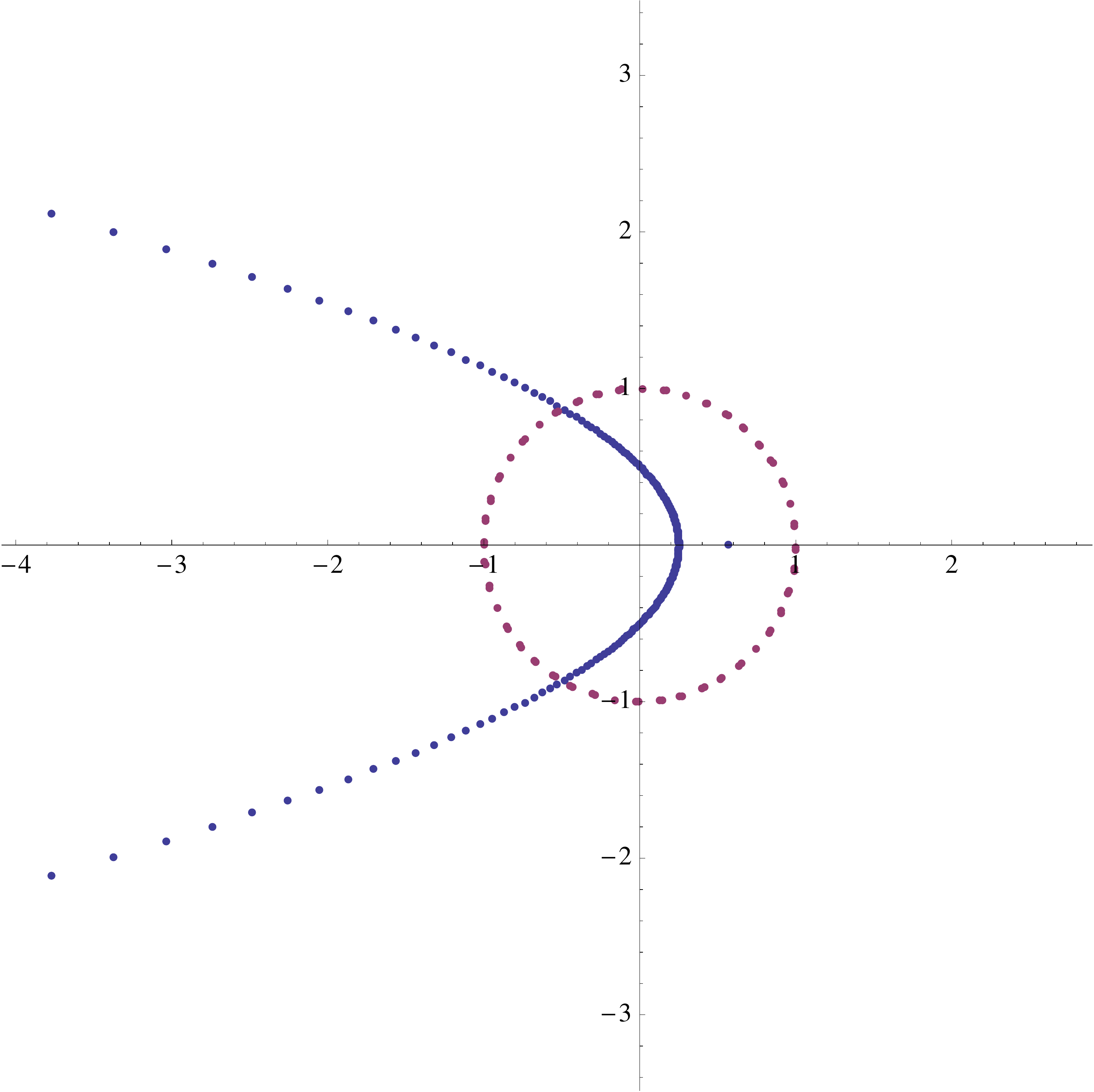}~ 
\includegraphics[width=4cm]{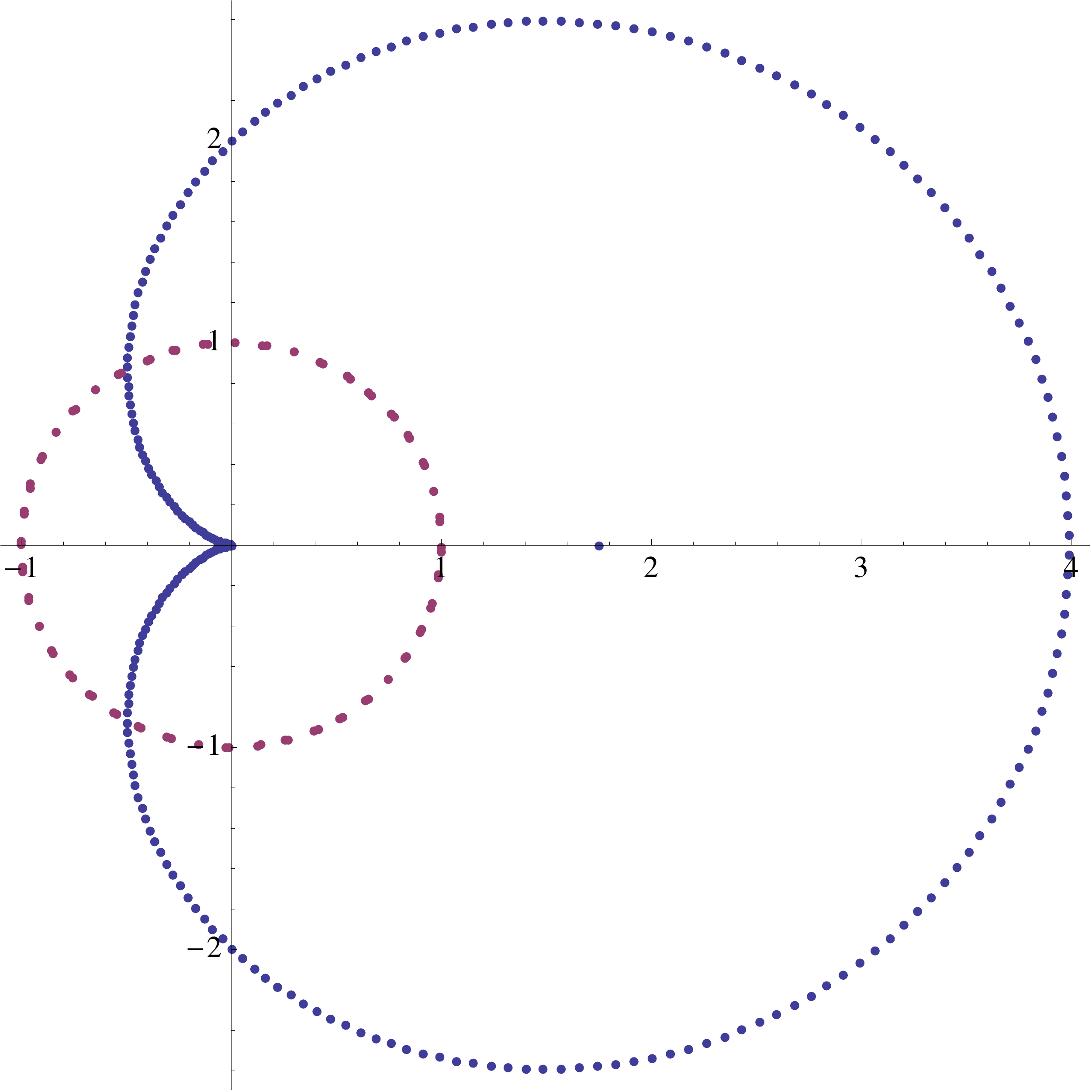}~~~~~~~~~
\includegraphics[width=5cm]{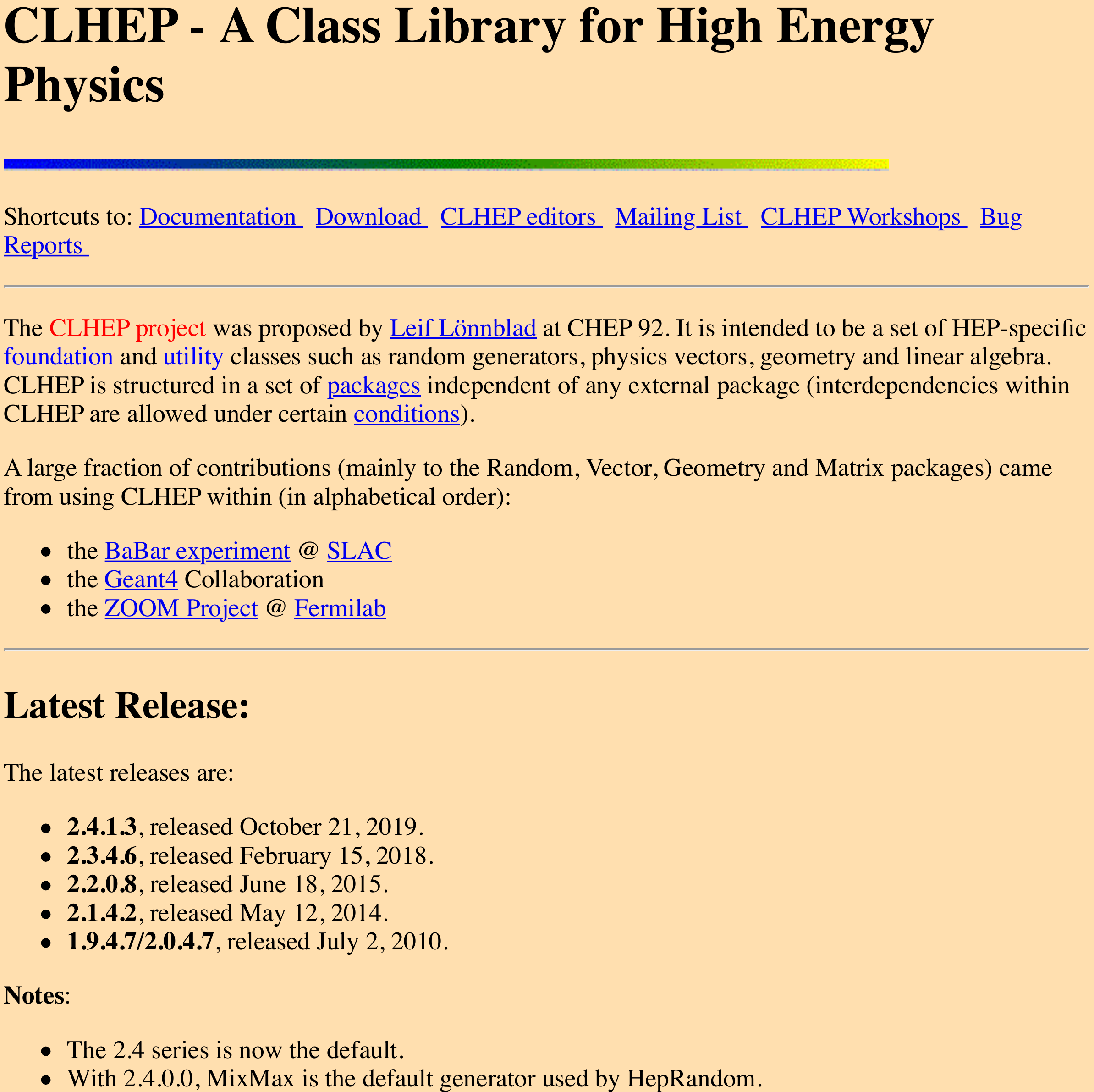}
\centering
\caption{The eigenvalue distribution of $T(N,s)$ and of $T^{-1}(N,s)$.  All eigenvalues are lying  outside of the unit circle. On the r.h.s.  is the MIXMAX generator implemented as the default engine  into  the Geant4/CLHEP toolkit at CERN.} 
\label{fig4}
\end{figure}
 It is defined recursively, 
since the matrix of size $N+1$ contains in it the matrix of the size $N$.  
The only variable entry in the matrix is $T_{32} =3+s$, where  $s$  
should be chosen such that to avoid eigenvalues lying on a unit circle.
In order to generate pseudo-random vectors  $x_n = T^n x$, one should 
 choose  the initial vector $x=(x_{1},...,x_{N})$, called the ``seed",  
 with at least one non-zero component to avoid fixed point of $T$, which is at the origin.
The eigenvalues of the $T$ matrix (\ref{eq:matrix}) are widely dispersed for all $N$,
see Fig.\ref{fig4}
from reference \cite{konstantin}. The spectrum  is "multi-scale", with trajectories exhibiting exponential instabilities at different scales \cite{yer1986a}. The spectrum of the operator $T(N,s)$
has two real eigenvalues for even $N$ and three for odd $N$, all the rest
of the eigenvalues are complex and lying on leaf-shaped curves.
It is seen that the spectrum tends to a universal limiting form as
$N$ tends  to infinity, and the complex eigenvalues  $ 1/\lambda$
(of the inverse operator) lie asymptotically on the cardioid  curve Fig.\ref{fig4} which has the representation
\be 
r(\phi) = 4 \cos^2(\phi/2)
\label{eq:curve}
\ee
in the polar coordinates $ \lambda = r \exp(i\phi)$. From the above 
analytical expression for eigenvalues it follows that the eigenvalues satisfying the 
condition $ 0 <  \vert \lambda_{\phi} \vert   < 1$  are in the range $-2\pi/3 <  \phi < 2\pi/3$ 
and the ones satisfying the 
condition $ 1 <  \vert \lambda_{\phi} \vert $ are in the interval $2\pi/3 <  \phi < 4\pi/3$.
One can conjecture that there exists a limiting infinite-dimensional dynamical system with continuous space coordinate and discrete time with the above spectrum.  
The entropy of the C-K system $T(N,s)$ can now be calculated for
large values of $N$  as an integral over eigenvalues (\ref{eq:curve}):
\be\label{linear}
h(T)= \sum_{\alpha   } \ln \vert {1 \over \lambda_{\alpha} }\vert = \sum_{-2\pi/3 <  \phi_i < 2\pi/3} \ln (4\cos^2(\phi_i/2)~ \rightarrow  ~N \int^{2\pi/3}_{-2\pi/3} \ln (4\cos^2(\phi/2){d\phi \over 2\pi}=
{2\over \pi}~ N
\ee
and to confirm  that the entropy  {\it increases linearly with the dimension $N$ of the operator} $T(N,s)$.
In the paper \cite{konstantin} the period of
the trajectories of the system $T(N,s)$ was found which is characterised by 
a prime number $p$\footnote{The
general theory of Galois field and the periods
of its elements can be found in \cite{mixmaxGalois,lnbook,nied,niki}.} .
In  \cite{konstantin} the necessary and sufficient criterion were
formulated for the sequence to be of the maximal possible period:
\be\label{period1}
\tau={p^N-1 \over p-1} \sim e^{(N-1)\ln p}.
\ee
{\it It follows then that the period of the trajectories exponentially increases with
the size of the  operator $T(N,s)$}. Thus the knowledge of the spectrum allows to
calculate the entropy (\ref{linear}) and the period (\ref{period1})   of the
trajectories.  The number of periodic trajectories of a period less than $\tau$ behaves as
\be\label{density1}
\pi(\tau) \sim \exp{({2 N \tau \over \pi})} /\tau.
\ee
In summary we have  the spectrum given by (\ref{eq:curve}), the entropy by (\ref{linear}),
the period on a rational sublattice  by (\ref{period1}) and the corresponding  density by (\ref{density1})
of the C-K system $T(N,s)$.

\section{\it Acknowledgement }
Preliminary versions of this work were presented at the Steklov Mathematical Institute (September 10, 2019) as well as the  CERN Theory Department and A. Alikhanian National Laboratory in Yerevan, where part of this work was completed.  I thank these institutions for their hospitality.   I would like to thank Luis Alvarez-Gaume for stimulating discussions, for kind hospitality at Simons Center for Geometry and Physics and providing to the author the references \cite{hejhal}, \cite{hejhal1} and \cite{hejhal2}. I would like to thank  H.Babujyan, R.Poghosyan and K.Savvidy for collaboration and enlightening  discussions.

\section{\it Appendix A}
\begin{figure}
 \centering
\includegraphics[width=9cm]{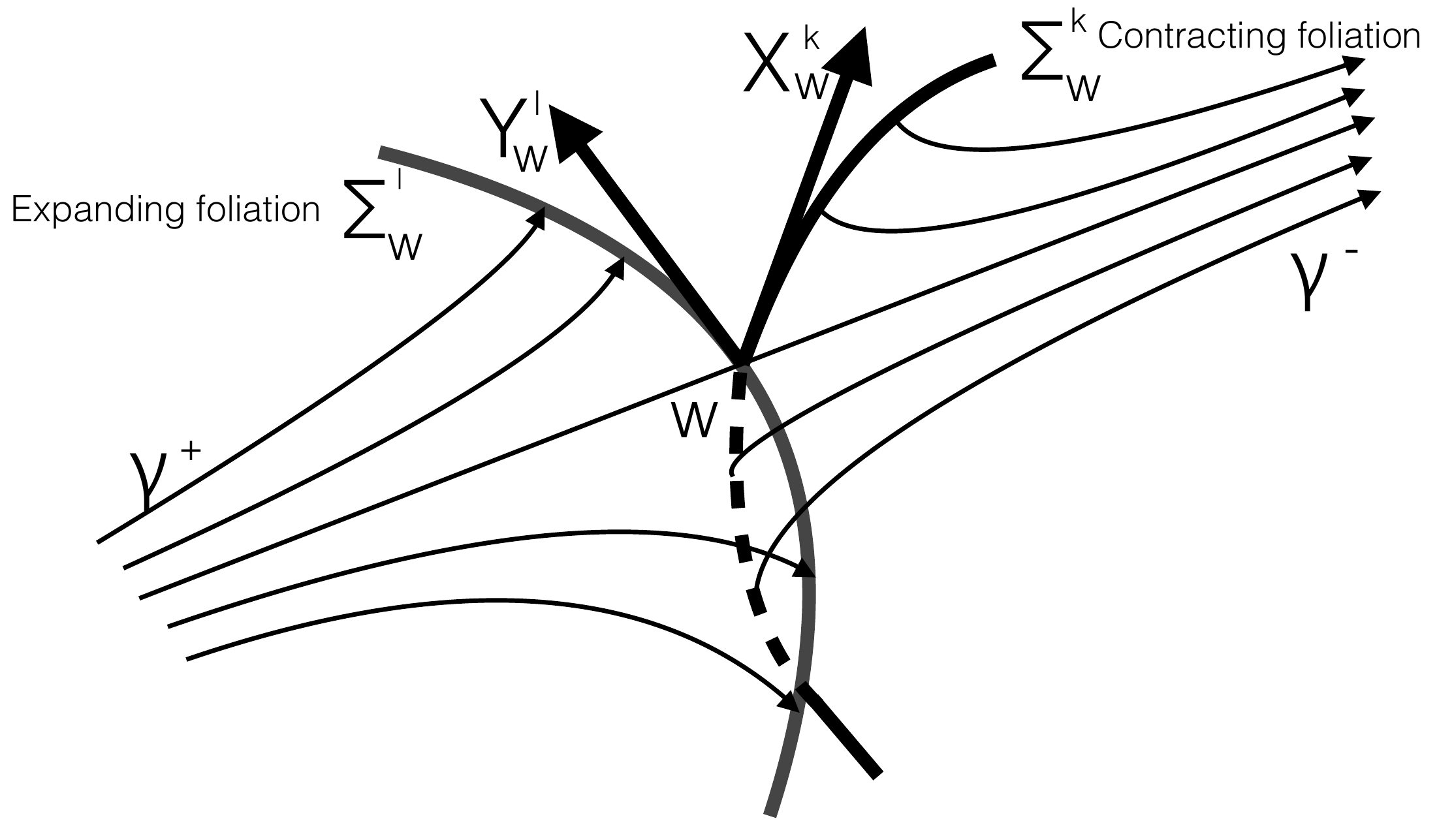}
\centering
\caption{ At each point $w$ of the C-system the tangent space $R^{m}_{w}$  
is  decomposable  into a direct sum of two linear spaces  $Y^l_{w} $ and $X^k_{w} $. 
The expanding and contracting geodesic flows
are $\gamma^+$ and $\gamma^-$. The expanding and 
contracting invariant  foliations  $\Sigma^l_{w} $ and $\Sigma^k_{w} $ 
{\it are transversal to the geodesic flows} and their corresponding tangent spaces 
are  $Y^l_{w} $ and $X^k_{w} $.  } 
\label{fig13}
\end{figure}
The systems with discrete time \cite{anosov} is defined as a 
 {\it cascade}  on the d-dimensional compact phase space $M^{d}$ is induced by 
the diffeomorphisms $T: M^d \rightarrow M^d$.  The iterations
are defined by  a repeated action of the operator  
$\{ T^n, -\infty < n < +\infty  \}$, where $n$ is an  integer number.  
The tangent space at 
the point $x \in M^d$ is  denoted by $R^d_{x}$ and the 
tangent vector bundle by $\CR(M^d)$. 
The diffeomorphism $\{T^n\}$ induces the mapping
of the tangent spaces $\tilde{T}^{n}: R^d_x \rightarrow R^d_{ T^{n}x}$.
The C-condition requires that the tangent space $R^{d}_{x}$ at each point $x$
of the d-dimensional phase space $M^{d}$ of the dynamical system $\{T^{n}\}$ 
should be decomposable  into a 
direct sum of the two linear spaces  $X^{k}_{x}$ and $Y^{l}_{x}$ with the following 
properties \cite{anosov}:
\bea\label{ccondition}
C1.&R^{d}_{x}= X^{k}_{x} \bigoplus Y^{l}_{x} ~~\\
C2.&~~a) \vert \tilde{T}^{n} \xi  \vert  \leq ~ a \vert   \xi \vert e^{-c n}~ for ~n \geq 0 ; ~
\vert \tilde{T}^{n} \xi  \vert  \geq~ b \vert \xi \vert e^{-c n} ~for~ n \leq 0,~~~\xi \in  X^{k}_{x}, \nn\\
&b) \vert \tilde{T}^{n} \eta  \vert  \geq~ b \vert \eta \vert e^{c n} ~~for~ n \geq 0;~
\vert \tilde{T}^{n} \eta  \vert  \leq ~ a \vert   \eta \vert e^{c n}~ for~ n \leq 0,~~~\eta \in Y^{l}_{x},\nn
\eea
where the constants a,b and c are positive and are the same for all $x \in M^d$ and all 
$\xi \in  X^{k}_{x}$, $\eta \in Y^{l}_{x}$.
The length $\vert ...\vert$ of the tangent vectors   $\xi $ and $  \eta $  
is defined by the Riemannian metric on $M^d$.
The linear spaces $X^{k}_{x}$ and $Y^{l}_{x}$ are invariant  with respect to 
the derivative  mapping  $\tilde{T}^{n} X^{k}_{x} = X^{k}_{T^n x}, ~
\tilde{T}^{n} Y^{l}_{x} = Y^{l}_{T^n x}$ and represent the {\it contracting and expanding 
linear spaces} (see Fig.\ref{fig13}).
The C-condition describes the behaviour of all trajectories $\tilde{T}^n \omega$ 
on the tangent vector bundle  $\omega \in R^{d}_{x}$. 
Anosov proved that the vector spaces  $X^{k}_{x}$ and $Y^{l}_{x}$ are continuous 
functions of the coordinate $x$ and that they are the target vector spaces  to 
the foliations $\Sigma^k$ and $\Sigma^l$ which are  the {\it  surfaces transversal to 
the trajectories}  $T^n x$ on $M^d$ (see Fig.\ref{fig13}). 
The contracting and expanding foliations  $\Sigma^k_x$ and $\Sigma^l_x$  are invariant 
with respect to the cascade $T^n$ in the sense that,  under the action of 
these transformations 
a foliation transforms into a foliation  \cite{anosov}.

\section{\it Appendix B}

\begin{figure}
 \centering
\includegraphics[width=7cm]{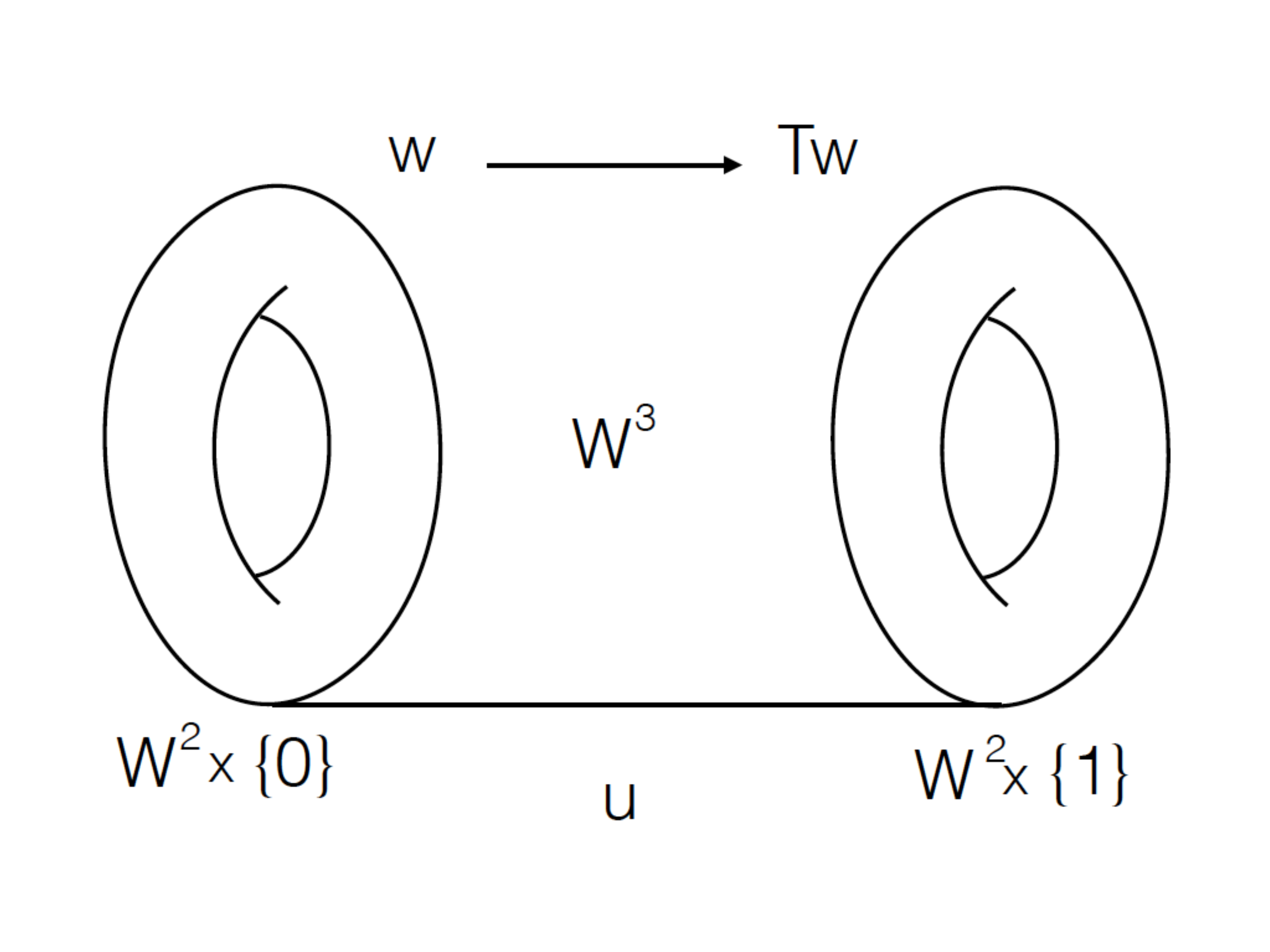}
\caption{The  identification of the  $W^2 \times \{0\}$ with $W^2 \times \{1\}$ by the 
formula $(w,1) \equiv  (Tw,0)$ of a cylinder $W^2 \times [0,1]$, where $[0,1]= 
\{  u~ \vert ~ 0 \leq u \leq 1 \}$.
The resulting compact manifold $W^{3}$ has a bundle structure 
with the base $S^1$ and fibres of the type $W^2$. The manifold $W^{3}$ 
has the local coordinates $\tilde{w}=(w^1,w^2,u)$ . 
} 
\label{fig14}
\end{figure}
In \cite{anosov} Anosov demonstrated how any C-cascade on a torus can be
embedded into a certain  C-flow. The embedding was defined by the identification 
(\ref{identification}) and the corresponding  C-flow on a smooth Riemannian manifold 
$W^{m+1}$ with the metric (\ref{metric}) was defined by the equations (\ref{velo}).
We are interested here to analyse the {\it geodesic flow} on the same Riemannian manifold $W^{m+1}$. 
The geodesic flow has different dynamics (\ref{geodesicflow} ) and  as we shall demonstrate below 
has very interesting hyperbolic components different from (\ref{velo}). 

Let us consider a C-cascade on a torus $W^m$ and increase its dimension m  by one unit
constructing a cylinder $W^m \times [0,1]$, where $[0,1]= \{  u~ \vert ~0 \leq u \leq 1 \}$,
and identifying $W^m \times \{0\}$ with $W^m \times \{1\}$ by the formula:
\be\label{identification}
(w,1) \equiv  (Tw,0).
\ee
Here T is diffeomorphism:
\bea\label{cmap1}
 w^i \rightarrow \sum T_{i,j} w^j,~~~~(mod ~1).
\eea
The resulting compact Riemannian manifold $W^{m+1}$ has a bundle structure
with the base $S^1$ and fibres of the type $W^m$. The manifold $W^{m+1}$
has the local coordinates $\tilde{w}=(w^1,...,w^m,u)$ shown on Fig.\ref{fig14}.
The C-flow $T^t$ on the manifold $W^{m+1}$ is defined by the equations \cite{anosov}
\be\label{velo}
{d  w^1 \over d t}=0~, ....,~ {d  w^m  \over d t} = 0,~ {d  u \over d t}=1.
\ee
For this flow the tangent space $R^{m+1}_{\tilde{w}}$ can be represented as
a direct sum of three subspaces:  contracting  and  expanding
 linear spaces  $X^k_{\tilde{w}} $,$ Y^l_{\tilde{w}} $   and $ Z_{\tilde{w}}$:
 \be
R^{m+1}_{\tilde{w}} = X^k_{\tilde{w}}  \oplus  Y^l_{\tilde{w}}  \oplus  Z_{\tilde{w}}.
\ee
The linear space  $X^k_{\tilde{w}} $ is tangent to the fibre  $W^m \times u$ and is parallel
to the eigenvectors corresponding to the eigenvalues which are lying inside the unit circle  $0 <  \vert \lambda_{\alpha} \vert   < 1$ and $Y^l_{\tilde{w}} $ is tangent to the fibre  $W^m \times u$ and is parallel to the eigenvectors  corresponding to the eigenvalues
which are lying outside of the unit circle $1 <\vert \lambda_{\beta}\vert$. $ Z_{\tilde{w}}$ is collinear to the phase space velocity (\ref{velo}). Under the derivative mapping of the (\ref{velo}) the vectors (\ref{eigenvectros}) from $X^k_{\tilde{w}} $ and  $Y^l_{\tilde{w}} $ are contracting  and expanding:
\be
\vert \tilde{T}^{t} v_1 \vert =  \lambda_2^{ t}~ \vert v_1 \vert,~~~~
\vert \tilde{T}^{t} v_2 \vert =  \lambda_1^{ t}~ \vert v_2 \vert.
\ee
This identification of contracting and expanding spaces proves  that (\ref{velo})
indeed defines a C-flow \cite{anosov}.

It is also interesting to analyse the {\it geodesic flow} on a  Riemannian manifold
$W^{m+1}$. The equations for the geodesic flow on $W^{m+1}$
\be
{d^2 \tilde{w}^{\mu} \over d t^2} +\Gamma^{\mu}_{\nu \rho}
{d  \tilde{w}^{\nu} \over d t } {d  \tilde{w}^{\rho} \over d t } =0
\ee
are different from the flow equations
defined by the equations (\ref{velo}) and our goal is to learn if the geodesic flow
has also the properties of the C-flow. The answer to this  question is not obvious and
requires investigation of the curvature structure of the manifold $W^{m+1}$. If all sectional
curvatures  are negative then geodesic flow defines a C-flow \cite{anosov}. For simplicity 
let us consider the automorphisms of  a two-dimensional torus which is defined by the $2 \times 2 $ matrix $T(2,0)$ (\ref{eq:matrix}).
The metric on the corresponding manifold $W^3$ can be defined as \cite{arnoldavez}
\be\label{metric}
ds^2 = e^{2u} [\lambda_1 d w^1 + (1-\lambda_1) d w^2]^2 +
e^{2u} [\lambda_2 d w^1 + (1-\lambda_2) d w^2]^2 +du^2 =\nn\\
 g_{\mu\nu} d\tilde{w}^{\mu} d\tilde{w}^{\nu},
\ee
where $0 < \lambda_2  < 1 <  \lambda_1$ are eigenvalues of the matrix $T(2,0)$ and
fulfil the relations $\lambda_1 \lambda_2 =1,\lambda_1+ \lambda_2=3$.
The metric  is invariant  under the transformation
\be\label{trans}
w^1 = 2 w^{'1} - w^{'2},~~~ w^2 = -w^{'1}_1 + w^{'2}, ~~~u  = u^{'}-1
\ee
and is therefore consistent with the identification (\ref{identification}). The metric
tensor has the form
\be\label{metric}
 g_{\mu\nu}(u)= \begin{pmatrix}
\lambda_1^{2 + 2 u} + \lambda_2^{2 + 2 u} & (1 - \lambda_1) \lambda_1^{1 + 2 u} + (1 - \lambda_2) \lambda_2^{1 + 2 u}& 0 \\
(1 - \lambda_1)\lambda_1^{1 + 2 u} + (1 - \lambda_2) \lambda_2^{1 + 2 u}&(1 - \lambda_1)^2 \lambda_1^{2 u} + (1 - \lambda_2)^2 \lambda_2^{2 u}&0\\
0&0&1\\
 \end{pmatrix}
 \ee
and the corresponding geodesic equations take the following form:
 \bea\label{geodesicflow}
&  \ddot{w}^1 + 2{(\lambda_1-1) \ln\lambda_1 \over \lambda_1+1} \dot{w^1}\dot{u}
-4 {(\lambda_1-1) \ln\lambda_1 \over \lambda_1+1} \dot{w^2}\dot{u}  =0
\nn\\
&  \ddot{w}^2 - 2{(\lambda_1-1) \ln\lambda_1 \over \lambda_1+1} \dot{w^2}\dot{u}
  - 4 {(\lambda_1-1) \ln\lambda_1 \over \lambda_1+1} \dot{w^1}\dot{u} =0\\
& \ddot{u} + {(1-\lambda^{4u+4}_1) \ln\lambda_1 \over \lambda^{2u+2}_1} \dot{w^1}\dot{w^1}
 + 2{(1+\lambda^{4u+3}_1)(\lambda_1-1) \ln\lambda_1 \over \lambda^{2u+2}_1} \dot{w^1}\dot{w^2}+\nn\\
& +  {(1-\lambda^{4u+2}_1)(\lambda_1-1)^2 \ln\lambda_1 \over \lambda^{2u+2}_1} \dot{w^2}\dot{w^2}=0.\nn
\eea
One can get convinced that these equations are invariant under the transformation (\ref{trans}).
 In order to study a stability of the geodesic flow one has to compute the sectional curvatures.
We shall choose the orthogonal frame in the directions of the linear spaces  $X^1_{\tilde{w}} , Y^1_{\tilde{w}} $   and $ Z_{\tilde{w}}$. The corresponding vectors are:
\be\label{eigenvectros}
v_1 = (\lambda_1-1, \lambda_1,0),~~~v_2=(\lambda_2 -1, \lambda_2,0),~~~ v_3=(0,0,1)
\ee
and in the metric (\ref{metric}) they have the lengths:
\be
\vert v_1  \vert^2= (\lambda_1 - \lambda_2)^2 \lambda_2^{2 u},~~~~
\vert v_2  \vert^2= (\lambda_1 - \lambda_2)^2 \lambda_1^{2 u},~~~~
\vert v_3 \vert^2 = 1.
\ee
The corresponding sectional curvatures are:
\bea
K_{12} = {R_{\mu\nu\lambda\rho} v^{\mu}_1 v^{\nu}_2  v^{\lambda}_1 v^{\rho}_2 \over
\vert v_1 \wedge v_2 \vert^2}=   \ln^2 \lambda_1
\nn\\
K_{13} = {R_{\mu\nu\lambda\rho} v^{\mu}_1 v^{\nu}_3  v^{\lambda}_1 v^{\rho}_3 \over
\vert v_1 \wedge v_3\vert^2}= -  \ln^2 \lambda_2
\\
K_{23} = {R_{\mu\nu\lambda\rho} v^{\mu}_2 v^{\nu}_3  v^{\lambda}_2 v^{\rho}_3 \over
\vert v_2 \wedge v_3 \vert^2}= -  \ln^2 \lambda_1. \nn
\eea
It follows from the above equations that the geodesic
flow is exponentially unstable on the planes (1,3) and (2,3)
and is stable in the plane (1,2). This behaviour is dual to the flow (\ref{velo}) which
is unstable in (1,2) plane and is stable in (1,3) and (2,3) planes.
 The scalar curvature is
\be
R= R_{\mu\nu\lambda\rho} g^{\mu\lambda}g ^{\nu\rho} = 2(K_{12} +K_{13}+K_{23}) =   - 2 
\ln^2 \lambda_1 = -2 h(T)^2 ,
\ee
where $h(T)$ is the entropy of the automorphism $T(2,0)$.

\section{\it Appendix C}

In a typical computer implementation of the automorphism \eqref{cmap1} 
the initial vector will have rational 
components $u_i=a_i/p$, where $a_i$ and $p$ are natural numbers.  
Therefore it is convenient to represent $u_i$ by its numerator $a_i$ in computer memory and define the iteration in terms of $a_i$
\cite{mixmaxGalois}:
\be
\label{eq:recP}
a_i \rightarrow\sum_{j=1}^N T_{ij} \, a_j ~\textrm{mod}~ p .
\ee
If the denominator p is taken to be a prime number \cite{mixmaxGalois}, 
then the recursion is realised on extended 
Galois field $GF[p^N]$  \cite{niki,nied} and 
allows to find the period of the trajectories in terms of p and the properties of the 
characteristic polynomial $P(x)$ of the matrix T \cite{mixmaxGalois}. If 
the characteristic polynomial $P(x)$ of matrix $T$ is primitive in the 
extended Galois  field $GF[p^N]$, then
\cite{mixmaxGalois,nied,lnbook}
\be\label{period}
 T^q = p_0~ \mathbb{I}~~\textrm{ where}~~  q=\frac{p^N-1} {p-1} ~,
\ee
where $p_0$ is a free term of the  polynomial $ P(x)$ and is a {\it primitive element} of $GF[p]$.
Since our matrix T has $p_0=Det T= 1$, the polynomial $ P(x)$ of T cannot be primitive. 
The solution suggested  in \cite{konstantin} is to define the necessary and 
sufficient conditions for the period $q$  
to attain its maximum  are the following:
\begin{enumerate}
\item[\bf{1.}] $T^q = \mathbb{I} ~(mod~ p) $,~~~where $q=\frac{p^N-1} {p-1}$
\item[\bf{2.}] $T^{q/r} \neq \mathbb{I} ~(mod~ p)$,~~~~ for any r which is a prime divisor of q .
\end{enumerate}
The first condition is equivalent to the requirement  that the characteristic polynomial is irreducible. 
The second condition can be checked if the integer factorisation of $q$ is available \cite{konstantin}, then
the period of the sequence is equal to (\ref{period}) and is independent of the seed. 
There are precisely $p-1$ distinct  trajectories which together fill up all states of the $GF[p^N]$ lattice: 
\be
 q~ (p-1) = p^N-1.
\ee
In \cite{konstantin} the actual value of p was  taken as $p=2^{61}-1$, 
the largest Mersenne number that fits into an 
unsigned integer on current 64-bit computer architectures.  For the matrix of the 
size $N=256$ the period in that case is  $q \approx 10^{4600}$.
The algorithm  which allows the efficient implementation of the generator 
in actual computer hardware, reducing the matrix multiplication to the O(N) operations
was found in \cite{konstantin}.
The other advantage of this implementation is that it allows to make  "jumps" into 
any point on a periodic trajectory 
without calculating all previous coordinates on a trajectory, which typically has a 
very large  period $q \approx 10^{4600}$.
This MIXMAX random number generator is currently made available 
in  a portable implementation  in the C language at hepforge.org \cite{hepforge}
and was implemented into the   Geant4/CLHEP and ROOT toolkits at CERN \cite{cern,root,geant}.

\section{\it Appendix D}

The most convenient way to calculate the entropy of a C-system automorphisms on a torus 
is to integrate over the phase space the logarithm of the volume expansion rate $\lambda(w)$ 
of a $l$-dimensional infinitesimal cube which is embedded  into the foliation  $\Sigma^{l}_w$. 
The derivative map
$\tilde{T}$ maps the linear space $Y^{l}_{w}$ into the $Y^{l}_{Tw}$ and if 
the rate of expansion of the volume of the $l$-dimensional cube is  $\lambda(w)$,
then \cite{anosov,sinai3,rokhlin2,sinai4,gines}
\be\label{biuty}
h(T) = \int_{W^m} \ln \lambda(w) d w.
\ee
Here the volume of the $W^m$ is normalised to 1. For the automorphisms 
on a torus (\ref{cmap}) the coefficient $\lambda(w)$  does not depends of the phase 
space coordinates $w$ and is equal to the product of eigenvalues 
$\{  \lambda_{\beta }  \} $  with modulus  larger than one (\ref{eigenvalues}): 
\be\label{more}
\lambda(w) = \prod^{l}_{\beta=1} \lambda_{\beta}
\ee
and obtain the formula (\ref{entropyofT}) for the entropy.

\section{\it Appendix E}

The entropy defines the variety  and richness of the periodic 
trajectories of the C-systems  \cite{anosov,bowen0,bowen,bowen1}.
The C-systems have a countable set of everywhere dense  
periodic trajectories \cite{anosov}.
The $E^m$ cover of the torus $W^m$ allows 
to translate every set of points on torus into a set of points on Euclidean space $E^m$ and 
the space of functions on torus into the periodic functions on $E^m$.
To every closed curve $\gamma$ on a torus corresponds a curve  $\phi: [0,1] \rightarrow E^m$
for which $\phi(0) = \phi(1)~ mod~1$ and if $\phi(1) - \phi(0)=(p_1,...,p_m)$, then the 
corresponding winding numbers on a torus are $p_i \in Z$.

Let us fix the integer number $N$, then the points on a torus with 
the coordinates having a denominator $N$ form a finite set $\{p_1/N,...,p_m/N \}$. The 
automorphism (\ref{mmatrix}) with integer entries transform this set 
of points into itself, therefore all these points belong to periodic trajectories.
Let  $w=(w_1,...,w_m)$ be a point of a trajectory with the period $n > 1$. Then 
 \be\label{periodictrajectories1}
 T^n w = w + p,
 \ee
 where $p$ is an integer vector. The above equation with 
 respect to $w$ has nonzero determinant, therefore the components 
 of $w$ are {\it rational}. 
 
Thus the periodic trajectories of the period $n$ of the automorphism $T$ are given 
by the solution of the equation (\ref{periodictrajectories1}),
where $p \in Z^m$ is an integer vector and $w=(w_1,...,w_m) \in W^m$.  
 As $p$ varies in $Z^m$ the solutions of the equation (\ref{periodictrajectories1}) determine
 a fundamental domain $D_n$ in the covering Euclidian space $E^m$  of the volume 
 $\mu(D_n)=1/\vert Det (T^n-1) \vert $. Therefore the number of all points $N_n$ on the 
 periodic trajectories of the period $n$  
 is given by  the corresponding inverse volume \cite{smale,sinai2,margulis,bowen0,bowen}: 
 \be\label{numbers}
N_n = \vert Det (T^n-1) \vert = \vert  \prod^{m}_{i=1}(\lambda^n_i -1) \vert .
 \ee
Using the theorem of Bowen \cite{bowen,bowen1} which states that the entropy  
of the automorphism $T$ can be  represented  in terms of $N_n$
defined in   (\ref{numbers}): 
\be\label{bowen}
h(T)= \lim_{n \rightarrow \infty}{1\over n}~ \ln  N_n ~~,  
\ee
one can derive the formula for the entropy (\ref{entropyofT}) for the automorphism $T$
in terms of its eigenvalues: 
\be
h(T)= \lim_{n \rightarrow \infty}{1\over n} \ln (\vert  \prod^{m}_{i=1}(\lambda^n_i -1) \vert) =
\sum_{\vert \lambda_{\beta} \vert > 1} \ln \vert \lambda_{\beta} \vert.
\ee
Let us now define the number of periodic trajectories of the period  $n$  by $\pi(n)$.
Then the number of all points $N_n$ on the 
 periodic trajectories of the period $n$  can be written in the following form:
 \be
 N_n = \sum_{l~ divi~ n  } l ~\pi(l) ,
 \ee
where $l$ divides $n$. Using again the Bowen result (\ref{bowen}) one can get 
\be\label{assimpto}
N_n =\sum_{l~ divi~ n  } l \pi(l) \sim e^{n h(T)}.
\ee
This result can be rephrased as a statement that the number of points 
on the periodic trajectories of the period n exponentially 
grows  with the entropy.

Excluding the periodic trajectories which divide n 
(for example $T^n w=T^{l_2}(T^{l_1}w)$, where $n=l_1 l_2$ and $T^{l_i}w =w$)  
one can get the number of periodic trajectories of period n 
which are not divisible.
For that one should represent  the $\pi(n)$ in the following form:
\be\label{density}
\pi(n) = {1\over n} \big( \sum_{l~ divi~ n  } l ~\pi(l) -
\sum_{l~ divi~ n,~ l <n  } l ~\pi(l) \big)
\ee
and from (\ref{density})  and (\ref{assimpto}) it follows that 
\be
\pi(n) \sim { e^{n h(T)} \over n}  \big( 1   -
{\sum_{l~ divi~ n,~ l <n  } l ~\pi(l) \over \sum_{l~ divi~ n  } l ~\pi(l)} \big) \sim { e^{n h(T)} \over n}, 
\ee
because  the ratio in the bracket is strictly smaller than one. This result tells that a
system with larger entropy  $\Delta h = h(T_1) - h(T_2) >0$ is more densely populated by the 
periodic trajectories of the same period $n$:
\be
{\pi_1(n) \over \pi_2(n)} \sim e^{n\ \Delta h}.
\ee
The next important result of the Bowen theorem \cite{bowen,bowen1} states that 
\be\label{ation9}
\int_{W^m} f(w) d\mu(w) = \lim_{n \rightarrow \infty}  {1\over N_n} \sum_{ w \in \Gamma_n} f(w),
\ee
where $\Gamma_n$ is a set of all points on the trajectories of period
$n$. 
The total number of points in the set $\Gamma_n$ we defined earlier as $N_n$.

This result has important consequences for the calculation of the 
integrals on the manifold $W^m$, because, as it follows from (\ref{ation9}), the integration 
reduces to the summation 
over all points of periodic trajectories. It is appealing to consider periodic trajectories 
of the period $n$ which is a prime number. Because every infinite subsequence of convergent  
sequence converges to the same limit we can consider in (\ref{ation9}) only terms 
with the prime periods.  In that case $N_n = n \pi(n)$ and the above formula becomes:
\be\label{integral}
\int_{W^m} f(w) d\mu(w) = \lim_{n \rightarrow \infty}  {1\over n \pi(n)} \sum^{\pi(n)}_{j=1}
\sum^{n-1}_{ i=0} f(T^i w_{j}),
\ee
where the summation is over all points of the trajectory $T^i w_{j}$ and  over all 
distinct  trajectories of period n which are enumerated by index $j$. The $w_{j}$ is the initial point of the 
trajectory $j$ \footnote{It appears to be a difficult mathematical problem to decide whether two  vectors 
$w_{1}$ and $w_{2}$ belong to the same or to distinct trajectories.}.  
From the above consideration it follows that the convergence is guaranteed 
if one sums over all trajectories of the same period $n$.    One can conjecture     
that all $\pi(n)$ trajectories at the very large period $n$  contribute  
equally into the sum (\ref{integral}),  
therefore the integral (\ref{integral}) can be reduced to a  sum over fixed trajectory
\be\label{reduce}
  {1\over n } \sum^{n-1}_{ i=0} f(T^i w).
\ee

\vfill

\end{document}